\documentclass[a4paper,twoside]{IEEEtran}

\pdfoutput=1

\usepackage[T1]{fontenc}
\usepackage{lineno}

\usepackage{amstext}
\usepackage{dsfont}
\usepackage{cite}
\usepackage[colorlinks=true, linkcolor=blue, urlcolor=blue,citecolor=blue]{hyperref}

\usepackage{orcidlink}

\usepackage{float}
\usepackage[centertags]{amsmath}
\usepackage{latexsym}
\usepackage{amsthm}
\usepackage{amssymb}
\usepackage{amsfonts}
\usepackage{amsbsy}
\usepackage[shortlabels]{enumitem}
\usepackage{url}
\usepackage{multirow}
\usepackage{xcolor}

\usepackage{bm}
\usepackage{balance}

\usepackage{tikz}

\usepackage{soul}

\def\rank{\mathop{\rm rank}}
\def\tr{\mathop{\rm tr}\nolimits}
\def\PsiB{\Psi_{\mathcal{B}}}

\def\perpE{{\perp_{\rm E}}}
\def\perpH{{\perp_{\rm H}}}
\def\perpS{{\perp_{\rm S}}}
\def\perpT{{\perp_{\rm T}}}

\def\wgt{\mathop{\rm wgt}}
\def\swt{\mathop{\rm swt}}
\def\dS{d_{\rm S}}

\def\ie{\textit{i.\,e.}}
\def\eg{\textit{e.\,g.}}

\theoremstyle{plain}
\newtheorem{thm}{Theorem}[section]
\newtheorem{lemma}[thm]{Lemma}
\newtheorem{prop}[thm]{Proposition}

\theoremstyle{definition}
\newtheorem{rem}[thm]{Remark}
\newtheorem{ex}[thm]{Example}
\numberwithin{equation}{section}

\newcommand{\Fq}{\mathbb{F}_{q}}

\newcommand{\C}{\mathbb{C}}
\newcommand{\F}{{\mathbb F}}
\newcommand{\G}{{\mathbb G}}
\newcommand{\HH}{{\mathbb H}}
\newcommand{\Tr}{{\rm Tr}}
\newcommand{\Span}{{\rm Span}}

\begin{document}

\title{Characterization of Nearly Self-Orthogonal Quasi-Twisted Codes and Related Quantum Codes}

\author{
  Martianus Frederic Ezerman${}^{\orcidlink{0000-0002-5851-2717}}$,
  Markus Grassl${}^{\orcidlink{0000-0002-3720-5195}}$, \IEEEmembership{Senior Member, IEEE},
  San Ling${}^{\orcidlink{0000-0002-1978-3557}}$, \IEEEmembership{Senior Member, IEEE},
  Ferruh~\"Ozbudak${}^{\orcidlink{0000-0002-1694-9283}}$, and
  Buket~\"Ozkaya${}^{\orcidlink{0000-0003-2658-5441}}$
  \thanks{Part of this work contains results presented at the 2019
    IEEE International Symposium on Information Theory \cite{ISIT19}.}
  \thanks{Martianus Frederic Ezerman and San Ling are with the School
    of Physical and Mathematical Sciences, Nanyang Technological
    University, 21 Nanyang Link, Singapore 637371, Singapore.}
  \thanks{Markus Grassl is with the International Centre for Theory of
    Quantum Technologies, University of Gdansk, 80-309 Gda\'nsk,
    Poland.}
  \thanks{Ferruh \"Ozbudak is with the Department of Mathematics,
    Faculty of Engineering and Natural Sciences, Sabanc{\i} University, 34956 \.{I}stanbul, Turkey.}
  \thanks{Buket \"Ozkaya is with the Institute of Applied Mathematics,
    Middle East Technical University, 06800 Ankara, Turkey.}
  \thanks{This work has been submitted to the IEEE for possible
    publication. Copyright may be transferred without notice, after
    which this version may no longer be accessible.}

}

\maketitle

\begin{abstract}
  Quasi-twisted codes are used here as the classical ingredients in
  the so-called Construction X for quantum error-control codes. The
  construction utilizes nearly self-orthogonal codes to design quantum
  stabilizer codes. We expand the choices of the inner product to also
  cover the symplectic and trace-symplectic inner products, in
  addition to the original Hermitian one. A refined lower bound on the
  minimum distance of the resulting quantum codes is established and
  illustrated. We report numerous record breaking quantum codes from
  our randomized search for inclusion in the updated online database.
\end{abstract}


\section{Introduction}\label{intro}

\IEEEPARstart{C}{entral} to building quantum computers of significant
scale is the ability to control quantum errors. The most widely
studied class of quantum error-correcting codes is that of the quantum
stabilizer codes.  It is now well-established that they can be
designed via classical additive or linear codes over finite fields
when certain orthogonality conditions are met. The literature on
stabilizer codes is large and can be traced back to seminal works done
around thirty years ago. Notable among them, Calderbank, Rains, Shor,
and Sloane in \cite{Calderbank1998} firmly established the connection
between the group of quantum error operators and classical coding
theory. The stabilizer formalism, introduced slightly earlier by
Gottesman in his PhD thesis \cite{Gottesman97}, then came to the
attention of wider communities of researchers beyond the initial
circle of pioneers in quantum information theory. The research area of
quantum error control has since grown tremendously. The article by
Ketkar, Klappenecker, Kumar, and Sarvepalli \cite{Ketkar} provides an
overview of various constructions of stabilizer codes.  A more recent
exposition highlighting links between quantum mechanics and discrete
mathematics is given by Grassl in \cite{Grassl2021}.

Classical linear cyclic codes and their generalizations, such as
nega-cyclic, quasi-cyclic, and quasi-twisted codes, have been fruitful
as classical ingredients in constructing quantum codes.  Cyclic-type
codes have nice algebraic structures that allow for characterization
of their orthogonality under suitable inner products.  Since there are
too many relevant papers on the topic to mention in this brief
introduction, we refer interested readers to the works of Galindo,
Hernando, and Matsumoto in \cite{Galindo2018} and its references as
starting points.

As derived by Lally in \cite[Section 2]{Lally03}, every quasi-cyclic
code corresponds to an additive cyclic code over the extension field
whose extension degree is the index of the quasi-cyclic code.  This
correspondence is then used by G\"uneri, \"Ozdemir, and Sol\'{e} in
\cite{Guneri2018} to formally establish that an index $\ell$, length
$m \ell$ quasi-cyclic code over $\mathbb{F}_q$ can be written as a
cyclic code of length $m$ over $\mathbb{F}_{q^{\ell}}$ via a basis of
the extension $\mathbb{F}_{q^{\ell}}$ over $\mathbb{F}_q$. The cyclic
code is only linear over $\mathbb{F}_q$, making it an additive cyclic
code over the extension field.

In the setting of this article, we mainly consider classical
quasi-cyclic/quasi-twisted codes of length $n$ over the field
$\F_{q^2}$ and quasi-cyclic/quasi-twisted codes of length $2n$ over
$\F_q$ using their symplectic representation.  In the literature, one
can also find both approaches.  The quantum twisted codes of
Bierbrauer and Edel in \cite{BE}, for example, are presented as coming
from additive cyclic codes of length $n$.

Our focus in this work is on Quantum Construction X, which was
originally proposed by Li\v{s}onek and Singh for the qubit setup in
\cite{LS01}. It starts with a nearly Hermitian self-orthogonal
classical linear code over $\mathbb{F}_{4}$ and uses the information
about the code's Hermitian hull to extend the length in such a way
that the extended code is self-orthogonal.  A qubit code can then be
designed and its parameters be inferred accordingly.  Already now, we
note that---unlike the situation for classical codes---Quantum
Construction X does not allow the use of an arbitrary auxiliary code
for the extension (see the main text for the choice of the auxiliary
code).  Generalization to $\mathbb{F}_{q^2}$-linear classical codes to
design $q$-ary quantum codes follows soon afterwards. Quantum
Construction X applied to cyclic and quasi-twisted classical codes as
ingredients yields many improved quantum codes, affirming its efficacy
as a general construction route.

This work continues along this route. We report the following contributions.
\begin{itemize}
	\item We generalize the Li\v{s}onek-Singh construction further
          by providing a complete treatment on nearly self-orthogonal
          quasi-twisted codes, covering the Euclidean, Hermitian, and
          (trace) symplectic inner products.  We provide constructive
          proofs for the results. This also results in
          improved bounds on the parameters.
        \item By examining the algebraic structure of quasi-twisted
          codes in combination with the respective inner products, we
          show how to design the dimension of the hulls.
	\item We use the gained insights to devise a randomized search
          for suitable quasi-twisted codes over small quadratic
          fields. We gladly report that the search has found numerous
          record-breaking $q$-ary quantum codes for $q \in
          \{2,3,4,5,7,8\}$. The codes, upon application of propagation
          rules, lead to many more improved codes.
\end{itemize}

This paper is organized as follows. Section \ref{sec:basics} provides
the necessary background on how quantum error-correcting codes can be
constructed using techniques from algebraic coding theory.  In Section
\ref{sec:ConstructionX}, Quantum Construction X is explained in
detail.  In Section \ref{sec:QT_section}, quasi-twisted codes and
their algebraic structure are studied in order to construct Hermitian
self-orthogonal or nearly Hermitian self-orthogonal codes. An
analogous investigation is carried out in Section
\ref{sec:QT_section-2} with respect to the symplectic inner
product. The codes are then used as main ingredients in Quantum
Construction X. Section \ref{sec:findings} presents the outcome of our
randomized search which yields many new parameters of quantum codes
improving the best-known ones.  Section \ref{sec:final_remarks}
contains some final remarks.  The two appendices explain the
relations among the inner products and algorithms to compute suitable
bases of vector spaces depending on a given inner product,
respectively.

\section{Background and Notation}\label{sec:basics}
Let $\F_q$ and $\F_{q^2}$ denote the finite fields with $q$ and $q^2$
elements, respectively, where $q$ is a prime power. For positive
integers $n$ and $k$ such that $k\leq n$, a \emph{linear code} $C$ of
length $n$ and dimension $k$ over $\F_{q^2}$ is a $k$-dimensional
subspace of $\F_{q^2}^n$. A linear code $C\subseteq \F_{q^2}^n$ is
said to have parameters $[n, k, d]_{q^2}$ if $C$ has length $n$,
dimension $k$, and minimum Hamming distance $d =
\min\bigl\{\wgt(\bm{c})\colon \bm{c}\in C\setminus\{\bm{0}\}\bigr\}$,
where $\wgt(\bm{c})$ denotes the Hamming weight of $\bm{c}\in C$. For
any non-empty set $S\subseteq\F_q^n$, we define its minimum non-zero
Hamming weight $\wgt(S)=\min\bigl\{\wgt(\bm{x})\colon\bm{x}\in
S\setminus\{\bm{0}\}\bigr\}$ and its minimum Hamming distance
$d(S)=\min\{\wgt(\bm{x}-\bm{y})\colon\bm{x},\bm{y}\in
S\mid\bm{x}\ne\bm{y}\}$.
(If the argument of the minimum is the empty
set, we define the quantities to be equal to $n$.)

For codes $C\subseteq\F_q^{2n}$ of even length, we split the vector
as $\bm{c}=(\bm{a}|\bm{b})$ with $\bm{a},\bm{b}\in\F_q^n$ and define
the \emph{symplectic weight} as
\begin{equation}
  \swt\bigl((\bm{a}|\bm{b})\bigr)=|\{i\colon i\in\{0,\ldots,n-1\}\mid(a_i,b_i)\ne(0,0)\}|.\label{eq:swt}
\end{equation}
For a set $S\subseteq\F_2^{2n}$, we denote the \emph{symplectic
minimum distance} by $\dS(S)=\min\{\swt(\bm{x}-\bm{y})\colon\bm{x},\bm{y}\in S
\mid \bm{x}\ne\bm{y}\}$.

An additive code $C$ over the alphabet $\F_q$ with $q=p^m$, $p$ prime,
is denoted by $(n,p^k,d)_{p^m}$. It is a set of vectors of length $n$
over the alphabet $\F_q$ that is closed with respect to addition, \ie,
an $\F_p$-linear subspace. Hence, its cardinality $p^k$ is an integral
power of $p$, but in general may not be an integral power of $q$.

\subsection{Inner Products and Dual Codes}
We consider duality with respect to various inner products. On the
space $\F_q^n$, the \emph{Euclidean inner product} is defined as
\begin{equation}\label{eq:Euclidean_IP}
  \bm{u}\cdot\bm{v}=\langle \bm{u},\bm{v}\rangle_{\rm E} := \sum_{i=0}^{n-1}u_iv_i,
\end{equation}
where
$\bm{u}=(u_0,u_1,\ldots,u_{n-1}),\bm{v}=(v_0,v_1,\ldots,v_{n-1})\in\F_q^n$. Given
an $[n,k,d]_q$-code $C$, its \emph{(Euclidean) dual} $C^\perpE=C^\perp$ is a
linear code of length $n$ and dimension $n-k$ defined as
\begin{equation}\label{eq:Euclidean_dual}
C^\perpE:=\{\bm{b}\in\F_q^n \mid \langle \bm{c},\bm{b}\rangle_{\rm E}=0, \forall \bm{c}\in C\}.
\end{equation}
The \emph{Euclidean hull} of $C$ is defined to be $C\cap C^\perpE$. A
code $C$ is \emph{Euclidean self-orthogonal} if it is equal to its
Euclidean hull, or equivalently, if $C\subseteq C^\perpE$, \ie, $C$ is
contained in its Euclidean dual.

For finite fields of size $q^2$, we define the
\emph{Hermitian inner product} on $\F_{q^2}^n$ as
\begin{equation}\label{eq:Hermitian_IP}
  \langle \bm{u},\bm{v}\rangle_{\rm H} := \sum_{i=0}^{n-1} u_iv_i^q.
\end{equation}
The \emph{Hermitian dual} $C^\perpH$ is given by
\begin{equation}\label{eq:Hermitian_dual}
C^\perpH:=\{\bm{b}\in\F_{q^2}^n \mid \langle \bm{c},\bm{b}\rangle_{\rm H}=0, \forall \bm{c}\in C\}.
\end{equation}
The \emph{Hermitian hull} of $C$ is defined to be $C\cap
C^\perpH$. The code $C$ is \emph{Hermitian self-orthogonal} if it is
equal to its Hermitian hull, or equivalently, if $C\subseteq
C^\perpH$, \ie, $C$ is contained in its Hermitian dual.  The conjugate
transpose of a matrix $M=\left(M_{i,j}\right)\in\F_{q^2}^{m\times n}$
is the matrix $M^*=\left(M_{j,i}^q\right)\in\F_{q^2}^{n\times m}$.

On the space $\F_q^{2n}$, we define the \emph{symplectic inner
product}
\begin{alignat}{5}
  \bigl\langle (\bm{a}|\bm{b}), (\bm{u}|\bm{v})\bigr\rangle_{\rm S}
    :={}&\langle\bm{a},\bm{v}\rangle_{\rm E} - \langle\bm{b},\bm{u}\rangle_{\rm E}\\
     ={}&\sum_{i=0}^{n-1} a_iv_i- b_iu_i,\label{eq:symplectic_IP}
\end{alignat}
where $\bm{a},\bm{b},\bm{u},\bm{v}\in\F_q^n$.  The \emph{symplectic
dual} $C^\perpS$, the \emph{symplectic hull} $C \cap C^\perpS$, and a
\emph{symplectic self-orthogonal} code $C\subseteq C^\perpS$ are
defined analogously.

Finally, for $q=p^m$, $p$ prime, the \emph{trace-symplectic inner
product} on $\F_{p^m}^{2n}$ is defined as
\begin{alignat}{5}
  \bigl\langle (\bm{a}|\bm{b}), (\bm{u}|\bm{v})\bigr\rangle_{\rm T}
    :={}& \tr\Bigl(\bigl\langle(\bm{a}|\bm{b}), (\bm{u}|\bm{v})\bigr\rangle_{\rm E}\Bigr)\label{eq:tracesymplectic_IP1}\\
  ={}&\sum_{i=0}^{n-1} \tr\bigl(a_iv_i- b_iu_i\bigr),\label{eq:tracesymplectic_IP2}
\end{alignat}
where $\bm{a},\bm{b},\bm{u},\bm{v}\in\F_{p^m}^n$ and
$\tr(x)=\sum_{i=0}^{m-1}x^{p^i}$ denotes the absolute trace of the field
extensions $\F_{p^m}/\F_p$. The \emph{trace-symplectic dual}
$C^\perpT$ is an additive code over $\F_{p^m}$, \ie, it is
$\F_p$-linear.

In Appendix \ref{sec:IP_relations} we discuss several relations among
these inner products.

\subsection{Quantum Codes}

A quantum error-correcting code $\mathcal{Q}$, denoted by
$(\!(n,K,d)\!)_q$ or $[\![n,k,d]\!]_q$, is a subspace of the $n$-fold
tensor product $(\C^q)^{\otimes n}$ of the complex vector space
$\C^q$. The code has dimension $\dim(\mathcal{Q})=K=q^k$ and minimum
distance $d$, \ie, any error acting on at most $d-1$ of the
tensor factors can be detected or has no effect on the code. For more
on quantum codes, see, for example, \cite{Grassl2021}.  Here we focus
on the so-called stabilizer codes based on codes over finite fields,
see \cite{Ketkar}.

The following construction can, \eg, be found in \cite[Theorem
  13]{Ketkar}.
\begin{prop}[trace-symplectic construction]\label{prop:trace_symp_constr}
Let $C\subseteq\F_q^{2n}$ be a trace-symplectic self-orthogonal additive
code, \ie, $C\subseteq C^\perpT$, of size $|C|=q^n/K$.  Then there
exists an $(\!(n,K,d(\mathcal{Q}))\!)_q$ quantum stabilizer code
$\mathcal{Q}$ with
\begin{alignat}{5}
  d(\mathcal{Q})=\begin{cases}
    \swt(C^\perpT),&\text{when $C=C^\perpT$};\\
    \swt(C^\perpT\setminus C)\ge\swt(C^\perpT),&\text{when $C\ne C^\perpT$}.
  \end{cases}
\end{alignat}
For $C\ne C^\perpT$, \ie, $K>1$, the code is called \emph{impure}
whenever $d(\mathcal{Q})>\swt(C^\perpT)$; otherwise the code is called
\emph{pure}.
\end{prop}
Note that the cardinality of an additive ($\F_p$-linear) code is an
integral power of $p$, but need not be an integral power of
$q$. Hence, the dimension of the resulting quantum code $\mathcal{Q}$
need not be an integral power of $q$ either.

The fact that $\tr(\gamma x)=0$ for all $\gamma\in\F_q$ is equivalent
to $x=0$, together with \eqref{eq:tracesymplectic_IP2} imply that
$C^\perpT=C^\perpS$ for $\F_q$-linear codes. Hence, the following is a
special case of the trace-symplectic construction.
\begin{prop}[symplectic construction]\label{prop:symplectic_constr}
Let $C\subseteq\F_q^{2n}$ be an $\F_q$-linear symplectic
self-orthogonal code with parameters $[2n,n-k]_q$.  Then there
exists an $[\![n,k,d(\mathcal{Q})]\!]_q$ quantum stabilizer code
$\mathcal{Q}$  with
\begin{alignat}{5}
  d(\mathcal{Q})=\begin{cases}
    \swt(C^\perpS),&\text{when $C=C^\perpS$};\\
    \swt(C^\perpS\setminus C)\ge\swt(C^\perpS),&\text{when $C\ne C^\perpS$}.
  \end{cases}
\end{alignat}
\end{prop}

A special case of the symplectic construction yields CSS codes
\cite{CalderbankShor,Steane1996}, see also \cite[Lemma 20]{Ketkar}.
\begin{prop}[CSS construction]\label{prop:CSS_constr}
Let $C_1,C_2\subseteq\F_q^n$ be two linear codes with parameters
$[n,k_1,d_1]_q$ and $[n,k_2,d_2]_q$ with $C_2^\perpE\subseteq C_1$.
Then there exists an $[\![n,k_1+k_2-n,d(\mathcal{Q})]\!]_q$ quantum
stabilizer code $\mathcal{Q}$ with
\begin{alignat}{5}
  d(\mathcal{Q})=\begin{cases}
    \min\{\wgt(C_1),\wgt(C_2)\},&\text{when $C_2^\perpE=C_1$};\\
    \min\{\wgt(C_1\setminus C_2^\perpE),\\
    \phantom{\min\{}
    \wgt(C_2\setminus C_1^\perpE)\},&\text{when $C_2^\perpE\ne C_1$}.
  \end{cases}
\end{alignat}
\end{prop}
\begin{IEEEproof}
  We apply the symplectic construction of Proposition
  \ref{prop:symplectic_constr} to the code $C=C_2^\perpE\times
  C_1^\perpE$ and compute $d(\mathcal{Q})$ accordingly.
\end{IEEEproof}

Most of our examples use $\F_{q^2}$-linear codes. For those, we can,
\eg, use more efficient algorithms to compute their minimum distance.
The construction of quantum codes is based on the following result.
\begin{prop}[Hermitian construction]\label{prop:Hermitian_constr}
  Let $C\subseteq\F_{q^2}^{n}$ be an $\F_{q^2}$-linear code with
    parameters $[n,k]_{q^2}$ that is Hermitian self-orthogonal, \ie,
    $C\subseteq C^\perpH$.  Then there exists an
  $[\![n,n-2k,d(\mathcal{Q})]\!]_q$ quantum stabilizer code
  $\mathcal{Q}$ with
\begin{alignat}{5}
  d(\mathcal{Q})=\begin{cases}
    d(C^\perpH),&\text{when $C=C^\perpH$};\\
    d(C^\perpH\setminus C)\ge d(C^\perpH),&\text{when $C\ne C^\perpH$}.
  \end{cases}
\end{alignat}
\end{prop}
\begin{IEEEproof}
We apply the inverse map $\Phi^{-1}$ (see \eqref{eq:Phi_vector})
to the code $C$, \ie, we expand the vectors in $\F_{q^2}^n$ into
vectors in $\F_q^{2n}$ by using a basis of $\F_{q^2}/\F_q$. The
resulting code $\Phi^{-1}(C)\subset\F_q^{2n}$ is symplectic
self-orthogonal by \eqref{eq:dualities}. The desired conclusion
follows by Proposition \ref{prop:symplectic_constr}, noting that
$\swt\left(\Phi^{-1}(\bm{v})\right)=\wgt(\bm{v})$.
\end{IEEEproof}

\section{Generalized Lison\v{e}k-Singh Construction}\label{sec:ConstructionX}
In this section, we refine and generalize the construction of quantum
stabilizer codes from nearly self-orthogonal codes given by
Lison\v{e}k and Singh in \cite[Theorem 2]{Lisonek2014}.  Like in
\cite[Theorem 2]{ISIT19}, we interchange the roles of $C$ and
$C^\perpH$ compared with those in \cite{Lisonek2014}.

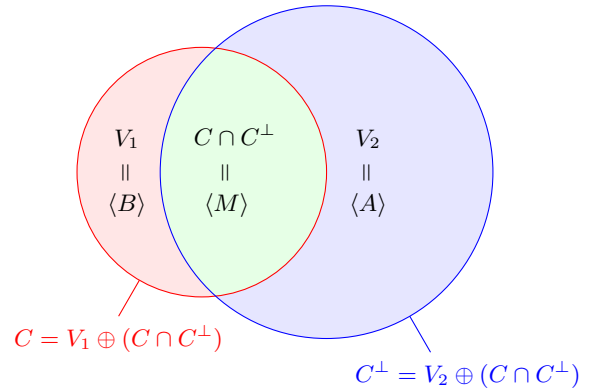
\begin{figure}[htb]
\centerline{
  \begin{tikzpicture}[scale=1.1]
  \small
  \begin{scope}
    \clip (-1,-2) rectangle (5,3)
          (25mm,10mm) circle (20mm);
    \fill[fill=red!10!white] (10mm,10mm) circle (15mm);
  \end{scope}
  \begin{scope}
    \clip (-1,-2) rectangle (5,3)
          ( 10mm,10mm) circle (15mm);
    \fill[fill=blue!10!white] (25mm,10mm) circle (20mm);
  \end{scope}
  \begin{scope}
    \clip (25mm,10mm) circle (20mm);
    \fill[fill=green!10!white] ( 10mm,10mm) circle (15mm);
  \end{scope}
  \draw[color=red] ( 10mm,10mm) circle (15mm);
  \draw[color=blue] (25mm,10mm) circle (20mm);
  \draw[color=red] (2.5mm,-2.99mm) -- (0mm,-7.32mm);
  \draw[color=red] (0mm,-7.32mm) node [below] {$C=V_1\oplus(C\cap C^\perp)$};
  \draw ( 1mm,14mm) node [text=black]{$V_1$};
  \draw ( 1mm,10mm) node [text=black,rotate=90]{$=$};
  \draw ( 1mm, 6mm) node [text=black]{$\langle B\rangle$};
  \draw (13mm,14.5mm) node [text=black]{$C\cap C\rlap{${}^\perp$}$};
  \draw (13mm,10mm) node [text=black,rotate=90]{$=$};
  \draw (13mm, 6mm) node [text=black]{$\langle M\rangle$};
  \draw (30mm,14mm) node [text=black]{$V_2$};
  \draw (30mm,10mm) node [text=black,rotate=90]{$=$};
  \draw (30mm, 6mm) node [text=black]{$\langle A\rangle$};
  \draw[color=blue] (42mm,-11.65mm) node [below]{$C^\perp=V_2\oplus(C\cap C^\perp)$};
  \draw[color=blue] (35mm,-7.32mm) -- (37.5mm,-11.65mm);
  \end{tikzpicture}%
}
\caption{Illustration of the different spaces related to the matrix in
  \eqref{eq:genmat_H} as well as the matrices in
  \eqref{eq:genmat_symp} and \eqref{eq:genmat_CSS}. The symbol
  ${}^\perp$ denotes the corresponding notion of duality, and the
  matrices $B$, $M$, and $A$ might be formed by several
  blocks.\label{fig:spaces_H}}
\end{figure}

\subsection{The Hermitian Case}
\begin{thm}\label{thm:X}
For an $\F_{q^2}$-linear code $C$ with parameters $[n,k]_{q^2}$, let
$e:=k - \dim(C \cap C^\perpH)$. Then there exists an
$[\![n+e,n-2k+e,d(\mathcal{Q})]\!]_q$ quantum stabilizer code
$\mathcal{Q}$ with
\begin{alignat}{5}
  d(\mathcal{Q}) &
  \geq\min \bigl\{
     &&\wgt\bigl(C^\perpH\setminus(C\cap C^\perpH)\bigr),\nonumber\\
     &&&\wgt\bigl((C + C^\perpH)\setminus C\bigr)+1\bigr\}\label{eq:d_impure}\\
 &\geq\min\bigl\{&&d(C^\perpH), d(C + C^\perpH)+1\bigr\},\label{eq:d_pure}\\
\noalign{\text{and}}
  d(\mathcal{Q})&
  \le\wgt\bigl(&&C^\perpH\setminus(C\cap C^\perpH)\bigr).\label{eq:d_max}
\end{alignat}
\end{thm}

\begin{IEEEproof}
Let $M\in\F_{q^2}^{s\times n}$ be a generator matrix for the Hermitian
hull $C\cap C^{\perpH}$ of dimensions $s=k-e$. Moreover, let
$B\in\F_{q^2}^{e\times n}$ be the matrix associated with a Hermitian
orthonormal basis of a vector space complement $V_1$ of the Hermitian
hull in the code $C$, which can be computed using the algorithm in the
proof of Lemma \ref{lemma:HermitianGramSchmidt} in Appendix
\ref{sec:Gram-Schmidt}.  Additionally, let
$A\in\F_{q^2}^{(n-k-s)\times n}$ be a generator matrix for the vector
space complement $V_2$ of the hull in $C^{\perpH}$. Finally, let
$\beta\in\F_{q^2}$ be an element of norm $-1$, \ie, $\beta^{q+1}=-1$.
Consider the following matrix $G$ (see also Fig. \ref{fig:spaces_H})
\begin{alignat}{5}
  \def\arraystretch{1.2}
    \begin{array}{r@{}r@{}cc@{}l@{}l}
      &\multirow{3}{*}{$\left(\rule{0pt}{25pt}\right.$}
      & A_{(n-k-s)\times n} & 0_{(n-k-s)\times e} &
       \multirow{3}{*}{$\left.\rule{0pt}{25pt}\right)$} &
       \multirow{2}{*}{$\left.\rule{0pt}{15pt}\right\}\text{for $C^{\perpH}$}$}\\
    \multirow{2}{*}{for $C\left\{\rule{0pt}{15pt}\right.$}
    && M_{s\times n} & 0_{s\times e}\\
    && B_{e\times n} & \beta I_{e\times e}\\[-3ex]
    && \underbrace{\hphantom{A_{(n-k-s)\times n}}}_n & \underbrace{\hphantom{0_{(n-k-s)\times e}}}_e
  \end{array}\label{eq:genmat_H}
\end{alignat}
where the subscripts indicate the dimensions of the matrices.  The
blocks $M$ and $B$ together generate the code $C$, and the blocks $M$
and $A$ together generate the code $C^{\perpH}$.  The three blocks
$A$, $M$, and $B$ together span the code $C+C^{\perpH}$ of dimensions
$n-s=n+e-k$, which is the Hermitian dual of the Hermitian hull.

As $B$ corresponds to a Hermitian orthonormal basis, $B B^*=I_e$, and
hence $\left(B\;\beta I_e\right)\left(B\;\beta
I_e\right)^*=0$. Moreover, as $M$ spans the Hermitian hull, $B
M^*=0$. Hence, the submatrix corresponding to the last two rows of
blocks in \eqref{eq:genmat_H} generates a Hermitian self-orthogonal
code $C'=[n+e,k,d']_{q^2}$, which we refer to as the \emph{Hermitian
self-orthogonal extension of $C$}. The matrix $A$ spans a subcode of
$C^\perpH$, and hence $AM^*=0$ and $BM^*=0$.  Therefore, the matrix
$G$ generates a subcode of the Hermitian dual of $C'$. The matrix $G$
has full rank, and hence is a generator matrix of ${C'}^\perpH$.

The quantum code $\mathcal{Q}$ is obtained by using the code $C'$ in
the Hermitian construction of Proposition \ref{prop:Hermitian_constr}.

In order to prove the lower bound on the minimum distance, we follow
the idea of Construction X \cite[Chap. 18, \S7.1]{MS78}. We have to
find the minimum weight of ${C'}^\perpH\setminus C'$, \ie, we can
ignore codewords in $C'$.  For a codeword
$\bm{c}'=(\bm{c}|\bm{0})\in{C'}^\perpH\setminus C'$ whose last $e$
coordinates are zero, the first $n$ coordinates form a codeword
$\bm{c}\in C^\perpH$. Since $\bm{c}'\notin C'$, it follows that
$\bm{c}\notin C$. This yields the first term in the lower bound
\eqref{eq:d_impure}.  For a codeword
$\bm{c}'=(\bm{c}|\bm{x})\in{C'}^\perpH\setminus C'$ with $\bm{x}\ne
0$, we have $\wgt(\bm{c}')\ge\wgt(\bm{c})+1$.  In this case,
$\bm{c}\in C+C^\perpH$, but $\bm{c}\notin C$, and hence,
$\wgt(\bm{c})\ge\wgt\bigr((C+C^\perpH)\setminus C\bigl)$, which yields
the second term in the lower bound \eqref{eq:d_impure}.

The lower bound \eqref{eq:d_pure} equals the lower bound on the
minimum distance of the code ${C'}^\perpH$. From the matrix in
\eqref{eq:genmat_H} it can be seen that ${C'}^\perpH$ is obtained by
applying Construction X to the code $C+C^\perpH$, its subcode
$C^\perpH$, and a trivial code $[e,e,1]_q$ as the auxiliary
code. It also follows trivially from \eqref{eq:d_impure} since, for
both terms, one simply considers the minimum weight of a larger set.

The upper bound follows from the fact that appending $e$ zeros to any
codeword in $\bm{c}\in C^\perpH\setminus(C\cap C^\perpH)$ yields a
codeword $\bm{c}'=(\bm{c}|\bm{0})\in{C'}^\perpH\setminus C'$.
\end{IEEEproof}
Note that in \cite{Lisonek2014} only the lower bound \eqref{eq:d_pure}
is given. The following example demonstrates that the bound
\eqref{eq:d_impure} can be strictly larger. The description of the classical codes in the example is expressed based on the algebraic structure of quasi-twisted codes in Section \ref{sec:QT_section}.

\begin{ex}\label{ex:impure}
  \def\w{\omega}
  Consider the following polynomials in $\F_4[x]$
  \begin{alignat*}{4}
  g_{1,0}(x)&{}=x^3 + \w^2x^2 + \w x + \w^2,\\
  g_{1,1}(x)&{}=\w x^{17} + \w x^{16} + \w^2x^{13} + \w x^{12} + \w^2x^{11}+ \w x^{10}\\
    &\qquad + x^8 + x^6 + \w x^5 + x^4 + \w x^3 + \w x^2 + \w x,\\
  g_{2,0}(x)&{}=0,\\
  g_{2,1}(x)&{}=x^{18} + \w^2x^{15} + \w x^{12} + x^9 + \w^2x^6 + \w x^3 + 1,
  \end{alignat*}
  where $\w$ is a primitive element of $\F_4$.  The two-generator
  quasi-twisted code of index $\ell=2$, co-index $m=21$, and
  $\lambda=\w^2$ with generators ${\bf g}_1=(g_{1,0},g_{1,1})$ and
  ${\bf g}_2=(g_{2,0},g_{2,1})$ is a $[42,21,7]_4$ code $C$. The
  Hermitian hull is $C\cap C^\perpH$ with parameters $[42,15,14]_4$,
  the Hermitian dual $C^\perpH$ has parameters $[42,21,11]_4$, and
  $C+C^\perpH$ has parameters $[42,27,7]_4$. The first terms of the
  weight enumerators are as follows:
  \begin{alignat*}{5}
    W_{C\cap C^\perpH}&{}=1+63 \,y^{14}+756 \,y^{16}+14112 \,y^{18}+\ldots\\
    W_{C}&{}=1+18 \,y^7 + 126 \,y^{10}+63 \,y^{11}+\ldots\\
    W_{C^\perpH}&{}=1+252 \,y^{11}+2079 \,y^{12}+11907 \,y^{13}+\ldots\\
    W_{C+C^\perpH}&{}=1+18 \,y^7+756 \,y^8+8442 \,y^9+\ldots
  \end{alignat*}
  This implies $\wgt\bigl(C^\perpH\setminus(C\cap
  C^\perpH)\bigr)=d(C^\perpH)=11$ and $\wgt\bigr((C+C^\perpH)\setminus
  C\bigr)=8>d(C+C^\perpH)=7$. For the quantum code, we get $9\le
  d(\mathcal{Q})\le 11$, while the lower bound in \eqref{eq:d_pure} is
  only $8$.

  Note that the improved lower bound need not be tight.  With a
  suitable choice of the complement of $C\cap C^\perpH$ in $C$ and an
  orthonormal basis $B$, we can achieve $d(\mathcal{Q})=11$, \ie, we
  get a $[\![48,6,11]\!]_2$ quantum code.

  When we interchange $C$ and $C^\perpH$, both having dimension $21$
  in this case, we only get $d(\mathcal{Q})=8$, since the lower and
  upper bounds are equal.
\end{ex}

\begin{rem}
  Starting with an $[n,k,d]_{q^2}$ code $C$, Theorem \ref{thm:X}
  yields an $[\![n+e,n-2k+e,d(\mathcal{Q})]\!]_q$ quantum code
  $\mathcal{Q}$.  When we start with the $[n,n-k,d^\perpH]_{q^2}$
  Hermitian dual code $C^\perpH$, the Hermitian hull $C\cap C^\perpH$,
  of dimension $s$, and its Hermitian dual $C+C^\perpH$, of dimension
  $n-s=n+e-k$, are unchanged.  In this case, the self-orthogonal
  extension of $C^\perpH$ requires the length to be increased by
  $\widetilde{e}=n-k-s=e+n-2k$. Defining $\widetilde{k}=n-k$, the
  Hermitian dual $C^\perpH$ yields a quantum code
  $\widetilde{\mathcal{Q}}$ with parameters
  $[\![n+\widetilde{e},n-2\widetilde{k}+\widetilde{e},d(\widetilde{\mathcal{Q}})]\!]_q
  =[\![n+e+(n-2k),e,d(\widetilde{\mathcal{Q}})]\!]_q$.

  If we assume that $2k<n$, then the length of the code
  $\widetilde{\mathcal{Q}}$ is increased by $n-2k$ compared to
  $\mathcal{Q}$, whereas the dimension is reduced by $n-2k$.  The
  minimum distance of $C+C^\perpH$ is not smaller than the minimum
  distances of the subcodes $C$ and $C^\perpH$. We can, therefore,
  expect that the weaker lower bound \eqref{eq:d_pure} is equal to the
  second term $d(C+C^\perpH)+1$, which is the same for both
  $\mathcal{Q}$ and $\widetilde{\mathcal{Q}}$.  Hence, it is plausible
  that starting with a code $C$ of rate at most $1/2$ yields a better
  quantum code compared to starting with the dual code. However,
  partially related to the improved lower bound \eqref{eq:d_impure},
  it cannot be excluded that starting with a code of rate larger than
  $1/2$ yields a quantum code with good parameters.

  Note that in Example \ref{ex:impure}, we start with a code of rate
  $1/2$. The length and dimension of both $\mathcal{Q}$ and
  $\widetilde{\mathcal{Q}}$ are the same, but one of the codes has
  only distance $7$, while the other has distance at least $9$.
\end{rem}

\subsection{The Symplectic Case}
Lison\v{e}k and Dastbasteh \cite{Lisonek2018} generalized the
construction of \cite{Lisonek2014} to additive codes over $\F_4$.
Here we present the general case of quantum codes from $\F_q$-linear
codes of length $2n$, with refined bounds on the minimum distance.
\begin{thm}\label{thm:X_symp}
  For an $\F_{q}$-linear code $C$ with parameters $[2n,k]_q$, let
  $2e:=k - \dim(C \cap C^\perpS)$. Then there exists an
  $[\![n+e,n-k+e,d(\mathcal{Q})]\!]_q$ quantum stabilizer code
  $\mathcal{Q}$ with
  \begin{alignat}{5}
    d(\mathcal{Q}) &
    \geq\min \bigl\{
       &&\swt\bigl(C^\perpS\setminus(C\cap C^\perpS)\bigr),\nonumber\\
       &&&\swt\bigl((C + C^\perpS)\setminus C\bigr)+1\bigr\}\label{eq:d_impure_symp}\\
   &\geq\min\bigl\{&&\dS(C^\perpS), \dS(C + C^\perpS)+1\bigr\},\label{eq:d_pure_symp}\\
  \noalign{\text{and}}
    d(\mathcal{Q})&
    \le\swt\bigl(&&C^\perpS\setminus(C\cap C^\perpS)\bigr).\label{eq:d_max_symp}
  \end{alignat}
\end{thm}

\begin{IEEEproof}
  Similar to the Hermitian case, we consider specific bases of the
  vector spaces as shown in Fig.~\ref{fig:spaces_H},
  replacing $C^\perpH$ by $C^\perpS$.

  Let $M=(M_1|M_2)\in\F_q^{s\times 2n}$ be a generator matrix for the
  symplectic hull $C\cap C^{\perpS}$ of dimensions $s=k-2e$.
  Moreover, let $B=(B_1|B_2)\in\F_q^{2e\times n}$ be the matrix
  associated with a basis of a vector space complement $V_1$ of the
  symplectic hull in the code $C$.  By Lemma
  \ref{lemma:SymplecticGramSchmidt} in Appendix
  \ref{sec:Gram-Schmidt}, there exists a basis consisting of $e$
  symplectic pairs $\{\bm{z}_{2i_1},\bm{z}_{2i_1+1}\}$, with $0\le
  i_1\le e-1$.  Let $(B_{11}|B_{12})\in\F_q^{e\times 2n}$ be the
  submatrix of $B$ with rows $\bm{z}_{2i_1}$ and let
  $(B_{21}|B_{22})\in\F_q^{e\times 2n}$ be the submatrix with rows
  $\bm{z}_{2i_i+1}$.  Additionally, let
  $A=(A_1|A_2)\in\F_q^{(2n-k-s)\times 2n}$ be a generator matrix for
  the vector space complement $V_2$ of the hull in
  $C^{\perpS}$. Consider the matrix
  given by
  \begin{alignat}{5}
    \def\arraystretch{1.2}
      \begin{array}{r@{}r@{}cc|cc@{}l@{}l}
        &\multirow{4}{*}{$\left(\rule{0pt}{30pt}\right.$}
        & A_1 & 0 & A_2 & 0 &
         \multirow{4}{*}{$\left.\rule{0pt}{30pt}\right)$} &
         \multirow{2}{*}{$\left.\rule{0pt}{15pt}\right\}\text{for $C^{\perpS}$}$}\\
      \multirow{3}{*}{for $C\left\{\rule{0pt}{22.5pt}\right.$}
      && M_1   & 0 & M_2   & 0  \\
      && B_{11} & I & B_{12} & 0  \\
      && B_{21} & 0 & B_{22} & -I \\[-3ex]
      && \underbrace{\hphantom{B_{12}}}_n &\multicolumn{1}{c}{\underbrace{\hphantom{0}}_e}
       & \underbrace{\hphantom{B_{22}}}_n &\underbrace{\hphantom{0}}_e
    \end{array}\label{eq:genmat_symp}
  \end{alignat}
  where, for clarity, we are not using subscripts to indicate the
  dimensions of the blocks as in \eqref{eq:genmat_H}.  Similar to the
  Hermitian case, the last three rows of blocks of the matrix $G$
  generate an $\F_q$-linear symplectic self-orthogonal code $C'$ with
  parameters $[2(n+e),k]_q$ which we refer to as the \emph{symplectic
  self-orthogonal extension of $C$}.

  The quantum code $\mathcal{Q}$ is obtained from the code $C'$ using
  the symplectic construction of Proposition
  \ref{prop:symplectic_constr}.

  Note that the codewords of the symplectic dual of the extended code
  are of the form $\bm{c}'=(\bm{c}_1,\bm{x}_1|\bm{c}_2,\bm{x}_2)$.
  When $(\bm{x}_1|\bm{x}_2)=(\bm{0}|\bm{0})\in\F_q^{2e}$, the codeword
  $\bm{c}=(\bm{c}_1|\bm{c}_2)\in C^\perpS$; otherwise,
  $\bm{c}=(\bm{c}_1|\bm{c}_2)\in C$ and $\swt(\bm{c}')\ge
  \swt(\bm{c})+1$. The reasoning concerning the bounds on the minimum
  distance is analogous to the proof of Theorem \ref{thm:X}.
\end{IEEEproof}

\begin{rem}
 Applying the inverse map $\Phi^{-1}$ (see \eqref{eq:Phi_vector})
 to an $\F_{q^2}$-linear code, we could also derive Theorem
 \ref{thm:X} as a corollary to Theorem \ref{thm:X_symp}.  The
 resulting symplectic self-dual extension would, however, in
 general not be an $\F_{q^2}$-linear code.
\end{rem}

While the CSS construction can be considered as a special case of the
symplectic construction, we also state a version of the generalized
Lison\v{e}k-Singh construction for this case.  It turns out that we do
not apply a Gram-Schmidt-like orthogonalization in this case.
\begin{thm}\label{thm:X_CSS}
  Let $C_1,C_2\subseteq\F_q^n$ be two linear codes with parameters
  $[n,k_1,d_1]_q$ and $[n,k_2,d_2]_q$. Furthermore, let
  $e=n-k_1-\dim(C_1\cap C_2^\perpE)=n-k_2-\dim(C_2\cap C_1^\perpE)$.
  Then there exists an $[\![n+e,k_1+k_2-n+e,d(\mathcal{Q})]\!]_q$
  quantum stabilizer code $\mathcal{Q}$ with
  \begin{alignat}{5}
    d(\mathcal{Q}) &
    \geq\min \bigl\{
       &&\wgt\bigl(C_1\setminus(C_1\cap C_2^\perpE)\bigr),\nonumber\\
       &&&\wgt\bigl(C_2\setminus(C_2\cap C_1^\perpE)\bigr),\nonumber\\
       &&&\wgt\bigl((C_1 + C_2^\perpE)\setminus C_2^\perpE\bigr)+1,\nonumber\\
       &&&\wgt\bigl((C_2 + C_1^\perpE)\setminus C_1^\perpE\bigr)+1\bigr\}\label{eq:d_impure_CSS}\\
   &\geq\min\bigl\{&&d_1,d_2, d(C_1 + C_2^\perpE)+1, d(C_2 + C_1^\perpE)+1\bigr\},\label{eq:d_pure_CSS}\\
  \noalign{\text{and}}
    d(\mathcal{Q})&
    \le\min\bigl\{&& d(C_1\setminus(C_1\cap C_2^\perpE),
                     d(C_2\setminus(C_2\cap C_1^\perpE) \bigr\}.\label{eq:d_max_CSS}
  \end{alignat}
\end{thm}
\begin{IEEEproof}
  For the proof, we start with the $\F_q$-linear code
  $C=C_2^\perpE\times C_1^\perpE$ with parameters $[2n,2n-k_1-k_2]_q$
  and use the scenario of Theorem \ref{thm:X_symp}.  Let
  $C_{12}=C_1\cap C_2^\perpE$ and $C_{21}=C_2\cap C_1^\perp$ be the
  relative hulls (cf. \cite{RelativeHull}) of the codes $C_1$ and
  $C_2$, with generator matrices $M_{12}$ and $M_{21}$,
  respectively. Then the symplectic hull of $C_2^\perpE\times
  C_2^\perpE$ is
  \begin{alignat*}{5}
    &(C_2^\perpE\times C_1^\perpE)\cap(C_2^\perpE\times C_1^\perpE)^\perpS\\
    &\qquad{}= (C_2^\perpE\times C_1^\perpE)\cap(C_1\times C_2)= & C_{12}\times C_{21}.
  \end{alignat*}
  Moreover, let $B_{12}$ be a generator matrix for the complement of
  $C_{12}$ in $C_2^\perpE$ and let $B_{12}$ be a generator matrix for the
  complement of $C_{21}$ in $C_1^\perpE$.  Both matrices have equal
  rank $e$. We define the matrix $E=B_{12} B_{21}^{\rm T}$. It has
  full rank, since $\langle B_{12}\rangle^\perpE\cap \langle
  B_{21}\rangle=\{\bm{0}\}$.  Finally, let $A_1$ be a generator matrix
  of the complement of $C_{12}$ in $C_1$ and be $A_2$ be a generator
  matrix of the complement of $C_{21}$ in $C_2$, \ie, the direct
  product of $A_1$ and $A_2$ generates the complement of the
  symplectic hull in $C_1\times C_2$.  Consider the following matrix
  $G$:
  \begin{alignat}{5}
    \def\arraystretch{1.2}
      \begin{array}{r@{}r@{}cc|cc@{}l@{}l}
        &\multirow{6}{*}{$\left(\rule{0pt}{45pt}\right.$}
        & A_1 & 0 & 0   & 0 &
         \multirow{6}{*}{$\left.\rule{0pt}{45pt}\right)$} &
         \multirow{2}{*}{$\left.\rule{0pt}{15pt}\right\}\text{for $C^{\perpS}$}$}\\
       && 0   & 0 & A_2 & 0\\
      \multirow{4}{*}{for $C\left\{\rule{0pt}{30pt}\right.$}
      && M_{12}   & 0 & 0       & 0  \\
      && 0       & 0 & M_{21}   & 0  \\
      && B_{12} & -E & 0 & 0  \\
      && 0     &  0 & B_{21} & I \\[-3ex]
      && \underbrace{\hphantom{B_{12}}}_n &\multicolumn{1}{c}{\underbrace{\hphantom{0}}_e}
       & \underbrace{\hphantom{B_{22}}}_n &\underbrace{\hphantom{0}}_e
    \end{array}\label{eq:genmat_CSS}
  \end{alignat}
  Similar to the symplectic case, now the last four rows of blocks of
  the matrix $G$ generate an $\F_q$-linear symplectic self-orthogonal
  code $C'$ with parameters $[2(n+e),2n-k_1-k_2]_q$ which we refer to
  as the \emph{symplectic CSS extension of $C=C_2^\perpE\times
  C_1^\perpE$}.

  The quantum code $\mathcal{Q}$ is obtained using the code $C'$ in
  the CSS construction of Proposition \ref{prop:CSS_constr}.  Note
  that the matrix in \eqref{eq:genmat_CSS} is a direct product of a
  block in the first half of columns and another block in the second
  half.

  The bounds on the minimum distance follow from
  \eqref{eq:d_impure_symp}--\eqref{eq:d_max_symp} based on the fact
  that, for a direct product of codes $A$ and $B$, $\swt(A\times
  B)=\min\{\wgt(A),\wgt(B)\}$.

\end{IEEEproof}

\subsection{The Trace-Symplectic Case}
For completeness, we also state the version of the generalized
Lison\v{e}k-Singh construction for the most general case of the
trace-symplectic construction. For quantum codes $[\![n,k,d]\!]_p$,
$p$ prime, this coincides with the symplectic case.  We will not
include this case in the random search for new quantum codes.

\begin{thm}\label{thm:X_trace_symp}
  For an additive code $C$ over $\F_{p^m}$ code $C$ with parameters
  $(2n,p^k)_q$, let $2e:=k - \dim_{\F_p}(C \cap C^\perpT)$. Then there
  exists an $[\![n+e',k',d(\mathcal{Q})]\!]_q$ quantum stabilizer code
  $\mathcal{Q}$, where $e'=\lceil e/m\rceil$ and $k'=n-(k-e)/m$. The
  bounds on the minimum distance $d(\mathcal{Q})$ are
  \begin{alignat}{5}
    d(\mathcal{Q}) &
    \geq\min \bigl\{
       &&\swt\bigl(C^\perpT\setminus(C\cap C^\perpT)\bigr),\nonumber\\
       &&&\swt\bigl((C + C^\perpT)\setminus C\bigr)+1\bigr\}\label{eq:d_impure_trace_symp}\\
   &\geq\min\bigl\{&&\dS(C^\perpT), \dS(C + C^\perpT)+1\bigr\},\label{eq:d_pure_trace_symp}\\
  \noalign{\text{and}}
    d(\mathcal{Q})&
    \le\swt\bigl(&&C^\perpT\setminus(C\cap C^\perpT)\bigr).\label{eq:d_max_trace_symp}
  \end{alignat}
\end{thm}

\begin{IEEEproof}
  Using the map $\PsiB$ defined in \eqref{eq:add_expansion} in
  Appendix~\ref{sec:IP_relations}, we obtain an $\F_p$-linear code
  $\Psi_{\mathcal B}(C)$ with parameters $[2nm,k,d]_p$.  By Theorem
  \ref{thm:X_symp}, we obtain a quantum code $\mathcal{Q}'$ with
  parameters $[\![2(mn+e),mn-k+e]\!]_p$, based on a symplectic extension
  $\PsiB(C)'$. To apply the inverse map $\PsiB^{-1}$, we extend the
  length of the code $\PsiB(C)'$ from $2(mn+e)$ to the next even
  multiple of $m$.  To do this, we take the direct product with any
  symplectic self-dual $\F_p$ linear code of suitable length, \eg, the
  code $C_0$ generated by $(I|0)$.  We can then apply Proposition
  \ref{prop:trace_symp_constr} to the additive code
  $\PsiB^{-1}\bigl(\PsiB(C)'\times C_0\bigr)$ to obtain a quantum code
  $\mathcal{Q}$ with parameters $[\![n+e',k',d(\mathcal{Q})]\!]_q$.

  The bounds on the minimum distance follow analogously to those in
  Theorem \ref{thm:X_symp}, noting that the direct product with the
  symplectic self-orthogonal code $C_0$ contributes both to the code
  and its trace-symplectic dual.
\end{IEEEproof}
As already noted after Theorem \ref{prop:trace_symp_constr}, the
dimension of the resulting quantum code from the trace-symplectic
construction need not be an integral power of $q$, that is $k'$ in
Theorem \ref{thm:X_trace_symp} need not be an integer.  The same
applies to the parameter $e$, but by extending the length of the code
$\mathcal{Q}'$, we obtain a quantum code whose length is a multiple of
$m$, preserving its dimension.

We illustrate the situation with an example, using quasi-cyclic codes
as described in the next section.
\begin{ex}\label{ex:additive_GF16}
  We consider the case $q=2^2$ and start with a binary quasi-cyclic
  $[188,73]_2$ code $C$ with index $\ell=4$ and four generators,
  given by
  \[
    \begin{array}{@{}l@{}}
    \mathbf{f}_1=100011000111011011101111,0,0,\\
    \quad  1100001011011011111110101101011101011100111111;\\[1ex]
    \mathbf{f}_2=  0,100011000111011011101111,0,\\
    \quad  0011010010010001111101111111111010100001101111;\\[1ex]
    \mathbf{f}_3=  0,0,100011000111011011101111,\\
    \quad  111000110000001011110110111101001010100011001;\\[1ex]
    \mathbf{f}_4=  0,0,0,\\
    \quad  11111111111111111111111111111111111111111111111,
    \end{array}
  \]
  where we list only the coefficients of the polynomials, starting with
  the constant term.  The symplectic hull $C\cap C^\perpS$, the
  symplectic dual $C^\perpS$, and the sum $C+C^\perpS$ have parameters
  $[188,69]_2$, $[188,115]_2$, and $[188,119]_2$, respectively.
  Using Theorem \ref{thm:X_symp} with $e=2$, we obtain a symplectic
  self-orthogonal $[192,73]_2$ code. As the length is a multiple of
  the extension degree, we do not need to take a direct product with a
  self-orthogonal code.

  Using the inverse map $\PsiB^{-1}$ with the self-dual basis
  $\mathcal{B}=\{\alpha,\alpha^2\}$ of $\F_4/\F_2$ with
  $\alpha^2=\alpha+1$, we obtain a symplectic self-orthogonal additive
  $(96,2^{73})_4$ code over $\F_4$.  The resulting quantum code has
  parameters $(\!(48,2^{23},12)\!)_4$, \ie, it has dimension $2^{23}=4^{23/2}$,
  and hence encodes a fractional number of ququads.

  The minimum distance of this code was determined as follows. The
  binary quasi-cyclic codes $C^\perpS$ and $C+C^\perpS$ of length
  $188$ with index $\ell=4$ correspond to cyclic additive codes of
  length $47$ over $\F_{16}$. Using Magma, we found that these codes
  have parameters $(47,2^{115},12)_{16}$ and $(47,2^{119},11)_{16}$,
  respectively, where the distance is with respect to the Hamming
  metric.  The CPU time was about $16.5$ days for each of the
  codes. The minimum distance of the quantum code is then derived
  using \eqref{eq:d_pure_trace_symp}.  To conclude this example, we
  note that we have later found a $[\![46,12,12]\!]_4$ code using
  the Hermitian construction (see Table \ref{tab:ququad}) that has a
  larger dimension and uses fewer ququads than the code from this
  example.
\end{ex}

\begin{ex}
There is a binary quasi-cyclic $[188,46]_2$ code with index $\ell=4$
and two generators:
\[
  \begin{array}{@{}l@{}}
    \mathbf{f}_1=1100101001001101100110001,0,\\
    \ 111101001110100110101111111011000010111011001,\\
    \ 01000101010100001001011001111101001000010110001;\\[1ex]
  \mathbf{f}_2=0,1100101001001101100110001,\\
    \ 11100100011110001101100011101111000000001001111,\\
    \ 10110110010000010010110010001101111000001111111.
  \end{array}
  \]
This code is symplectic self-orthogonal and yields a
$[\![98,46]\!]_2$ qubit quantum code.  As a ququad code, we obtain a
$[\![47,24,8]\!]_4$ quantum code, whose classical stabilizer code is
an additive cyclic $(47,2^{46},26)_{16}$ code.
\end{ex}

\section{Quasi-Twisted Codes with Designed Hermitian Hulls}\label{sec:QT_section}

Let $m$ and $\ell$ be positive integers such that $\gcd(m,q)=1$ and
let $\lambda$ be a non-zero element in $\F_q$. The
$\lambda$-constashift operator applied on a vector
$(v_0,\ldots,v_{m-1}) \in \Fq^m$ gives $(\lambda
v_{m-1},v_0,\ldots,v_{m-2}) \in \Fq^m$. A linear code $C$ of length
$m\ell$ over $\Fq$ is called $\lambda$-quasi-twisted ($\lambda$-QT)
code of index $\ell$ if it is invariant under the
$\lambda$-constashift of codewords by $\ell$ positions and $\ell$ is
the smallest number with this property. In particular, if $\ell=1$,
then $C$ is a $\lambda$-constacyclic code, and if $\lambda = 1$ or
$q=2$, then $C$ is a quasi-cyclic (QC) code of index $\ell$. If we
view any codeword of $C$ as an $m \times \ell$ array
\begin{equation}\label{array}
{\bf c}= \left(
  \begin{array}{ccc}
    c_{0,0} & \ldots & c_{0,\ell-1} \\
    \vdots & \ddots  & \vdots \\
    c_{m-1,0} & \ldots & c_{m-1,\ell-1}
  \end{array}
\right),
\end{equation}
then being invariant under the $\lambda$-constashift by $\ell$ units
in $\Fq^{m\ell}$ amounts to being closed under the
$\lambda$-constashift of rows in $\Fq^{m\times\ell}$. The first row is
replaced by $\lambda$ times the last row, and rows with indices $0$ to
$m-2$ are moved down by one row.  The resulting array is
\begin{equation*}
{\bf c}'= \left(
  \begin{array}{ccc}
    \lambda c_{m-1,0} & \ldots & \lambda c_{m-1,\ell-1} \\
    c_{0,0} & \ldots & c_{0,\ell-1} \\
    \vdots & \ddots  & \vdots \\
    c_{m-2,0} & \ldots & c_{m-2,\ell-1}
  \end{array}
\right).
\end{equation*}

Consider the principal ideal $I=\langle x^m-\lambda \rangle$ of
$\Fq[x]$ and define the quotient ring $R:=\Fq[x]/I$. If $T$ denotes
the shift-by-one operator on $\Fq^{m\ell}$, let us denote its action on
$v\in \F_q^{m\ell}$ by $T\circ v$. Then $\Fq^{m\ell}$ has an
$\Fq[x]$-module structure given by the multiplication
\begin{alignat*}{5}
\Fq[x] \times \F_{q}^{m\ell} & {}\longrightarrow{}&& \F_{q}^{m\ell}\\
(a(x),v) & {}\longmapsto{} && a(T^\ell)\circ v .
\end{alignat*}
Note that for $a(x)=x^m-1$, we have
\[
a(T^\ell)\circ v = (T^{m\ell})\circ v - v=0.
\]
Hence, the ideal $I$ annihilates
$\Fq^{m\ell}$ and we can view $\F_q^{m\ell}$ as an
$R$-module. Therefore, a QC code $C\subset \F_q^{m\ell}$ of index
$\ell$ is an $R$-submodule of $\F_q^{m\ell}$.  To an element $c\in
\Fq^{m\times \ell} \cong \Fq^{m\ell}$ as in \eqref{array}, we
associate an element $\bm{c}(x) \in R^\ell$ written as
\begin{equation} \label{associate-1}
\bm{c}(x):=(c_0(x),c_1(x),\ldots ,c_{\ell-1}(x)) \in R^\ell ,
\end{equation}
where, for each $0\leq t \leq \ell-1$,
\begin{equation}\label{columns}
c_t(x):= c_{0,t}+c_{1,t}x+c_{2,t}x^2+\ldots +
c_{m-1,t}x^{m-1} \in R.
\end{equation}
Then, the following map is an $R$-module isomorphism.
\begin{alignat}{5}
\phi\colon&& \F_q^{m\ell}\kern2cm &{} \longrightarrow{} && R^\ell \nonumber \\
&\bm{c}=&\left(
  \begin{array}{ccc}
    c_{00} & \ldots & c_{0,\ell-1} \\
    \vdots & \ddots & \vdots \\
    c_{m-1,0} & \ldots & c_{m-1,\ell-1} \\
  \end{array}
\right) & {}\longmapsto{} && \bm{c}(x) . \label{identification-1}
\end{alignat}
For $\ell=1$, this is the classical polynomial representation of a
$\lambda$-constacyclic code. Note that the $\lambda$-constashift of
rows in $\Fq^{m\times\ell}$ corresponds to componentwise
multiplication by $x$ in $R^\ell$. Thus, a $q$-ary $\lambda$-QT code
$C$ of length $m\ell$ and index $\ell$ is an $R$-submodule in
$R^{\ell}$.

Now we consider $\lambda$-QT codes and their duals with respect to the
Hermitian inner product over $\F_{q^2}$. We describe the decomposition
of a $\lambda$-QT code over $\F_{q^2}$ into shorter codes over field
extensions of $\F_{q^2}$. We refer the reader to \cite{SJU2017} for
further details and for a more general description.

Recall that the reciprocal of a polynomial $f(x)\in\F_{q^2}[x]$ is
defined as $f^*(x):=x^{\deg(f)}f(x^{-1})$. The \emph{conjugate} of a
polynomial $f(x) = \sum_{i=0}^n f_i x^i$ over $\F_{q^2}$ is
$\bar{f}(x) = \sum_{i=0}^n \bar{f}_i x^i = \sum_{i=0}^n f_i^q x^i$,
where $\bar{\ }\colon \F_{q^2} \rightarrow \F_{q^2}$ is the field
automorphism that maps $\beta \mapsto \beta^q$, for all $\beta \in
\F_{q^2}$. Hence, the \emph{conjugate-reciprocal} of
$f(x)\in\F_{q^2}[x]$ is defined to be
$f^\dagger(x):=\bar{f}^*(x)=x^{\deg(f)}\bar{f}(x^{-1})$.

Henceforth, we assume that $\lambda^q = \lambda^{-1}$, that is,
$\lambda^{q+1}=1$, and consider such $\lambda$-QT codes over
$\F_{q^2}$. Observe that $f(x) = -\lambda f^\dagger(x)$ when
$f(x)=x^m-\lambda$ and $\lambda^q = \lambda^{-1}$.  By a slight abuse
of notation, we identify a polynomial with its normalized version,
and refer to $f(x)=x^m-\lambda$ as being
\emph{self-conjugate-reciprocal} in this case.  More generally, we use
the terminology self-(conjugate)-reciprocal when a polynomial is
proportional to its (conjugate)-reciprocal polynomial.
We factor the polynomial $x^m-\lambda$ into irreducible polynomials in
$\F_{q^2}[x]$ as
\begin{equation}\label{irreducibles}
x^{m}-\lambda=g_{1}(x)\cdots g_{s}(x) \, h_{1}(x)h_{1}^\dagger(x)\cdots h_{r}(x)h_{r}^\dagger(x),
\end{equation}
where the polynomials $g_{i}$ are self-conjugate-reciprocal where\-as
$h_{j}$ and $h_{j}^\dagger$ are conjugate-reciprocal pairs, for all
$i,j$.  Since $m$ is relatively prime to $q$, there are no repeating
factors in \eqref{irreducibles}. By the Chinese Remainder Theorem
(CRT) we have the following ring isomorphism:
\begin{alignat}{5} \label{CRT-1}
R \cong{}& \left( \bigoplus_{i=1}^{s} \F_{q^2}[x]/\langle g_{i}(x) \rangle \right)\nonumber\\
 &\oplus\left( \bigoplus_{j=1}^{r} \Bigl( \F_{q^2}[x]/\langle h_{j}(x) \rangle
      \oplus \F_{q^2}[x]/\langle h_{j}^\dagger(x) \rangle \Bigr) \right).
\end{alignat}

Let $t$ be the smallest divisor of $q^2-1$ with $\lambda^t=1$ and let
$\alpha$ be a primitive $(tm)^\text{th}$ root of unity such that
$\alpha^m=\lambda$. Then, $\xi:=\alpha^t$ is a primitive $m^\text{th}$
root of unity and the roots of $x^m-\lambda$ are of the form $\alpha,
\alpha\xi, \ldots, \alpha\xi^{m-1}$. Since each $g_i(x)$, $h_{j}(x)$,
and $h_{j}^\dagger(x)$ divide $x^m-\lambda$, their roots are of the
form $\alpha\xi^k = \alpha^{1+kt}$, for some $0\leq k \leq m-1$. For
each $i\in\{1,\ldots, s\}$, let $u_i$ be the smallest nonnegative integer
such that $g_i(\alpha\xi^{u_i})=0$. For each $j\in\{1,\ldots, r\}$, let
$v_j$ be the smallest nonnegative integer such that
$h_j(\alpha\xi^{v_j})=h_{j}^\dagger(\alpha^{-q}\xi^{-qv_j})=0$. Since
all factors in (\ref{irreducibles}) are irreducible, the direct
summands in (\ref{CRT-1}) are isomorphic to field extensions of
$\F_{q^2}$. Let $\G_{i}=\F_{q^2}[x]/\langle g_{i}(x) \rangle$,
$\HH_{j}'=\F_{q^2}[x]/\langle h_{j}(x) \rangle$ and
$\HH_{j}''=\F_{q^2}[x]/\langle h_{j}^\dagger(x) \rangle$ denote those
extensions, for each $i$ and $j$, respectively. By the CRT, the
decomposition of $R$ in (\ref{CRT-1}) now becomes
\begin{alignat}{5}\label{CRT-2}
R \cong{}&\left( \bigoplus_{i=1}^{s} \G_{i} \right) \oplus \left( \bigoplus_{j=1}^{r} \Bigl( \HH_{j}' \oplus \HH_{j}'' \Bigr) \right)\\
f(x) \mapsto{}& \Bigl(\big[ f(\alpha\xi^{u_i})\big]_{1\leq i\leq s},\nonumber\\
     &\qquad\big[ f(\alpha\xi^{v_j})\big]_{1\leq j\leq r},
            \big[ f(\alpha^{-q}\xi^{-qv_j})\big]_{1\leq j\leq r}\Bigr).\nonumber
\end{alignat}
This implies that
\begin{alignat}{5} \label{CRT-3}
R^{\ell}\cong{}& \left(\bigoplus_{i=1}^{s} \G_i^{\ell}\right) \oplus \left(\bigoplus_{j=1}^{r} (\HH'_{j})^{\ell} \oplus (\HH''_{j})^{\ell}\right)\\
\mathbf{f}(x) \mapsto{}& \Bigl(\big[ \mathbf{f}(\alpha\xi^{u_i})\big]_{1\leq i\leq s},\nonumber\\
      &\qquad \big[ \mathbf{f}(\alpha\xi^{v_j})\big]_{1\leq j\leq r},
              \big[ \mathbf{f}(\alpha^{-q}\xi^{-qv_j})\big]_{1\leq j\leq r}\Bigr),\nonumber
\end{alignat}
where $\mathbf{f}(a)$ denotes the componentwise evaluation at $a$, for
any $\mathbf{f}(x)=\bigl(f_{0}(x),\ldots ,f_{\ell-1}(x)\bigr)\in
R^{\ell}$. Hence, as an $R$-submodule of $R^{\ell}$, the decomposition
of a $\lambda$-QT code $C\subseteq R^{\ell}$ is given by
\begin{equation} \label{constituents}
C\cong\left( \bigoplus_{i=1}^{s} C_i \right) \oplus \left( \bigoplus_{j=1}^{r} \Bigl( C_{j}' \oplus C_{j}'' \Bigr) \right).
\end{equation}
Here, all $C_i$ are $\G_i$-linear codes of length $\ell$, for all
$i\in\{1,\ldots,s\}$, while the $C'_{j}$ and $C''_{j}$ are $\HH'_j$- and
$\HH''_j$-linear codes of length $\ell$, respectively, for all
$j\in\{1,\ldots,r\}$. We call these linear codes of length $\ell$ over
various extensions of $\F_{q^2}$ the {\it constituents} of $C$.  Note
that in \eqref{CRT-2}, we are evaluating $f(x)$ at various powers of
the primitive $(tm)^\text{th}$ root of unity $\alpha\in\F_{q^{2\nu}}$,
where $\F_{q^{2\nu}}$ is an extension field of $\F_{q^2}$ that
contains all the fields $\mathbb{G}_i$, $\mathbb{H}_j'$, and
$\mathbb{H}_j''$.

Let $C\subseteq R^\ell$ be generated by $\{\mathbf{f}_1(x),\ldots,
\mathbf{f}_z(x)\}$, where $\mathbf{f}_b(x)=\bigl(f_{b,0}(x),\ldots
,f_{b,\ell-1}(x)\bigr)\in R^{\ell}$, for each $1\leq b \leq z$.  Then,
for $1\leq i \leq s$ and $1\leq j \leq r$, we have
\begin{alignat}{5}
   C_i &{}= \Span_{\G_i}\bigl\{\bigl(f_{b,0}(\alpha\xi^{u_i}),\ldots, f_{b,\ell-1}(\alpha\xi^{u_i})\bigr)\colon 1\leq b \leq z\bigr\},&\nonumber\\
 C'_{j} &{}=\Span_{\HH'_j}\bigl\{\bigl(f_{b,0}(\alpha\xi^{v_j}),\ldots, f_{b,\ell-1}(\alpha\xi^{v_j})\bigr)\colon 1\leq b \leq z \bigr\},& \nonumber \\
C''_{j} &{}= \Span_{\HH''_j}\bigl\{\bigl(f_{b,0}(\alpha^{-q}\xi^{-qv_j}),\ldots, f_{b,\ell-1}(\alpha^{-q}\xi^{-qv_j})\bigr)\colon\nonumber\\
              &&\llap{$1\leq b \leq z \bigr\}.$}\nonumber\\\label{eq:explicit_constituents}
\end{alignat}

Conversely, let $C_i\subseteq\G_i^{\ell}$,
$C'_j\subseteq(\HH_j')^{\ell}$ and $C''_j\subseteq(\HH_j'')^{\ell}$ be
arbitrary linear codes, for each $i\in \{1,\ldots ,s\}$ and each $j\in
\{1,\ldots, r\}$, respectively. Then, an arbitrary codeword
$\bm{c}$ in the corresponding $\lambda$-QT code $C$ over
$\F_{q^2}$ described as in (\ref{constituents}) can be written as an
$m\times \ell$ array like in \eqref{array} such that each row of
$\bm{c}$ is of the form (cf.  \cite[Thm.~5.1]{LS01})
\begin{alignat}{5}
\bm{c}_g  = \frac{1}{m} \bigg(
  &\sum\limits_{i=1}^{s} \Tr_{\G_{i}/\F_{q^2}} \big( \delta_{i,t} \alpha^{-g} \xi^{-g u_{i}} \big)\nonumber\\
  &+ \sum\limits_{j=1}^{r} \Bigl[\Tr_{\HH_j'/\F_{q^2}}\big(\delta'_{j,t}\alpha^{-g}\xi^{-gv_j} \big)\nonumber\\
  &\qquad+\Tr_{\HH_j''/\F_{q^2}} \big(\delta''_{j,t}\alpha^{gq}\xi^{gqv_j} \big)\Bigr] \bigg)_{0\leq t \leq \ell-1},\label{eq:trace_codeword}
\end{alignat}
for $0\leq g \leq m-1$, where
$\boldsymbol{\delta}_i=(\delta_{i,0},\ldots ,\delta_{i,\ell-1}) \in
C_i$, for all $i$, $\boldsymbol{\delta}_j'=(\delta'_{j,0},\ldots
,\delta'_{j,\ell-1}) \in C_j'$ and
$\boldsymbol{\delta}_j''=(\delta''_{j,0},\ldots ,\delta''_{j,\ell-1}) \in
C_j''$, for all $j$. Since $\frac{1}{m}C=C$, we cancel $\frac{1}{m}$
out. Note that, in this representation, the quasi-twisted shift by
$\ell$ units corresponds to multiplication by $\xi^{-1}$.

For each $1\leq i \leq s$, let $\G_i^{\ell}$ be equipped with the
Hermitian inner product over the field $\G_i$, which is an even degree
extension of $\F_q$. For $1\leq j \leq r$, let $(\HH_j')^{\ell}$ and
$(\HH_j'')^{\ell}$ be equipped with the usual Euclidean inner
product. Observe that $\HH_j'\cong\HH_j''$ follows from the fact that
$\F_{q^2}(\alpha\xi^a)=\F_{q^2}(\alpha^{-1}\xi^{-a})=\F_{q^2}(\alpha^{-q}\xi^{-qa})$. If
$h_j(\alpha\xi^a)=0$ for some $a\in\{0,\ldots,m-1\}$, then
$\HH_j'=\F_{q^2}(\alpha\xi^a)$ and $\alpha^{-q}\xi^{-qa}$ is a root of
$h_j^\dagger(x)$, for each $1\leq j \leq r$. Note, however, that
$\HH_j'$ and $\HH_j''$ have different defining polynomials.

The CRT decomposition of the Hermitian dual of a $\lambda$-QT code,
which is a $\lambda^{-q}$-QT code (see \cite[Proposition
  6.2]{SJU2017}), is given as follows (see also \cite{SJU2017},
  \cite[Proposition 2]{LvLW20b}):
\begin{prop}\label{duality}
Let $C$ be a $\lambda$-QT code over $\F_{q^2}$ with the CRT
decomposition as in (\ref{constituents}). Then its Hermitian dual code
$C^\perpH$ satisfies
\begin{equation}\label{dual}
C^\perpH\cong\left( \bigoplus_{i=1}^s C_i^\perpH \right) \oplus \left( \bigoplus_{j=1}^r \Bigl( (C_j'')^\perpE \oplus (C_j')^\perpE \Bigr)\right),
\end{equation}
where  $\bot_{\rm H}$ denotes the Hermitian dual on $\G_i^\ell$, for
all $1\leq i \leq s$, and  $\bot_{\rm E}$ denotes the Euclidean dual
on $(\HH_j')^{\ell} \cong (\HH_j'')^{\ell}$, for all $1\leq j \leq r$.
\end{prop}
As $\lambda^{q+1}=1$, both $C$ and $C^\perpH$ are $\lambda$-QT codes,
and we can now characterize Hermitian self-orthogonal $\lambda$-QT
codes based on their constituents.

\begin{thm} \label{SO_criteria}
Let $C$ be a $\lambda$-QT code of length $m\ell$ over $\F_{q^2}$ whose
CRT decomposition is as in (\ref{constituents}) and
$\lambda^{q+1}=1$. Then $C$ is Hermitian self-orthogonal if and only
if $C_i$ is Hermitian self-orthogonal over $\G_i$, for all $1\leq i
\leq s$, and $C_j'' \subseteq (C_j')^\perpE$ (equivalently,
$C_j' \subseteq (C_j'')^\perpE$) over $\HH_j'\cong \HH_j''$,
for all $1\leq j \leq r$.
\end{thm}
\begin{IEEEproof}
Immediate from the CRT decompositions of $C$ in \eqref{constituents}
and of its Hermitian dual $C^\perpH$ in (\ref{dual}).
\end{IEEEproof}

Note that in both Proposition \ref{duality} and Theorem
\ref{SO_criteria}, the Euclidean dual $(C_j')^\perpE$ of a
code $C_j'$ over  $\HH_j'$  is considered as a code over the
isomorphic field $\HH_j''$.  In terms of the polynomial representation
of the fields, the isomorphism is given by (see \cite[before (6.4)]{SJU2017})
\begin{alignat}{5}
  \widehat{\ }\colon
    && \F_{q^2}[x]/\langle h_j(x)\rangle&{}\longrightarrow\F_{q^2}[x]/\langle  h_j^\dagger(x)\rangle\label{eq:iso_hat}\\
    && f(x)=\sum_i c_i x^i &{}\longmapsto \widehat{f}(x)=\sum_i c_i^q x^{-i}.\nonumber
\end{alignat}
Recall that the constituent code $C_j'$ over $\HH_j'$ is obtained by
evaluating $f(x)$ at $\alpha\xi^{v_j}$, while the constituent code
$C_j''$ over $\HH_j''$ is obtained by evaluating $\widehat{f}(x)$ at
$\alpha^{-q}\xi^{-qv_j}=(\alpha\xi^{v_j})^{-q}$. In combination with
\eqref{eq:iso_hat}, we obtain
\begin{alignat}{5}
  \widehat{f}\bigl((\alpha\xi^{v_j})^{-q}\bigr)
  &{}=\sum_i c_i^q (\alpha\xi^{v_j})^{iq}
     =\left(\sum_i c_i(\alpha\xi^{v_j})^i\right)^q\nonumber\\
  &{}=\left(f(\alpha\xi^{v_j})\right)^q.
\end{alignat}
Therefore, when we use the very same field to represent codes $C_j'$
over $\HH_j'$ and codes $C_j''$ over $\HH_j''$, we have to apply the
conjugation map $\beta\mapsto\bar{\beta}=\beta^q$ to the code $C_j'$.

As it was done in Theorem \ref{SO_criteria} above for Hermitian
self-orthogonality, we use the CRT decomposition of the $\lambda$-QT
code $C$ in \eqref{constituents} and its Hermitian dual $C^\perpH$ in
\eqref{dual} to characterize nearly Hermitian self-orthogonal
$\lambda$-QT codes.

Let $k$ denote the dimension of $C$ over $\F_{q^2}$. Clearly,
$C^\perpH$ has dimension $m\ell-k$. For all $1\leq i \leq s$, let
$e_i:=[\G_i : \F_{q^2}]=\deg (g_i(x))$. Let $e_j:=[\HH_j' :
  \F_{q^2}]=[\HH_j'' : \F_{q^2}]=\deg( h_j(x)) =
\deg\left(h_j^\dagger(x)\right)$, for all $1\leq j \leq r$. Then we
have
\begin{alignat}{5}
  & m =\sum_{i=1}^s e_i + \sum_{j=1}^r 2 e_j\qquad\text{and}\nonumber\\
  & k = \sum_{i=1}^s e_i \dim_{\G_i} (C_i) + \sum_{j=1}^r  e_j  \left(\dim_{\HH_j'} (C_j') + \dim_{\HH_j''} (C_j'')\right).\label{dimension}
\end{alignat}
By using \eqref{dimension} and Proposition \ref{duality}, we derive
that the dimension of the Hermitian hull $C\cap C^\perpH$ is
\begin{alignat}{5}
  &\dim(C\cap C^\perpH)=\sum_{i=1}^s e_i \dim_{\G_i} (C_i \cap C_i^\perpH)\nonumber\\
  &\qquad+\sum_{j=1}^r e_j \left( \dim_{\HH_j'} (C_j' \cap C_j''^\perpE) + \dim_{\HH_j''} (C_j'' \cap C_j'^\perpE)\right).\label{hull-dimension}
\end{alignat}
Assuming $C$ to be Hermitian self-orthogonal is equivalent to saying
that $C=C\cap C^\perpH$, where (\ref{dimension}) and
(\ref{hull-dimension}) also coincide. In order to use Theorem
\ref{thm:X} with $\lambda$-QT codes in the desired way, we assume that
all constituent codes except for one $C_i$, for an $i$ such that
$1\leq i \leq s$, or for one pair of constituents $C_j'$ and $C_j''$,
for a $j$ such that $1\leq j \leq r$, satisfy the requirements of
Theorem \ref{SO_criteria}. If we set
\[
k_i:=\dim_{\G_i}(C_i) - \dim_{\G_i}(C_i \cap C_i^\perpH)
\]
for the chosen constituent $C_i$, and all the remaining
constituents $C_{i'}$, $C_j'$, $C_j''$ agree with the conditions in
Theorem \ref{SO_criteria}, for all $1\leq i' \leq s$, with $i'\neq i$,
and for $1\leq j \leq r$, then we obtain $e=k - \dim(C \cap
C^\perpH)=e_i k_i$ easily by subtracting
(\ref{hull-dimension}) from (\ref{dimension}). Similarly, if we set
\begin{alignat}{5}
k_j:=&{}\dim_{\HH_j'} (C_j')-\dim_{\HH_j'} (C_j' \cap C_j''^\perpE)\nonumber\\
    &{}+\dim_{\HH_j''}(C_j'')-\dim_{\HH_j''} (C_j'' \cap C_j'^\perpE)\label{eq:k_j}
\end{alignat}
for the chosen pair of constituents $C_j'$ and $C_j''$, for a $j$ such
that $1\leq j \leq r$, and all the remaining constituents agree with
the conditions in Theorem \ref{SO_criteria}, then we obtain $e=k -
\dim(C \cap C^\perpH)=e_j k_j$.

To show that $k_j$ is an even number, assume that $G_j'$ and $G_j''$
are generator matrices for the codes $C_j'$ and $C_j''$.  Then any
codeword $\bm{c}\in C_j'$ can be written as $\bm{c}=\bm{v} G_j'$. The
codeword is in ${C_j''}^{\perp_{\rm E}}$ if and only if $\bm{c}
{G_j''}^T =\bm{0}$, or equivalently $\bm{v}G_j'
{G_j''}^T=\bm{0}$. Hence, the dimension of $C_j'\cap
{C_j''}^{\perp_{\rm E}}$ is equal to the dimension of the nullspace of
the matrix $G_j' {G_j''}^T$, which is given by
$\dim_{\HH_j'}(C_j')-\rank(G_j' {G_j''}^T)$. This implies that
\begin{equation}\label{eq:rank1}
\dim_{\HH_j'} (C_j')-\dim_{\HH_j'} (C_j' \cap C_j''^\perpE)=\rank(G_j' {G_j''}^T).
\end{equation}
Interchanging the role of $C_j'$ and $C_j''$, we obtain
\begin{equation}\label{eq:rank2}
\dim_{\HH_j''} (C_j'')-\dim_{\HH_j''} (C_j'' \cap C_j'^\perpE)=\rank(G_j'' {G_j'}^T).
\end{equation}
To complete the argument, we recall that $\rank(G_j' {G_j''}^T) =
\rank\bigl((G_j' {G_j''}^T)^T\bigr)=\rank(G_j'' {G_j'}^T)$, \ie,
\eqref{eq:rank1} and \eqref{eq:rank2} are equal.  Then \eqref{eq:k_j}
implies $k_j=2\rank(G_j' {G_j''}^T)$.

\begin{ex}
In the qubit case, our search yields an \emph{optimal}
$[\![22,6,6]\!]_2$ code, which is \emph{strictly better} than the
prior best-known $[\![22,6,5]\!]_2$ code. Let us describe a quaternary
QC code (\ie, $\lambda=1$), which gives rise to the better code. Let
$\omega$ be a primitive element in $\F_4$, $m=7$ and $\ell=3$. Over $\F_4$,
the factorization of $x^7-1$ (cf. (\ref{irreducibles})) is
\[
x^7-1 = (x + 1)(x^3 + x + 1)(x^3 + x^2 + 1).
\]
The first factor $x+1$ is self-conjugate-reciprocal and the remaining
two factors form a conjugate-reciprocal pair. Hence, we have $s=r=1$
such that $\G_1=\F_4$ and $\HH'_1=\HH''_1=\F_{64}$. Assuming this
ordering, consider the following constituents $C_1$, $C_1'$, $C_1''$
of length $3$ whose generator matrices, corresponding to the ordered
factors, are
\begin{alignat}{5}
   G_1&{}:= \begin{pmatrix}  1 & 0 & \omega \\ 0 & 1 & 0\end{pmatrix},\nonumber\\
  G_1'&{}:= \begin{pmatrix}  1 &  \xi^7 & \xi^8 \end{pmatrix},\nonumber\\
 G_1''&{}:=  \begin{pmatrix}  1 &  \xi^{13} & \xi^{56} \end{pmatrix},\label{eq:qubit_cons}
\end{alignat}
where $\xi$ is a primitive element of $\F_{64}$ with minimal
polynomial $\mu_\xi(x)=x^3 + x^2 + x + \omega$ over $\F_4$.

The QC code $C\subseteq\F_4^{21}$ of index $3$ with those constituents
has dimension $8$, by (\ref{dimension}). If $R=\F_4[x]/\langle
x^7-1\rangle$, then, as an $R$-submodule in $R^3$, $C$ is generated by
\begin{alignat*}{5}
\mathbf{f}_1(x)&{}=\bigl(f_{1,0}(x), f_{1,1}(x),f_{1,2}(x)\bigr)\quad\text{and}\\
\mathbf{f}_2(x)&{}=\bigl(f_{2,0}(x), f_{2,1}(x),f_{2,2}(x)\bigr),
\end{alignat*}
with
\begin{align*}
f_{1,0}(x)={}&1, \\
f_{1,1}(x)={}&x^6 + x^3 + \omega^2x,\\
f_{1,2}(x)={}&x^6 + x^5 + \omega^2x^4 + \omega x^3 + \omega x^2 + \omega x + \omega^2,\\
f_{2,0}(x)={}&x^6,\\
f_{2,1}(x)={}&\omega^2x^6 + \omega x^5 + \omega^2x^4 + \omega^2x^3 + \omega x^2 + \omega^2x,\\
f_{2,2}(x)={}&\omega^2x^6 + x^5 + x^4 + \omega^2x^3 + \omega x^2 + \omega x + \omega .
\end{align*}
The generator polynomials are found by applying
\eqref{eq:trace_codeword} to the constituents given in
\eqref{eq:qubit_cons} followed by the map $\phi$ in
\eqref{identification-1}. Conversely, one obtains the constituents
listed in \eqref{eq:qubit_cons} by evaluating $\mathbf{f}_1(x)$ and
$\mathbf{f}_2(x)$ at $1$, $\xi^9$, $\xi^{45}$, respectively. All
constituents, except for $C_1$, satisfy the requirements of Theorem
\ref{SO_criteria}, whereas $\dim (C_1 \cap C_1^\perpH)=1$, implying
$e=1$. Hence, $C \cap C^\perpH$ has dimension $7$. The
Hermitian dual of $C$ is a $[21,13,6]_4$ code, which attains the
best-known distance for a quaternary code of this length and
dimension, and we have $d(C + C^\perpH)=5$. Thus, by Theorem
\ref{thm:X}, we obtain a $[\![22,6,6]\!]_2$ stabilizer code.

The propagation rules in Proposition \ref{prop:propagation} yield
$[\![23,6,6]\!]_2$, which also improves on the prior best-known
$[\![23,6,5]\!]_2$ code.
\end{ex}

\section{Quasi-Twisted Codes with Designed Symplectic Hulls}\label{sec:QT_section-2}

In this section, we consider QT codes of length $2m\ell$ and index
$2\ell$ over $\Fq$ with respect to the symplectic inner product
defined in \eqref{eq:symplectic_IP}. We characterize symplectic
self-orthogonal and nearly symplectic self-orthogonal QT codes using
their constituents, as we did in the previous section. We distinguish
two cases: first we look at $\lambda$-QT codes where $\lambda=\pm 1$,
and then we consider $(\lambda_1,\lambda_2)$-QT codes where the left
half of the codewords belong to a $\lambda_1$-QT code of length
$m\ell$ and index $\ell$ over $\Fq$, and the right half of the
codewords belong to a $\lambda_2$-QT code of length $m\ell$ and index
$\ell$ over $\Fq$ such that $\lambda_1 \neq \pm 1\ne\lambda_2$.

\subsection{Quasi-Cyclic and Quasi-Negacyclic Codes}
We assume the notation so far and define
\begin{alignat}{5}
\tau\colon&& \F_q^{2m\ell} &{}\longrightarrow{}&& \F_q^{2m\ell}\label{swap map}  \\
&&(\bm{u}_1|\bm{u}_2) &{}\longmapsto{} && (\bm{u}_2|-\bm{u}_1),\nonumber
\end{alignat}
where $\bm{u}_1,\bm{u}_2 \in \F_q^{m\ell}$. Then,
\begin{alignat*}{5}
  \langle (\bm{u}_1|\bm{u}_2),(\bm{v}_1|\bm{v}_2)\rangle_{\rm S}
   ={} & \langle \bm{u}_1,\bm{v}_2\rangle_{\rm E} - \langle \bm{u}_2,\bm{v}_1\rangle_{\rm E} \\
   ={} & \langle (\bm{u}_1|\bm{u}_2),\tau(\bm{v}_1|\bm{v}_2)\rangle_{\rm E} \\
   ={} & -\langle \tau(\bm{u}_1|\bm{u}_2),(\bm{v}_1|\bm{v}_2)\rangle_{\rm E}.
\end{alignat*}
After this preparation, it is easy to observe that the symplectic dual
$C^\perpS$ of a given $q$-ary $\lambda$-QT code $C$ of index $2\ell$
satisfies $C^\perpS=\tau(C^\perpE) = \tau(C)^\perpE$, making it a
$\lambda^{-1}$-QT code of index $2\ell$. In order to talk about
symplectic self-orthogonality, we have to restrict ourselves to the
case $\lambda=\lambda^{-1}$, \ie, $\lambda=\pm 1$. Recall that $1$-QT
codes correspond to QC codes and we call $(-1)$-QT codes
\emph{quasi-negacyclic} (QN) codes.

Observe that $f(x) = - f^*(x)$ when $f(x)=x^m-1$ and $f(x) =  f^*(x)$ when $f(x)=x^m+1$. We factor the polynomials $x^m\pm 1$ into irreducible polynomials in $\F_{q}[x]$ as
\begin{equation}\label{irreducibles-2}
x^{m}\pm 1=g_{1}(x)\cdots g_{s}(x) h_{1}(x)h_{1}^*(x)\cdots h_{r}(x)h_{r}^*(x),
\end{equation}
where the polynomials $g_{i}$ are self-reciprocal where\-as $h_{j}$ and
$h_{j}^*$ are reciprocal pairs, for all $i,j$.  Since $m$ is
relatively prime to $q$, there are no repeating factors in
(\ref{irreducibles}). By setting $R:=\Fq[x]/\langle x^m\pm 1 \rangle$
and using the CRT, we have
\begin{alignat}{5}
R \cong{}& \left( \bigoplus_{i=1}^{s} \F_{q}[x]/\langle g_{i}(x) \rangle \right)\nonumber\\
&\quad\oplus \left( \bigoplus_{j=1}^{r} \Bigl( \F_{q}[x]/\langle h_{j}(x) \rangle \oplus \F_{q}[x]/\langle h_{j}^*(x) \rangle \Bigr) \right). \label{CRT-S}
\end{alignat}

Let $\alpha=1$ if we consider $x^m-1$, otherwise let $\alpha$ be a
primitive $(2m)^\text{th}$ root of unity such that $\alpha^m=-1$. Let
$\xi$ be a primitive $m^\text{th}$ root of unity, with $\xi=\alpha^2$
for $\alpha\ne 1$. Then the roots of $x^m\pm 1$ are of the form
$\alpha, \alpha\xi, \ldots, \alpha\xi^{m-1}$. Since each $g_i(x)$,
$h_{j}(x)$, and $h_{j}^*(x)$ divide $x^m\pm 1$, their roots are of the
form $\alpha\xi^k$, for some $0\leq k \leq m-1$. For each
$i\in\{1,\ldots,s\}$, let $u_i$ be the smallest nonnegative integer
such that $g_i(\alpha\xi^{u_i})=0$. For each $j\in\{1,\ldots, r\}$,
let $v_j$ be the smallest nonnegative integer such that
$h_j(\alpha\xi^{v_j})=h_{j}^*(\alpha^{-1}\xi^{-v_j})=0$. Similarly,
let $\G_{i}=\F_{q}[x]/\langle g_{i}(x) \rangle$,
$\HH_{j}'=\F_{q}[x]/\langle h_{j}(x) \rangle$ and
$\HH_{j}''=\F_{q}[x]/\langle h_{j}^*(x) \rangle$ denote the respective
field extensions, for each $i$ and $j$. By the CRT, the analogs of
\eqref{CRT-2}, \eqref{CRT-3}, \eqref{constituents},
\eqref{eq:explicit_constituents}, and \eqref{eq:trace_codeword} easily
follow, where the evaluation at $\alpha^{-q}\xi^{-qv_j}$ over
$\HH_{j}''$ is replaced by the evaluation at $\alpha^{-1}\xi^{-v_j}$
in this setup.

Let the decomposition of a QC or QN code $C\subseteq R^{2\ell}$ be given by
\begin{equation}\label{constituents-2}
C\cong\left( \bigoplus_{i=1}^{s} C_i \right) \oplus \left( \bigoplus_{j=1}^{r} \Bigl( C_{j}' \oplus C_{j}'' \Bigr) \right),
\end{equation}
where the codes $C_i$ are the $\G_i$-linear constituents of length
$2\ell$, for all $i\in\{1,\ldots,s\}$, while the $C'_{j}$ and $C''_{j}$ are
the $\HH'_j$- and $\HH''_j$-linear constituents of length $2\ell$,
respectively, for all $j\in\{1,\ldots,r\}$. For each $1\leq i \leq s$, let
$\G_i^{2\ell}$ be equipped with the inner product $\langle
\bm{u},\bm{v} \rangle_{\G_i} = \langle \bm{u},\bm{v}
\rangle_{\rm E}$ if $g_i(x)=x\pm 1$ and $\langle \bm{u},\bm{v}
\rangle_{\G_i} = \langle \bm{u},\bm{v} \rangle_{\rm H}$
otherwise. For $1\leq j \leq r$, let $(\HH_j')^{2\ell}$ and
$(\HH_j'')^{2\ell}$ be equipped with the usual Euclidean inner
product. Observe that $\HH_j'\cong\HH_j''$ follows from the fact that
$\F_{q}(\alpha\xi^a)=\F_{q}(\alpha^{-1}\xi^{-a})$.

The (Euclidean) dual of a QC (respectively, QN) code is again a QC (respectively, QN) code with the decomposition (see \cite[Proposition 7.3.5]{GLO2021})
\begin{equation}\label{dual-2}
C^\perpE\cong\left( \bigoplus_{i=1}^s C_i^{\bot_{\rm \G_i}} \right) \oplus \left( \bigoplus_{j=1}^r \Bigl( (C_j'')^\perpE \oplus (C_j')^\perpE \Bigr)\right).
\end{equation}
Extending the map $\tau$ given in (\ref{swap map}) canonically to the
vector spaces $\G_i^{2\ell}$, ${\HH_j'}^{2\ell}$, and
${\HH_j'}^{2\ell}$ and applying these maps componentwise to
\eqref{dual-2}, we obtain the following result.

\begin{prop}\label{symplectic duality}
Let $C$ be a QC or QN code over $\F_{q}$ with the CRT decomposition as in (\ref{constituents-2}). Then its symplectic dual $C^\perpS$ satisfies
\begin{equation}\label{dual-3}
C^\perpS\cong\left( \bigoplus_{i=1}^s C_i^{\bot_{\rm S_i}} \right) \oplus \left( \bigoplus_{j=1}^r \Bigl( (C_j'')^\perpS \oplus (C_j')^\perpS \Bigr)\right),
\end{equation}
where  $C_i^{\bot_{\rm S_i}} = \tau(C_i)^{\bot_{\rm \G_i}} $ on $\G_i^{2\ell}$, for all $1\leq i \leq s$, and  $\bot_{\rm S}$ denotes the symplectic dual on $(\HH_j')^{2\ell} = (\HH_j'')^{2\ell}$, for all $1\leq j \leq r$.
\end{prop}

Note that, in contrast to the situation of the Hermitian inner product
in Section \ref{sec:QT_section}, no adjustment is required for the
symplectic inner product when we use the very same representation of
the isomorphic fields $\HH_j'$ and $\HH_j''$. The code $C'_j$ is
obtained by evaluation at $\alpha\xi^{v_j}$, and the code $C''_j$ by
evaluation at $\alpha^{-1}\xi^{-v_j}$. In combination with the
``conjugation'' map $\bar{\ }: x\mapsto x^{-1}$ on $R$ and the
Hermitian inner product on $R^\ell$ used in Proposition 3.2 of
Ref.~\cite{LS01}, we obtain the usual Euclidean and symplectic inner
products between the codes $C_j'$ and $C_j''$ expressed in the very
same field.

\begin{rem}
For $1\leq i \leq s$, $C_i^{\bot_{\rm S_i}} = \tau(C_i)^{\bot_{\rm \G_i}} =C_i^\perpS $ if $\G_i^{2\ell}$ is equipped with the Euclidean inner product. Otherwise, $C_i^{\bot_{\rm S_i}} = \tau(C_i)^{\bot_{\rm \G_i}} =C_i^{\bot_{\rm SH}} $, where
\[
\langle (\bm{u}_1|\bm{u}_2),(\bm{v}_1|\bm{v}_2)\rangle_{\rm SH} : =  \langle \bm{u}_1,\bm{v}_2\rangle_{\rm H} - \langle \bm{u}_2,\bm{v}_1\rangle_{\rm H},
\]
whenever $\G_i^{2\ell}$ is equipped with the Hermitian inner product.
\end{rem}

We are now ready to characterize symplectic self-orthogonal QC and QN codes using their constituents.

\begin{thm} \label{SO_criteria-2}
Let $C$ be a QC or QN code of length $2m\ell$ over $\F_{q}$ whose CRT
decomposition is as in (\ref{constituents-2}). Then $C$ is symplectic
self-orthogonal if and only if $C_i$ is self-orthogonal with respect
to the inner product $\langle\_,\_\rangle_{\rm S_i}$ over $\G_i$, for
all $1\leq i \leq s$, and $C_j'' \subseteq (C_j')^\perpS$
(equivalently, $C_j' \subseteq (C_j'')^\perpS$) over $\HH_j' =
\HH_j''$, for all $1\leq j \leq r$.
\end{thm}
\begin{IEEEproof}
Immediate from the CRT decompositions of $C$ in \eqref{constituents-2} and of its symplectic dual $C^\perpS$ in \eqref{dual-3}.
\end{IEEEproof}

As it was done in Section \ref{sec:QT_section} above for nearly
Hermitian self-orthogonality, we use the CRT decompositions of a QC or
QN code $C$ in \eqref{constituents-2} and its symplectic dual
$C^\perpS$ in \eqref{dual-3} to characterize nearly symplectic
self-orthogonal QC and QN codes.

Let $k$ denote the dimension of $C$ over $\F_{q}$ so that $C^\perpS$
has dimension $2m\ell-k$. Analogously, for $1\leq i \leq s$, let
$e_i:=[\G_i : \F_{q}]=\deg (g_i(x))$ and let $e_j:=[\HH_j' :
  \F_{q}]=[\HH_j'' : \F_{q}]=\deg( h_j(x)) = \deg(h_j^*(x))$, for all
$1\leq j \leq r$. Then (\ref{dimension}) holds for $k$, and by using
Proposition \ref{symplectic duality}, we obtain
(cf. (\ref{hull-dimension})) that the dimension of the symplectic hull
$C\cap C^\perpS$ is
\begin{alignat}{5}
  &\dim(C\cap C^\perpS)=\sum_{i=1}^s e_i \dim_{\G_i} (C_i \cap C_i^{\bot_{\rm  S_i}})\nonumber\\
  &\quad +\sum_{j=1}^r e_j \left( \dim_{\HH_j'} (C_j' \cap C_j''^\perpS) + \dim_{\HH_j''} (C_j'' \cap C_j'^\perpS)\right).\label{hull-dimension-2}
\end{alignat}
If $C$ is symplectic self-orthogonal, then $C=C\cap C^{\bot_{\rm
    S}}$. If $C$ is not symplectic self-orthogonal, then $k - \dim(C
\cap C^\perpS)$ must be an even number. To see this, we know that, for
any $\bm{c}\in C\setminus\left( C \cap C^\perpS\right)$, there exists
another element $\bm{c'}\in C\setminus\left(C \cap C^\perpS\right)$
such that $\langle \bm{c},\bm{c'}\rangle_{\rm S}\neq 0$, since
$\langle \bm{c},\bm{c}\rangle_{\rm S}=\langle
\bm{c'},\bm{c'}\rangle_{\rm S}=0$. Note that $\langle
\bm{c},\bm{c'}\rangle_{\rm S}\neq 0$ implies $\langle
\bm{c},\bm{c'}\rangle_{\rm S}=-\langle \bm{c'},\bm{c}\rangle_{\rm
  S}\neq 0$, ensuring $\bm{c'}\notin C \cap C^\perpS$.  Moreover,
$\bm{c}$ and $\bm{c}'$ are linearly independent, as $\langle
\bm{c},\beta\bm{c}\rangle_{\rm S}=0$ for all $\beta\in\F_q$. One can
continue inductively and show that $k - \dim(C \cap C^\perpS)=2e$, for
some nonnegative integer $e$.

Now we assume that all constituent codes except for one $C_i$, with
$i$ such that $1\leq i \leq s$, or for one pair of constituents $C_j'$
and $C_j''$, for a $j$ such that$1\leq j \leq r$, satisfy the
requirements of Theorem \ref{SO_criteria-2}. If we set
\begin{equation}\label{eq:k_i_symplectic}
2k_i:=\dim_{\G_i}(C_i) - \dim_{\G_i}(C_i \cap C_i^{\bot_{\rm S_i}})
\end{equation}
for the chosen constituent $C_i$, and all the remaining constituents
$C_{i'}$, $C_j'$, $C_j''$ agree with the conditions in Theorem
\ref{SO_criteria-2}, for all $1\leq i' \leq s$, with $i'\neq i$, and
for $1\leq j \leq r$, then we easily get $2e=k - \dim(C \cap
C^{\bot_{\rm S}})=2e_i k_i$. Note that the discussion above extends to
the field $\G_i$, showing that the right hand side of
\eqref{eq:k_i_symplectic} is even.  Similarly, if we set
\begin{alignat}{5}
2k_j:={}&\dim_{\HH_j'} (C_j')-\dim_{\HH_j'} (C_j' \cap C_j''^\perpS)\nonumber\\
     &{}+\dim_{\HH_j''}(C_j'')-\dim_{\HH_j''} (C_j'' \cap C_j'^\perpS)\label{eq:k_j_symplectic}
\end{alignat}
for the chosen pair of constituents $C_j'$, $C_j''$, for a $j$ in the
range $1\leq j \leq r$, and all the remaining constituents agree with
the conditions in Theorem \ref{SO_criteria-2}, then we obtain $2e=k -
\dim(C \cap C^\perpS)=2e_j k_j$. Similar to the Hermitian case, one
can show that the right hand side of \eqref{eq:k_j_symplectic} is
even.

\subsection{$(\lambda_1,\lambda_2)$-Quasi-Twisted Codes}

For $\lambda_1,\lambda_2 \in \F_q\setminus\{0\}$, we now consider the
case when the given $q$-ary $(\lambda_1,\lambda_2)$-QT code $C$ of
length $2m\ell$ has codewords which are invariant under the
$\lambda_1$-constashift by $\ell$ units in the left half and invariant
under the $\lambda_2$-constashift by $\ell$ units in the right half,
applying both shifts simultaneously.
This is a particular case of generalized quasi-twisted codes,
  sometimes referred to as multi-twisted codes.  Note that if
$\lambda_1=\lambda_2=\lambda$, then we simply have a $\lambda$-QT code
of length $2m\ell$ and index $2\ell$ over $\Fq$ and we go back to the
above case, which requires $\lambda=\pm 1$. Now assume that $\lambda_1
\neq \lambda_2$. Then, the symplectic dual $C^\perpS = \tau(C^\perpE)$
is a $(\lambda_2^{-1},\lambda_1^{-1})$-QT code in $\F_q^{2m\ell}$. In
order to talk about symplectic self-orthogonality, we require
$\lambda_1=\lambda_2^{-1}$ (equivalently,
$\lambda_2=\lambda_1^{-1}$). Hence, we consider $q$-ary
$(\lambda,\lambda^{-1})$-QT codes such that $\lambda \in
\F_q\setminus\{0,1,-1\}$.

Note that, if $C$ is a $(\lambda,\lambda^{-1})$-QT code, then by
applying the constashift $m\ell$ times, we have
\begin{alignat*}{5}
&\bm{c}=(\bm{c}_1|\bm{c}_2) \in C\\
&\quad\Longrightarrow (\lambda\bm{c}_1|\lambda^{-1}\bm{c}_2) \in C\\
&\quad\Longrightarrow  (-\lambda^2\bm{c}_1|-\bm{c}_2) \in C\\
&\quad\Longrightarrow (\bm{c}_1|\bm{c}_2) + (-\lambda^2\bm{c}_1|-\bm{c}_2) = (1-\lambda^2)(\bm{c}_1|\bm{0})\in C\\
&\quad\Longrightarrow (\bm{c}_1|\bm{0})\in C.
\end{alignat*}
Another implication is that clearly $(\bm{0}|\bm{c}_2)\in C$,
resulting in $C=C_1 \times C_2$, where $C_1$ and $C_2$ are
$\lambda$-QT and $\lambda^{-1}$-QT codes of length $m\ell$ and index
$\ell$, respectively. Then, the symplectic dual of $C$ must be of the
form $C^\perpS = \tau(C^\perpE)=C_2^\perpE \times C_1^\perpE$. Hence,
$C\subseteq C^\perpS$ if and only if $C_1 \subseteq C_2^\perpE$
(equivalently, $C_2 \subseteq C_1^\perpE$), which is the case of the
CSS construction in Proposition \ref{prop:CSS_constr}.

Now let us characterize such pairs of $\lambda$-QT and
$\lambda^{-1}$-QT codes, which can be used in the CSS construction, in
terms of their constituents.  We follow the notation in \cite{LO2019}
and refer the reader to \cite{Y} for the proofs of the following
results, and for the general repeated-root case (\ie, when $\gcd(m,q)>
1$). We factor the polynomial $x^m-\lambda$ into irreducible
polynomials in $\F_q[x]$ as
\begin{equation}\label{irreducibles-3}
x^m-\lambda=f_1(x)f_2(x)\ldots f_y(x).
\end{equation}
Since $m$ is relatively prime to $q$, there are no repeating factors
in (\ref{irreducibles-3}). By setting $R:=\F_q[x]/\langle
x^m-\lambda\rangle$ and using the CRT, we have the ring isomorphism
\begin{equation} \label{CRT-4}
R \cong \bigoplus_{i=1}^{y} \F_q[x]/\langle f_i(x)\rangle .
\end{equation}

Let $t$ be the smallest divisor of $q-1$ with
$\lambda^t=1$.  We know that the roots of $x^m-\lambda$ are of the
form $\alpha, \alpha\xi, \ldots, \alpha\xi^{m-1}$, where $\alpha$ is a
primitive $(tm)^\text{th}$ root of unity such that $\alpha^m=\lambda$ and
$\xi:=\alpha^t$ is a primitive $m^\text{th}$ root of unity.  For each
$i\in\{1,\ldots, y\}$, let $u_i$ be the smallest nonnegative integer such
that $f_i(\alpha\xi^{u_i})=0$. Since the polynomials $f_i(x)$ are
irreducible, the direct summands in (\ref{CRT-4}) can be viewed as
field extensions of $\F_q$, obtained by adjoining the element
$\alpha\xi^{u_i}$. If we set $\F_i:=\Fq(\alpha\xi^{u_i})\cong
\Fq[x]/\langle f_i(x) \rangle$, for each $1\leq i \leq y$, then
$\big[\F_i : \F_q\big]=\mbox{deg}(f_i)$ and the analogs of
(\ref{CRT-2}) and (\ref{CRT-3}) immediately follow. Hence, a
$\lambda$-QT code $C\subset R^\ell$ can be viewed as an $R$-submodule
of $R^\ell$ and decomposes as
\begin{equation} \label{constituents-3}
C \cong C_1\oplus \cdots  \oplus C_{y},
\end{equation}
where $C_i$ is the constituent code of length $\ell$ over $\F_i$, for each $i$.

Given a $\lambda$-QT code $C\subset R^\ell$, its (Euclidean) dual
$C^\perpE$ is a $\lambda^{-1}$-QT code, which can be viewed as
an $R^{-1}$-submodule of $(R^{-1})^\ell$, where
$R^{-1}=\Fq[x]/{\langle x^{m}-\lambda^{-1}\rangle}$. Therefore, $C$
and $C^\perpE$ are not defined over the same ring for $\lambda
\neq \pm 1$.  Consider the identification
\begin{alignat}{5}
R^{-1} & {}\cong{} && R \nonumber\\
x & {}\leftrightarrow{} && x^{-1},\label{inversion}
\end{alignat}
where $x^{-1}=\lambda^{-1}x^{m-1}$ in $R$ and $x^{-1}=\lambda x^{m-1}$
in $R^{-1}$. By \eqref{inversion}, $R$ and $R^{-1}$ are isomorphic. We
define the following map (cf. \cite[Definition 6]{Y})
\begin{alignat}{9}
\phi\colon && (R^{-1})^\ell & {}\longrightarrow{} && R^\ell\nonumber\\
&&\bigl(a_0(x),\ldots,a_{\ell-1}(x)\bigr) & \longmapsto && \bigl(a_0(x^{-1}),\ldots,a_{\ell-1}(x^{-1})\bigr), \label{inversion-2}
\end{alignat}
which is clearly a bijection by (\ref{inversion}) and $\phi^{-1}=\phi$. Therefore, $\phi$ induces a one-to-one correspondence between the $R^{-1}$-submodules of $(R^{-1})^\ell$ and the $R$-submodules of $R^\ell$ (see \cite[Proposition 2]{Y}). Hence, given a $\lambda$-QT code $C\subseteq R^\ell$, its dual $C^\perpE\subseteq (R^{-1})^\ell$ has an isomorphic copy $\phi(C^\perpE)\subseteq R^\ell$ (cf. \cite[Theorem 2]{Y}).

If $f(x) = x^m-\lambda$, then $x^m-\lambda^{-1} = -\lambda^{-1}f^*(x)$. Hence, by (\ref{irreducibles-3}), we get
\begin{equation}\label{irreducibles-rec}
x^{m}-\lambda^{-1}=(-\lambda^{-1})f_1^*(x)f_2^*(x)\ldots f_y^*(x),
\end{equation}
For all $i$, if $f_i(x)$ is irreducible, then $f_i^*(x)$ is also irreducible. Therefore, (\ref{irreducibles-rec}) is the factorization of $x^m-\lambda^{-1}$ into irreducibles in $\Fq[x]$. Analogously, we get
\begin{equation} \label{CRT-5}
R^{-1}\cong \bigoplus_{i=1}^{y} \F_q[x]/\langle f_i^*(x)\rangle \cong \bigoplus_{i=1}^{y} \F_i^\star,
\end{equation}
where $\F_i^\star\cong \Fq[x]/\langle f_i^*(x) \rangle$ is an extension of $\Fq$, following the same arguments as before. Hence, $C^\perpE$ decomposes as
\begin{equation} \label{constituents-5}
C^\perpE\cong C_1^\star\oplus \cdots  \oplus C_{y}^\star
\end{equation}
such that $C_i^\star$ is the constituent code in $(\F_i^\star)^{\ell}$, for each $i\in\{1,\ldots,y\}$.

Note that when $\lambda\neq \pm 1$, the polynomials $x^{m}-\lambda$ and $x^{m}-\lambda^{-1}$ are relatively prime over $\Fq$. Therefore, the irreducible polynomials $f_i(x)$ and $f_j^*(x)$ given in (\ref{irreducibles-3}) and (\ref{irreducibles-rec}) are pairwise coprime for all $1 \leq i,j \leq y$. Thus, no irreducible factor is an associate of its reciprocal and no reciprocal pair exists in the factorization
of $x^{m}-\lambda$ and $x^{m}-\lambda^{-1}$, unlike in the case $\lambda=\pm 1$.

Recall that $\F_i = \Fq(\alpha\xi^{u_i})$, for each $i\in\{1,\ldots,y\}$,
where $u_i$ is the smallest nonnegative integer such that
$f_i(\alpha\xi^{u_i})=0$. Then, $\F_i^\star =
\Fq(\alpha^{-1}\xi^{-u_i})$, for each $i\in\{1,\ldots,y\}$, since
$f_i^*(\alpha^{-1}\xi^{-u_i})=0$. One can easily observe that $\F_i$
and $\F_i^\star$ are isomorphic since they are extension fields of
$\F_q$ of the same degree.
Using the representation $\F_i=\F_q[x]/\langle f_i(x)\rangle$ and
$\F_i^\star=\F_q[x]/\langle f_i^*(x)\rangle$, the isomorphism is given
by (cf. \eqref{inversion} and \eqref{inversion-2} as well as \cite[Definition 8]{Y})
\begin{alignat}{5}
\phi_i \colon&& \F_q[x]/\langle f_i(x)\rangle &{} \longrightarrow  \F_q[x]/\langle f_i^*(x)\rangle\nonumber\\
  && a(x)+\langle f_i(x)\rangle&{} \longmapsto  a(x^{-1})+\langle f_i^*(x)\rangle.\label{eq:map_phi_i}
\end{alignat}
The isomorphism $\phi_i$ naturally extends to $\ell$-tuples.

With this preparation, we are ready to state the following result and
we refer to \cite[Theorem 4]{Y} for its proof.
\begin{prop}\label{prop:EuclideanDual}
Given $\lambda\in \Fq\setminus \{0,1,-1\}$ and a $\lambda$-QT code
$C\cong C_{1} \oplus \cdots  \oplus C_{y}\subseteq R^\ell$, its
Euclidean dual decomposes as
\[
C^\perpE\cong C_1^\star\oplus \cdots  \oplus C_{y}^\star\subseteq (R^{-1})^\ell,
\]
with an isomorphic copy
\[
\phi(C^\perpE)\cong \bigoplus_{i=1}^y C_i^{\bot_{\rm  E}}=\bigoplus_{i=1}^y \phi_i\left(C_{i}^\star\right)\subseteq R^\ell.
\]
\end{prop}

Now we can characterize a pair of a $\lambda$-QT code and a
$\lambda^{-1}$-QT code, where one is contained in the dual of the
other.
\begin{thm}\label{SO_criteria-3}
Let $\lambda\in \Fq\setminus \{0,1,-1\}$. Given a $\lambda$-QT code $C_1\cong C_{1,1} \oplus \cdots  \oplus C_{1,y}\subseteq R^\ell$ with $C_1^\perpE\cong C_{1,1}^\star\oplus \cdots  \oplus C_{1,y}^\star\subseteq (R^{-1})^\ell,$ and a $\lambda^{-1}$-QT code $C_2\cong C_{2,1} \oplus \cdots  \oplus C_{2,y}\subseteq (R^{-1})^\ell$ with $C_2^\perpE\cong C_{2,1}^\star\oplus \cdots  \oplus C_{2,y}^\star\subseteq R^\ell,$ then $C_1 \subseteq C_2^\perpE$ (equivalently, $C_2 \subseteq C_1^\perpE$) if and only if $C_{1,i} \subseteq C_{2,i}^\star = \phi_i(C_{2,i}^\perpE)$ (equivalently, $C_{2,i} \subseteq C_{1,i}^\star = \phi_i(C_{1,i}^\perpE)$), for each $i\in \{1,\ldots,y\}$.
\end{thm}
\begin{IEEEproof}
Immediate from the CRT decompositions of $C$ in \eqref{constituents-3}
and of its Euclidean dual given in Proposition \ref{prop:EuclideanDual}.
\end{IEEEproof}
Note that the map $\phi_i$ in \eqref{eq:map_phi_i} corresponds to the
evaluation at $\alpha^{-1}\xi^{-u_i}$ to obtain the constituent code
$C_{2,i}^\star$.  Hence, we can directly compare the codes $C_{1,i}$
and $C_{2,i}^\star$ when using the identical representation of the
isomorphic fields $\F_i$ and $\F_i^\star$, similar to the symplectic
case.

Let $k_1$ and $k_2$ denote the dimensions of $C_1$ and $C_2$ over $\F_{q}$, respectively, so that $C=C_1\times C_2$ has dimension $k_1+k_2$ and $C^\perpS$ has dimension $2m\ell-k_1-k_2$. For $1\leq i \leq y$, let $e_i:=[\F_i : \F_{q}]=\deg (f_i(x))=\deg (f_i^*(x))=[\F_i ^\star: \F_{q}]$. Then
\begin{equation*}
\dim(C\cap C^\perpS) = \dim(C_1\cap C_2^\perpE) + \dim(C_2\cap C_1^\perpE).
\end{equation*}
By using Theorem \ref{SO_criteria-3}, we obtain
\begin{alignat*}{5}\label{hull-dimension-3}
\dim(C_1\cap C_2^\perpE)  &{}=  \sum_{i=1}^y e_i \dim_{\F_i} (C_{1,i} \cap C_{2,i}^\star),\\
 \dim(C_2\cap C_1^\perpE) &{}= \sum_{i=1}^y e_i \dim_{\F_i^\star} (C_{2,i} \cap C_{1,i}^\star).
\end{alignat*}
Hence,
\begin{alignat*}{5}
  & \dim(C) - \dim(C \cap C^\perpS)\\
  &\quad= \dim(C_1) - \dim(C_1\cap C_2^\perpE)\\
  &\quad\qquad+ \dim(C_2) - \dim(C_2\cap C_1^\perpE)\\
  &\quad= 2\sum_{i=1}^y e_i \left(\dim_{\F_i}(C_{1,i}) -\dim_{\F_i} (C_{1,i} \cap C_{2,i}^\star)\right)\\
  &\quad= 2\sum_{i=1}^y e_i \left(\dim_{\F_i^\star}(C_{2,i}) - \dim_{\F_i^\star} (C_{2,i} \cap C_{1,i}^\star)\right),
\end{alignat*}
where $\dim_{\F_i}(C_{1,i}) -\dim_{\F_i} (C_{1,i} \cap
C_{2,i}^\star)=\dim_{\F_i^\star}(C_{2,i}) - \dim_{\F_i^\star} (C_{2,i}
\cap C_{1,i}^\star)$ holds, for each $i$, by \eqref{eq:rank1},
\eqref{eq:rank2}, and the fact that the isomorphism $\phi_i$ preserves
the dimension.

Now we assume that all constituent codes, except for one pair of
constituents $C_{1,i}$ and $C_{2,i}^\star$ (equivalently, $C_{2,i}$
and $C_{1,i}^\star$), for an $i$ in the range $1\leq i \leq y$,
satisfy the requirements of Theorem \ref{SO_criteria-3}. If we set
\begin{alignat*}{5}
\kappa_i &{}=\dim_{\F_i}(C_{1,i}) -\dim_{\F_i} (C_{1,i} \cap C_{2,i}^\star)\\
&{}= \dim_{\F_i^\star}(C_{2,i}) - \dim_{\F_i^\star} (C_{2,i} \cap C_{1,i}^\star)
\end{alignat*}
for the chosen pair of constituents $C_{1,i}$, $C_{2,i}^\star$
(equivalently, $C_{2,i}$, $C_{1,i}^\star$), and all the remaining pairs
of constituents $C_{1,i'}$, $C_{2,i'}^\star$ (equivalently, $C_{2,i'}$,
$C_{1,i'}^\star$) agree with the conditions in Theorem
\ref{SO_criteria-3}, for all $1\leq i' \leq y$, with $i'\neq i$, then
we get $2e=k_1+k_2 - \dim(C \cap C^\perpS)=2e_i \kappa_i$ easily.

\section{New Quantum Codes}\label{sec:findings}
This section presents the outcomes of random searches over constituent
codes and the resulting quantum stabilizer codes, which we have
conducted using the computer algebra system MAGMA \cite{Magma}.  All
the results have been included in the online tables \cite{codetables},
which now not only include qubit codes up to length $256$, but also
codes for $q\in\{3,4,5,7,8\}$ up to length $100$ each.

\subsection{General Remarks}
We have scanned the literature for recent results on good quantum
codes constructed via cyclic-type classical codes.  Table
\ref{table:codelit} lists such quantum codes whose parameters either
coincide with or are better than our results.  The table does not
include the results by Bierbrauer and Edel \cite{BE} on quantum
twisted codes. Their parameters can be found online \cite{BE:tables},
and we have included constructions for those with the currently best
parameters in our online tables as well.  These quantum twisted codes
are based on $\F_q$-linear cyclic codes over $\F_{q^2}$.  Bierbrauer
and Edel also consider a particular version of Construction X with
$e=1$, which they refer to as \emph{standard lengthening}.

Table \ref{table:codelit} is naturally not claimed to be complete, and
we invite others to submit related results.  For codes with matching
parameters, sometimes we may include our codes in the updated online
database \cite{codetables} due to the availability of their explicit
construction in the form of MAGMA routines.  We have, for example, not
been able to reproduce the $[\![70,48,6]\!]_2$ code in \cite[after
  Ex. 1]{GuanLLLS2022}, but we have found an alternative construction.
The entry $[\![32,26,5]\!]_5$ in \cite[Table 7]{GuanLLLS2022} is
incorrect; the printed parameters violate the quantum Singleton bound,
and the quantum code corresponding to the data of the classical code
has only dimension $20$.

\begin{rem}
A number of works in the literature suggest to apply Construction X
\emph{directly} to nearly Euclidean self-orthogonal codes, basically
replacing the Hermitian inner product in Theorem \ref{thm:X} by the
Euclidean inner product.  The authors explicitly omit a proof.
Consider, however, the repetition code of length $4$ over $\F_3$ with
generator matrix $G=(1\,1\,1\,1)$. This code cannot be extended to a
Euclidean self-orthogonal code of length $5$; only to a code of length
$6$.  Hence, the length is increased by $2$, while the codimension of
the (trivial) Euclidean hull in the code is $1$. Moreover, extending
the length of the code to make it Euclidean self-orthogonal using
linearly dependent columns impacts the bound on the minimum distance
of its dual.
\end{rem}

\begin{table}[ht!]
	\caption{Parameters of some good quantum codes built via
		cyclic-type classical codes known prior to our work. They
		are listed here for comparison and record
		keeping.}\label{table:codelit}
	\def\arraystretch{1.15}
	\begin{center}
$\begin{array}{| l | l |@{\,}| l | l |}
	\hline
	\text{qubit} & \multicolumn{1}{c|@{\,}|}{\text{reference}} & %
	\text{qubit} & \multicolumn{1}{c|}{\text{reference}} \\
	\hline &&&\\[-2.7ex]\hline
	[\![22, 2,7]\!]_2 & \text{\cite[Table 1]{DastShiv2023}}  &          [\![56,45,4]\!]_2 &  \text{\cite[Ex. 3.8.3]{RDThesis}} \\{}
	[\![37,17,6]\!]_2 & \text{\cite[Table 1]{DastShiv2023}} &           [\![60, 6,13]\!]_2 & \text{\cite[Table 2]{GuanLLM20222}} \\{}
	[\![38,27,4]\!]_2 &\text{\cite[Ex. 3.8.3]{RDThesis}} &              [\![60,11,12]\!]_2 & \text{\cite[Table 2]{GuanLLM20222}} \\{}
        [\![42, 7,10]\!]_2 & \text{\cite[Table 3]{GuanLLM20222}} &          [\![60,38,6]\!]_2 &  \text{\cite[Ex. 6.5]{AbduBP23}} \\{}
	[\![42,14,8]\!]_2 & \text{\cite[Rem. 2]{GuanLLLS2022}} &            [\![62, 6,14]\!]_2 & \text{\cite[Table 2]{GuanLLM20222}} \\{}
	[\![44,33,4]\!]_2 &  \text{\cite[Ex. 3.8.3]{RDThesis}} &            [\![63,42,6]\!]_2 &  \text{\cite[Ex. 2]{LiuGDZ2023}} \\{}
        [\![45, 5,11]\!]_2 & \text{\cite[Table 2]{GuanLLM20222}} &          [\![64,35,8]\!]_2 & \text{\cite[Table 1]{DastShiv2023}} \\{}
	[\![45,21,7]\!]_2 &  \text{\cite[Ex. 3]{LvLW20b}}  &                [\![70,42,7]\!]_2 &  \text{\cite[Ex. 2]{GuanLLLS2022}} \\{}
	[\![46,35,4]\!]_2 &  \text{\cite[Ex. 3.8.3]{RDThesis}} &            [\![70,48,6]\!]_2 & \text{\cite[Ex. 3]{WangLLS20} and} \\{}
        [\![48, 3,12]\!]_2 & \text{\cite[Table 2]{GuanLLM20222}} &                            & \text{\cite[after Ex. 1]{GuanLLLS2022}} \\{}
	[\![48,37,4]\!]_2 & \text{\cite[Ex. 3.8.3]{RDThesis}} &             [\![73,18,13]\!]_2 & \text{\cite[Table 2]{GuanLLM20222}} \\{}
        [\![49, 4,12]\!]_2 & \text{\cite[Table 2]{GuanLLM20222}} &          [\![78,25,12]\!]_2 & \text{\cite[Table 2]{GuanLLM20222}} \\{}
	[\![50,39,4]\!]_2 &   \text{\cite[Ex. 3.8.3]{RDThesis}} &           [\![82,42,9]\!]_2 &  \text{\cite[Ex. 1]{GuanLLLS2022}} \\{}
        [\![51, 9,11]\!]_2 & \text{\cite[Table 2]{GuanLLM20222}} &          [\![102,70,7]\!]_2 &  \text{\cite[Ex. 2]{LvLW20b}}  \\{}
	[\![51,35,5]\!]_2 & \text{\cite[Ex. 4]{LiuGDZ2023}} &               [\![170,148,5]\!]_2 & \text{\cite[Rem. 2]{GuanLLLS2022}} \\{}
	[\![52,16,10]\!]_2 &  \text{\cite[Table 1]{DastShiv2023}} &         [\![241,193,8]\!]_2 &  \text{\cite[Ex. 3.8.7]{RDThesis}} \\{}
	[\![52,24,8]\!]_2 &  \text{\cite[Table 1]{DastShiv2023}} &          [\![105,80,6]\!]_2 & \text{\cite[Table 2]{GuanLLM20222}} \\{}
        [\![56,11,11]\!]_2 & \text{\cite[Table 2]{GuanLLM20222}} &&\\

        \hline &&&\\[-2.7ex]\hline
	\text{qutrit} & \multicolumn{1}{c|@{\,}|}{\text{reference}} &
        \text{qutrit} & \multicolumn{1}{c|}{\text{reference}} \\
	\hline &&&\\[-2.7ex]\hline
	[\![13, 6,4]\!]_3 & \text{\cite{LvLW20}} &                [\![32,10,8]\!]_3 & \text{\cite[Table 5]{GuanLLLS2022}} \\{}
	[\![14, 2,6]\!]_3 & \text{\cite{LvLW20}} &                [\![34,16,7]\!]_3 & \text{\cite[Table 5]{GuanLLLS2022}} \\{}
	[\![16, 6,5]\!]_3 & \text{\cite[Table 5]{GuanLLLS2022}} & [\![44,24,7]\!]_3 & \text{\cite[Table 5]{GuanLLLS2022}} \\{}
	[\![22,11,5]\!]_3 & \text{\cite{LvLW20}} &                [\![46,22,8]\!]_3 & \text{\cite[Table 5]{GuanLLLS2022}} \\{}
	[\![23,12,5]\!]_3 & \text{\cite{LvLW20}} &&\\
	
	\hline &&&\\[-2.7ex]\hline
	\text{ququad} & \multicolumn{1}{c|@{\,}|}{\text{reference}} &
        \text{ququad} & \multicolumn{1}{c|}{\text{reference}} \\
	\hline &&&\\[-2.7ex]\hline
	[\![11,0,6]\!]_4 & \text{\cite{LvLW20}}   & [\![22,10,6]\!]_4 & \text{\cite[Table 6]{GuanLLLS2022}} \\{}
	[\![15,3,6]\!]_4  &  \text{\cite{LvLW20}} & [\![58,42,6]\!]_4 &  \text{\cite[Table 4]{LvLW20b}} \\{}
        [\![21,11,5]\!]_4 &  \text{\cite{LvLW20}} &&\\

	\hline &&&\\[-2.7ex]\hline
	\text{ququint} & \multicolumn{1}{c|@{\,}|}{\text{reference}} & & \\
	\hline &&&\\[-2.7ex]\hline
	[\![22,10,6]\!]_5 &  \text{\cite[Table 7]{GuanLLLS2022}} && \\
	\hline
\end{array}$
\end{center}
\end{table}

When filling the online tables, we apply the following propagation
rules which can be found in \cite[Theorem 6]{Calderbank1998} for the
case of qubits ($q=2$) as well as in \cite[Theorem 4.1]{Grassl2021}.
\begin{prop}\label{prop:propagation}
Let $\mathcal{Q}$ be an $[\![n,k,d]\!]_q$ code. Then the following
quantum codes exist:
\begin{enumerate}
  \item[i)] $[\![n,k-1, \geq d]\!]_q$ by \emph{subcode construction},
    if $k > 1$ or if $k=1$ and the initial code $\mathcal{Q}$ is
    pure.
  \item[ii)] $[\![n+1,k, \geq d]\!]_q$ by \emph{lengthening}, if $k > 0$.
  \item[iii)] $[\![n-1,k, \geq d-1]\!]_q$ by \emph{puncturing}, if $n
    \geq 2$.
  \item[iv)] $[\![n-1,k+1,d-1]\!]_q$ when the code is pure.
\end{enumerate}
\end{prop}
Unlike the situation of shortening for classical codes, the existence
of a quantum code $[\![n,k,d]\!]_q$ does not necessarily imply the
existence of a code $[\![n-1,k-1,d]\!]_q$. However, using the
so-called \emph{puncture code} by Eric Rains \cite{Rai99:nonbinary},
one can derive sufficient conditions when shortening is possible.

We have used techniques similar as those for quantum MDS codes
\cite{GrRo15} to investigate the puncture code of quantum twisted
codes \cite{BE} and of our new quantum codes. We found that in
particular for relatively small distance and growing alphabet size
$q$, shortening quantum twisted codes to many lengths is possible.

\subsection{Comments on the Randomized Search}
In our randomized search, we first fix the type of inner product, the
field size, the co-index $m$, and the cyclic shift parameter $\lambda$
according to the conditions in Sections \ref{sec:QT_section} and
\ref{sec:QT_section-2}.  Codes with the same co-index $m$ share the
same factorization of the polynomial $x^m-\lambda$ and hence, the
corresponding extension fields.  Note that the monomial equivalence
for consta-cyclic codes (see, \eg, \cite{Dast2024}) need not be
compatible with the inner product.

Using the criteria on orthogonality derived in Sections
\ref{sec:QT_section} and \ref{sec:QT_section-2}, we randomly generate
constituent codes of length $\ell$ yielding self-orthogonal QC/QT
codes with index $\ell$.  We then modify one of the constituents to
obtain an almost self-orthogonal code.  Modifying more than one
constituent usually results in a larger parameter $e$, which makes it
less likely to obtain improved codes.

Most of our new codes are based on the Hermitian construction, using
$\F_{q^2}$-linear codes.  One of the reasons is the fact that when
computing the minimum distance of such a code, it is sufficient to
consider only one of the $q^2-1$ scalar multiples of each
codeword. For the symplectic case and for additive codes, this factor
reduces to $q-1$ and $p-1$, respectively, when $q=p^m$.  Moreover,
there is only a restricted notion of information sets for additive
codes \cite{WhGr06}.  We have many more candidates for which the
verification of the improved minimum distance takes too long.
Computing the minimum distance of the code $[\![116,0,26]\!]_2$, for
example, took about $94$ years of CPU time.  Another reason is that
for random linear codes the chances to have large Hamming distance are
higher than for random additive codes, see, \eg, \cite{WilHor2022}.
This, of course, does not exclude that there are additive codes with
better parameters than $\F_{q^2}$-linear codes.

For CSS codes, the expected minimum distance is also lower since the
codes are defined over a smaller alphabet of size $q$ compared to
$\F_{q^2}$.  We have not found CSS codes with better parameters.

In our search, we have focused on the pure minimum distance, \ie, the
minimum distance of the classical dual-containing code of rate $\ge
1/2$.  Hence, with the exception of Example \ref{ex:impure} and the
code $[\![32,6,10]\!]_4$ in Table \ref{tab:ququad}, all codes are
pure.

Note that for $e>1$, the self-orthogonal extension is not unique.
Depending on the particular choices of bases for the code and for the
complement of the hull in the code, the resulting quantum code can
have a larger minimum distance compared to the guaranteed lower bound.
Such an explicit optimization was carried out in
\cite{LvLW20b,WangLLS20}. We use a deterministic implementation of the
Gram-Schmidt orthonormalization described in Appendix
\ref{sec:Gram-Schmidt}, which nonetheless may result in improved
codes.  We also note that in order to apply the bounds in, \eg,
Theorem \ref{thm:X}, one has to know or compute the minimum distances
of the codes $C^\perpH$ and $C+C^\perpH$.  In many cases, computing
the minimum distance of each of these codes is almost as complex as
directly computing the minimum distance of the quantum code.

\subsection{Tables}
Table \ref{tab:symplectic} contains parameters of new quantum codes
that are obtained using classical QC/QT codes over $\F_q$ (symplectic
case). We list the shift constant $\lambda$ as well as the index
$\ell$ and co-index $m$. Hence, the QC/QT code has length $\ell m$,
and with the extension parameter $e$ we have $2n=\ell m+2e$.
Moreover, we list the number of improved codes that can be derived
using Proposition \ref{prop:propagation}.  For some of the qubit
codes, we also indicate the total CPU time spent to compute the
minimum distance with MAGMA, using up to $400$ cores.

Tables \ref{tab:qubit}--\ref{tab:quoct} contain parameters of new
quantum codes that are obtained from classical QC/QT codes over
$\F_{q^2}$ (Hermitian case).  We do not list the parameters of
  improved codes that can be derived using the propagation rules.
More details can be found in the online tables \cite{codetables}.

\belowcaptionskip0.1\belowcaptionskip

\begin{table}[ht!]
  \caption{Symplectic case using linear codes of length $2n$ over
    $\F_q$.\\ We also list the number of derived codes obtained using
    Proposition \ref{prop:propagation}. The column `CPU time' refers
    to the total time spent on computing the minimum distance with
    MAGMA, using multiple cores.\label{tab:symplectic}}
  \def\arraystretch{1.145}
    \begin{center}
      $\begin{array}{|l|c|ccc|c|c|}
        \hline
        \multicolumn{1}{|c|}{\text{code}} & e & \lambda & \ell & m &\text{derived}&\text{CPU time}\\
        \hline&&&&&&\\[-2.7ex]\hline
     [\![ 56,  34, 6]\!]_2 & 1 & 1 & 2 & 55 & 1 & \\{}
     [\![ 59,  21, 9]\!]_2 & 4 & 1 & 1 & 55 & 3 &\\{}
     [\![ 64,  36, 8]\!]_2 & 1 & 1 & 1 & 63 & 3 &\\{}
     [\![ 73,   9,15]\!]_2 & 0 & 1 & 2 & 73 & 6 & \text{$522$ days} \\{}
     [\![ 73, 37, 9]\!]_2 & 0 & 1 & 1 & 73 & 5 &\\{}
     [\![ 90, 59, 7]\!]_2 & 0 & 1 & 4 & 45 &   & \\{}
     [\![105,  59, 9]\!]_2 & 0 & 1 & 10 & 35 & 2 & \text{$532$ days} \\{}
     [\![141, 118, 5]\!]_2 & 0 & 1 & 6 & 47  & & \\{}
     [\![158, 117, 8]\!]_2 & 0 & 1 & 4 & 69 & 1 & \text{$196$ days}\\{}
     [\![159, 118, 8]\!]_2 & 1 & 1 & 4 & 69 &  & \text{$201$ days}\\{}
     [\![219, 170, 8]\!]_2 & 9 & 1 & 12 & 35 &  4 & \text{$6.6$ years}\\{}
     [\![237, 198, 7]\!]_2 & 0 & 1 & 6 & 79  & 23 & \text{$53$ days}\\{}
     [\![254, 219, 6]\!]_2 & 1 & 1 & 22 & 23  & 1 & \\
  \hline
     [\![33, 7, 9]\!]_3 & 0 & 1 & 6 & 11 & 1 & \\
  \hline
     [\![11,  0, 6]\!]_4 & 0 & 1 & 2 & 11  & & \\{}
     [\![22,  5, 7]\!]_4 & 1 & 1 & 6 & 7  & & \\{}
     [\![32, 25, 4]\!]_4 & 1 & 1 & 2 & 31 & & \\{}
     [\![43, 21, 8]\!]_4 & 0 & 1 & 2 & 43 & 1 & \\{}
     [\![57, 39, 6]\!]_4 & 0 & 1 & 6 & 19 & 3 & \\{}
     [\![59, 37, 7]\!]_4 & 2 & 1 & 6 & 19 & 2 & \\{}
     [\![70, 52, 6]\!]_4 & 0 & 1 & 4 & 35 & 10& \\{}
     [\![90, 69, 6]\!]_4 & 6 & 1 & 24 & 7 & 4 & \\{}
     [\![91, 70, 6]\!]_4 & 6 & 1 & 10 & 17 & 4 & \\{}
     [\![93, 72, 6]\!]_4 & 0 & 1 & 6 & 31 & 8 & \\{}
     [\![95, 75, 6]\!]_4 & 0 & 1 & 10 & 19 & 15 & \\
  \hline
    [\![24, 11,6]\!]_5 & 0 & 1 & 1 & 24 &  & \\{}
    [\![31,  7,10]\!]_5 & 0 & -1 & 2 & 31 & 2 & \\{}
    [\![31,  9, 9]\!]_5 & 0 & -1 & 2 & 31 &  & \\{}
    [\![35, 15, 8]\!]_5 & 2 & -1 & 2 & 33 & 2 & \\{}
    [\![39, 16,9]\!]_5 & 0 & -1 & 1 & 39 &  & \\{}
    [\![59, 36, 8]\!]_5 & 2 & 1 & 6 & 19 &  & \\{}
    [\![68, 48, 7]\!]_5 & 2 & 1 & 4 & 33 & 5 & \\{}
    [\![71, 50, 7]\!]_5 & 5 & -1 & 4 & 33 & 3 & \\
  \hline
    [\![ 33, 14, 8]\!]_7 & 1 & 1 & 2 & 32 & &\\
  \hline
      [\![21,  4, 8]\!]_8 & 0 & 1 & 2 & 21 & & \\{}
      [\![71, 56, 6]\!]_8 & 2 & 1 & 6 & 23 & & \\
  \hline
  \end{array}$
  \end{center}
\end{table}

\begin{table}[ht!]
  \caption{Qubit codes from linear codes over $\F_4=\F_2(\omega)$
    with $\omega^2=\omega+1$\label{tab:qubit}}
   \def\arraystretch{1.145}
   \arraycolsep0.9\arraycolsep
        \begin{center}
	$\begin{array}{|@{\ }l|c|ccc|@{\,}|@{\ }l|c|ccc|}
            \hline
		\multicolumn{1}{|c|}{\text{code}} & e & \lambda & \ell & m &
		\multicolumn{1}{c|}{\text{code}} & e & \lambda & \ell & m\\
                \hline&&&&&&&&&\\[-2.7ex]\hline
[\![ 22,  6, 6 ]\!]_2 & 1 & 1 & 3 & 7 &                [\![ 74,  2, 17 ]\!]_2 & 0 & 1 & 2 & 37 \\{}
[\![ 25, 11, 5 ]\!]_2 & 0 & 1 & 5 & 5 &                [\![ 74, 36, 9 ]\!]_2 & 0 & \omega^2 & 2 & 37 \\{}
[\![ 28,  6, 7 ]\!]_2 & 0 & 1 & 4 & 7 &                [\![ 75,  3, 17 ]\!]_2 & 0 & 1 & 3 & 25 \\{}
[\![ 29,  7, 7 ]\!]_2 & 1 & 1 & 4 & 7 &                [\![ 75, 27, 11 ]\!]_2 & 0 & 1 & 3 & 25 \\{}
[\![ 29, 15, 5 ]\!]_2 & 1 & 1 & 4 & 7 &                [\![ 75, 31, 10 ]\!]_2 & 0 & 1 & 3 & 25 \\{}
[\![ 31, 13, 6 ]\!]_2 & 1 & 1 & 2 & 15 &               [\![ 75, 37, 9 ]\!]_2 & 1 & \omega^2 & 2 & 37 \\{}
[\![ 32,  8, 8 ]\!]_2 & 2 & 1 & 2 & 15 &               [\![ 76,  4, 17]\!]_2 & 1 & 1 & 3 & 25 \\{}
[\![ 35,  9, 8 ]\!]_2 & 1 & \omega^2 & 2 & 17 &        [\![ 76,  6, 16 ]\!]_2 & 1 & 1 & 3 & 25 \\{}
[\![ 37, 21, 5 ]\!]_2 & 2 & 1 & 7 & 5 &                [\![ 76, 30, 10 ]\!]_2 & 1 & 1 & 5 & 15 \\{}
[\![ 39,  3, 11 ]\!]_2 & 0 & 1 & 3 & 13 &              [\![ 78,  4, 17 ]\!]_2 & 0 & 1 & 2 & 39 \\{}
[\![ 39, 19, 6 ]\!]_2 & 1 & \omega & 2 & 19 &          [\![ 78, 30, 11 ]\!]_2 & 0 & \omega & 2 & 39 \\{}
[\![ 40,  2, 11 ]\!]_2 & 1 & 1 & 3 & 13 &              [\![ 78, 40, 9 ]\!]_2 & 0 & 1 & 2 & 39 \\{}
[\![ 41, 13, 8 ]\!]_2 & 2 & 1 & 3 & 13 &               [\![ 78, 48, 7 ]\!]_2 & 0 & \omega^2 & 2 & 39 \\{}
[\![ 41, 25, 5 ]\!]_2 & 2 & 1 & 3 & 13 &               [\![ 79, 3, 18 ]\!]_2 & 1 & 1 & 6 & 13 \\{}
[\![ 42, 10, 9 ]\!]_2 & 0 & 1 & 2 & 21 &               [\![ 79, 5, 17 ]\!]_2 & 4 & 1 & 3 & 25 \\{}
[\![ 42, 14, 8 ]\!]_2 & 0 & 1 & 2 & 21 &               [\![ 79, 43, 8 ]\!]_2 & 4 & 1 & 5 & 15 \\{}
[\![ 43,  7, 10 ]\!]_2 & 1 & 1 & 2 & 21 &              [\![ 81, 51, 7 ]\!]_2 & 3 & \omega^2 & 2 & 39  \\{}
[\![ 43,  9, 9 ]\!]_2 & 1 & 1 & 2 & 21 &               [\![ 82, 2, 19 ]\!]_2 & 0 & 1 & 2 & 41 \\{}
[\![ 43, 15, 8 ]\!]_2 & 1 & 1 & 2 & 21 &               [\![ 82, 42, 9 ]\!]_2 & 0 & \omega & 2 & 41  \\{}
[\![ 43, 27, 5 ]\!]_2 & 1 & 1 & 2 & 21 &               [\![ 90, 48, 9 ]\!]_2 & 0 & \omega^2 & 6 & 15 \\{}
[\![ 44,  8, 10 ]\!]_2 & 2 & 1 & 2 & 21 &              [\![ 94, 0, 22 ]\!]_2 & 0 & 1 & 2 & 47 \\{}
[\![ 44, 16, 8 ]\!]_2 & 2 & 1 & 2 & 21 &               [\![ 96, 0, 22 ]\!]_2 & 2 & 1 & 2 & 47 \\{}
[\![ 44, 24, 6 ]\!]_2 & 2 & 1 & 2 & 21 &               [\![ 98, 0, 24 ]\!]_2 & 0 & 1 & 2 & 49 \\{}
[\![ 45,  7, 10 ]\!]_2 & 0 & 1 & 3 & 15 &              [\![ 98, 54, 9 ]\!]_2 & 7 & 1 & 13 & 7 \\{}
[\![ 45, 11, 9 ]\!]_2 & 0 & 1 & 3 & 15 &               [\![ 100,   2, 22 ]\!]_2 & 0 & \omega^2 & 4 & 25 \\{}
[\![ 45, 21, 7 ]\!]_2 & 3 & \omega^2 & 2 & 21 &        [\![ 100, 52, 10 ]\!]_2 & 0 & \omega^2 & 4 & 25 \\{}
[\![ 46,  4, 11 ]\!]_2 & 1 & 1 & 3 & 15 &              [\![100, 56, 9 ]\!]_2 & 0 & 1 & 4 & 25  \\{}
[\![ 48, 6, 11 ]\!]_2 & 6 & \omega^2 & 2 & 21 &        [\![ 102, 54, 10 ]\!]_2 & 0 & 1 & 2 & 51 \\{}
[\![ 46,  8, 10 ]\!]_2 & 1 & 1 & 3 & 15 &              [\![ 102, 58, 9 ]\!]_2 & 0 & 1 & 2 & 51\\{}
[\![ 46, 12, 9 ]\!]_2 & 1 & 1 & 3 & 15 &               [\![103, 59, 9 ]\!]_2 & 1 & \omega^2 & 6 & 17  \\{}
[\![ 47, 31, 5 ]\!]_2 & 2 & 1 & 3 & 15 &               [\![ 105, 57, 10 ]\!]_2 & 0 & \omega^2 & 5 & 21 \\{}
[\![ 48, 18, 8 ]\!]_2 & 6 & \omega^2 & 2 & 21 &        [\![107, 61, 9 ]\!]_2 & 2 & \omega^2 & 3 & 35  \\{}
[\![ 49,  9, 10 ]\!]_2 & 0 & 1 & 7 & 7 &               [\![108, 62, 9 ]\!]_2 & 3 & 1 & 5 & 21  \\{}
[\![ 49, 13, 9 ]\!]_2 & 0 & 1 & 7 & 7 &                [\![109, 63, 9 ]\!]_2 & 5 & \omega & 8 & 13  \\{}
[\![ 50,  2, 13 ]\!]_2 & 0 & 1 & 2 & 25 &              [\![110, 64, 9 ]\!]_2 & 0 & \omega & 2 & 55  \\{}
[\![ 50, 14, 9 ]\!]_2 & 0 & 1 & 10 & 5 &               [\![113, 29, 18]\!]_2 & 0 & 1 & 1 & 113  \\{}
[\![ 51, 29, 6 ]\!]_2 & 2 & 1 & 7 & 7 &                [\![ 116,   0, 26 ]\!]_2 & 1 & 1 & 5 & 23 \\{}
[\![ 51, 35, 5 ]\!]_2 & 0 & 1 & 3 & 17 &               [\![ 122,  76, 9 ]\!]_2 & 5 & 1 & 9 & 13 \\{}
[\![ 53, 27, 7 ]\!]_2 & 1 & 1 & 4 & 13 &               [\![ 125,  79, 9 ]\!]_2 & 0 & 1 & 5 & 25 \\{}
[\![ 56, 28, 7 ]\!]_2 & 0 & 1 & 8 & 7 &                [\![ 126,  80, 9 ]\!]_2 & 3 & 1 & 3 & 41 \\{}
[\![ 57, 11, 11 ]\!]_2 & 1 & 1 & 8 & 7 &               [\![ 137,  95, 8 ]\!]_2 & 1 & 1 & 8 & 17 \\{}
[\![ 60, 24, 9 ]\!]_2 & 0 & 1 & 4 & 15 &               [\![ 137, 115, 5 ]\!]_2 & 5 & 1 & 4 & 33 \\{}
[\![ 60, 32, 7 ]\!]_2 & 0 & 1 & 4 & 15 &               [\![ 138, 116, 5 ]\!]_2 & 0 & 1 & 6 & 23 \\{}
[\![ 61, 11, 12 ]\!]_2 & 1 & 1 & 4 & 15 &              [\![ 141,  95, 9 ]\!]_2 & 0 & 1 & 3 & 47 \\{}
[\![ 61, 15, 11 ]\!]_2 & 1 & 1 & 4 & 15 &              [\![ 151, 109, 8 ]\!]_2 & 3 & 1 & 4 & 37 \\{}
[\![ 61, 19, 10 ]\!]_2 & 1 & 1 & 4 & 15 &              [\![164,142, 5 ]\!]_2 & 0 & \omega & 4 & 41 \\{}
[\![ 62, 22, 10]\!]_2 & 0 & \omega^2 & 2 & 31 &        [\![ 165, 123, 8 ]\!]_2 & 1 & 1 & 4 & 41 \\{}
[\![ 67, 39, 7 ]\!]_2 & 2 & 1 & 5 & 13 &               [\![ 220, 174, 8 ]\!]_2 & 0 & \omega & 4 & 55 \\{}
[\![ 68, 18, 12 ]\!]_2 & 0 & \omega & 4 & 17 &         [\![ 221, 177, 8 ]\!]_2 & 0 & \omega^2 & 1 & 221 \\{}
[\![ 70,  4, 16 ]\!]_2 & 0 & 1 & 2 & 35 &              [\![ 232, 184, 8 ]\!]_2 & 2 & 1 & 10 & 23 \\{}
[\![ 70,  8, 15 ]\!]_2 & 2 & \omega & 4 & 17 &         [\![ 234, 186, 8 ]\!]_2 & 0 & 1 & 6 & 39 \\{}
[\![ 70, 14, 13 ]\!]_2 & 0 & 1 & 10 & 7 &              [\![ 235, 189, 8 ]\!]_2 & 0 & 1 & 5 & 47 \\{}
[\![ 70, 32, 9 ]\!]_2 & 0 & \omega^2 & 2 & 35 &        [\![ 238, 190, 8 ]\!]_2 & 4 & 1 & 18 & 13 \\{}
[\![ 70, 38, 8 ]\!]_2 & 0 & 1 & 2 & 35 &               [\![ 241, 217, 5 ]\!]_2 & 0 & 1 & 1 & 241 \\{}
[\![ 70, 42, 7 ]\!]_2 & 2 & 1 & 4 & 17 &               [\![ 246, 206, 7 ]\!]_2 & 0 & 1 & 6 & 41 \\{}
[\![ 71,  9, 15 ]\!]_2 & 3 & \omega & 4 & 17 &         [\![ 247, 199, 8 ]\!]_2 & 0 & \omega & 1 & 247 \\{}
[\![ 71, 19, 12 ]\!]_2 & 1 & 1 & 2 & 35 &              [\![ 249, 201, 8 ]\!]_2 & 3 & 1 & 6 & 41 \\{}
[\![ 71, 41, 7 ]\!]_2 & 3 & 1 & 4 & 17 &               [\![ 249, 205, 7 ]\!]_2 & 2 & \omega & 19 & 13 \\{}
[\![ 71, 43, 7 ]\!]_2 & 1 & 1 & 2 & 35 &               [\![ 250, 202, 8 ]\!]_2 & 0 & 1 & 10 & 25 \\{}
[\![ 72, 24, 11 ]\!]_2 & 4 & \omega & 4 & 17&          [\![ 252, 216, 6 ]\!]_2 & 5 & 1 & 19 & 13 \\
\hline
	  \end{array}$
        \end{center}
\end{table}

\begin{table}[ht!]
  \caption{Qutrit codes from linear codes over $\F_9=\F_3(\omega)$
    with $\omega^2=\omega+1$\label{tab:qutrit}}
   \def\arraystretch{1.145}
        \begin{center}
	$\begin{array}{|@{\quad}l|c|ccc|@{\,}|@{\quad}l|c|ccc|}
            \hline
		\multicolumn{1}{|c|}{\text{code}} & e & \lambda & \ell & m &
		\multicolumn{1}{c|}{\text{code}} & e & \lambda & \ell & m\\
                \hline&&&&&&&&&\\[-2.7ex]\hline
[\![ 14, 2, 6 ]\!]_3  & 0 & 1 & 2 & 7 &                 [\![ 44, 20, 8 ]\!]_3 & 0 & 1 & 2 & 22 \\{}
[\![ 16, 4, 6 ]\!]_3  & 0 & 1 & 2 & 8 &                 [\![ 44, 24, 7 ]\!]_3 & 0 & 1 & 2 & 22\\{}
[\![ 17,  7, 5 ]\!]_3 & 1 & 1 & 2 & 8&                  [\![ 45, 21, 8 ]\!]_3 & 1 & 1 & 2 & 22\\{}
[\![ 18, 2, 7 ]\!]_3  & 2 & 1 & 2 & 8 &                 [\![ 46, 12, 11 ]\!]_3 & 2 & 1 & 4 & 11\\{}
[\![ 18, 8, 5 ]\!]_3 & 2 & 1 & 2 & 7 &                  [\![ 46, 22, 8 ]\!]_3 & 0 & -1 & 2 & 23\\{}
[\![ 20, 2, 8 ]\!]_3  & 0 & 1 & 2 & 10   &              [\![ 47, 23, 8 ]\!]_3 & 1 & -1 & 2 & 23\\{}
[\![ 20, 4, 7 ]\!]_3  & 0 & \omega^2 & 2 & 10 &         [\![ 48, 24, 8 ]\!]_3 & 0 & 1 & 3 & 16 \\{}
[\![ 22,  8, 6 ]\!]_3 & 2 & -1 & 2 & 10&                [\![ 50, 2, 16 ]\!]_3  & 0 & 1 & 2 & 25 \\{}
[\![ 26, 6, 8 ]\!]_3  & 0 & 1 & 2 & 13 &                [\![ 50, 4, 15 ]\!]_3  & 0 & 1 & 2 & 25 \\{}
[\![ 28, 6, 9 ]\!]_3  & 2 & \omega^2 & 2 & 13 &         [\![ 51, 1, 16 ]\!]_3  & 0 & 1 & 3 & 17 \\{}
[\![ 31, 3, 10 ]\!]_3  & 1 & 1 & 6 & 5   &              [\![ 52, 0, 17 ]\!]_3  & 1 & 1 & 3 & 17 \\{}
[\![ 33, 3, 11 ]\!]_3  & 0 & 1 & 3 & 11   &             [\![ 52, 24, 9 ]\!]_3 & 0 & \omega^2 & 2 & 27\\{}
[\![ 33, 11, 8 ]\!]_3 & 0 & 1 & 3 & 11 &                [\![ 53,  1, 17 ]\!]_3 & 0 & \omega^6 & 1 & 53\\{}
[\![ 34, 8, 9 ]\!]_3  & 4 & \omega^2 & 3 & 10   &       [\![ 54,  0, 18 ]\!]_3 & 1 & \omega^6 & 1 & 53\\{}
[\![ 34, 12, 8 ]\!]_3 & 0 & 1 & 3 & 11 &                [\![ 54, 2, 16 ]\!]_3  & 4 & 1 & 2 & 25 \\{}
[\![ 34,16, 7 ]\!]_3  & 0 & 1 & 2 & 17 &                [\![ 55, 3, 16 ]\!]_3  & 0 & 1 & 5 & 11 \\{}
[\![ 36, 10, 9 ]\!]_3 & 3 & 1 & 3 & 11&                 [\![56, 0, 17 ]\!]_3 & 4 & 1 & 4 & 3 \\{}
[\![ 37, 7, 10 ]\!]_3  & 2 & 1 & 7 & 5   &              [\![ 56, 2,16 ]\!]_3  & 1 & 1 & 5 & 11 \\{}
[\![ 39, 7, 11 ]\!]_3  & 0 & \omega^6 & 3 & 13   &      [\![ 57, 1, 17 ]\!]_3  & 2 & 1 & 5 & 11 \\{}
[\![ 39, 9, 10 ]\!]_3  & 0 & -1 & 3 & 13  &             [\![ 58, 0, 18 ]\!]_3  & 0 & 1 & 2 & 29 \\{}
[\![39, 13, 9 ]\!]_3 & 0 & -1 & 3 &  13 &               [\![ 58, 2, 18 ]\!]_3  & 0 & 1 & 2 & 29 \\{}
[\![ 40, 2, 13 ]\!]_3  & 1 & 1 & 3 & 13   &             [\![ 58, 30, 9 ]\!]_3 & 0 & 1 & 2 & 29\\{}
[\![ 40, 4, 12 ]\!]_3  & 0 & 1 & 2 & 20   &             [\![ 60,  0, 19 ]\!]_3 & 2 & 1 & 2 & 29\\{}
[\![ 40, 8, 11 ]\!]_3  & 0 & \omega^2 & 2 & 20   &      [\![ 68, 0, 20 ]\!]_3  & 0 & 1 & 2 & 34 \\{}
[\![ 42, 12, 10 ]\!]_3 & 0 & \omega^2 & 3 & 14&         [\![ 69, 35, 10 ]\!]_3 & 1 & 1 & 4 & 17 \\{}
[\![ 44, 2, 14 ]\!]_3  & 0 & 1 & 2 & 22 &               [\![ 73, 25, 14 ]\!]_3  & 0 & 1 & 1 & 73 \\{}
[\![ 44, 14, 10 ]\!]_3 & 0 & 1 & 4 & 11 &               [\![ 74, 56, 6 ]\!]_3  & 0 & 1 & 2 & 37 \\
\hline
	  \end{array}$
        \end{center}
\end{table}

\begin{table}[ht!]
  \caption{Ququad codes from linear codes over $\F_{16}=\F_2(\omega)$
    with $\omega^4=\omega+1$\label{tab:ququad}}
  \def\arraystretch{1.145}
  \arraycolsep0.9\arraycolsep
        \begin{center}
	$\begin{array}{|@{\ }l|c|ccc|@{\,}|@{\ }l|c|ccc|}
            \hline
		\multicolumn{1}{|c|}{\text{code}} & e & \lambda & \ell & m &
		\multicolumn{1}{c|}{\text{code}} & e & \lambda & \ell & m\\
                \hline&&&&&&&&&\\[-2.7ex]\hline
[\![ 23,  7, 7 ]\!]_4 & 2 & 1 & 3 & 7 &               [\![ 47, 17, 10 ]\!]_4 & 2 & \omega^6 & 3 & 15 \\{}
[\![ 27,  7, 8 ]\!]_4 & 0 & 1 & 3 & 9 &               [\![ 47, 21, 9 ]\!]_4 & 2 & \omega^6 & 3 & 15 \\{}
[\![ 28,  8, 8 ]\!]_4 & 0 & 1 & 4 & 7 &               [\![ 47, 25, 8 ]\!]_4  & 2 & 1 & 3 & 15 \\{}
[\![ 29,  9, 8 ]\!]_4 & 1 & \omega^{12} & 4 & 7 &      [\![ 49, 11, 12 ]\!]_4 & 0 & 1 & 7 & 7\\{}
[\![ 30, 10, 8 ]\!]_4 & 0 & 1 & 2 & 15 &              [\![ 51,  1, 18 ]\!]_4 & 0 & \omega^6 & 1 & 51 \\{}
[\![ 30, 20, 5 ]\!]_4 & 0 & 1 & 2 & 15 &              [\![ 51, 5, 17 ]\!]_4 & 0 & \omega^6 & 1 & 51 \\{}
[\![ 31,  5, 10 ]\!]_4 & 1 & 1 & 2 & 15 &             [\![ 51, 9, 15 ]\!]_4 & 0 & \omega^6 & 1 & 51 \\{}
[\![ 31, 11, 8 ]\!]_4  & 1 & 1 & 2 & 15 &             [\![ 51, 13, 14 ]\!]_4 & 0 & \omega^6 & 1 & 51 \\{}
[\![ 32,  6, 10 ]\!]_4 & 4 & \omega^{12} & 4 & 7 &     [\![ 51, 15, 13 ]\!]_4 & 0 & \omega^6 & 1 & 51 \\{}
[\![ 33, 13, 8 ]\!]_4  & 0 & 1 & 3 & 11 &             [\![ 51, 17, 12 ]\!]_4 & 0 & \omega^6 & 1 & 51 \\{}
[\![ 34,  6, 11 ]\!]_4 & 0 & 1 & 2 & 17 &             [\![ 51, 21, 11 ]\!]_4 & 0 & \omega^6 & 1 & 51 \\{}
[\![ 34,  8, 10 ]\!]_4 & 0 & 1 & 2 & 17 &             [\![ 51, 25, 10 ]\!]_4 & 0 & \omega^6 & 1 & 51 \\{}
[\![ 34, 10, 9 ]\!]_4 & 0 & 1 & 2 & 17 &              [\![ 51, 27, 9 ]\!]_4 & 0 & 1 & 1 & 51 \\{}
[\![ 34, 14, 8 ]\!]_4 & 0 & 1 & 2 & 17 &              [\![ 51, 33, 7 ]\!]_4 & 0 & \omega^6 & 1 & 51 \\{}
[\![ 35,  3, 12 ]\!]_4 & 0 & 1 & 5 & 7 &              [\![ 51, 37, 6 ]\!]_4 & 0 & \omega^6 & 1 & 51 \\{}
[\![ 35,  9, 10 ]\!]_4 & 1 & 1 & 2 & 17 &             [\![ 52, 12, 14 ]\!]_4 & 1 & \omega^6 & 1 & 51 \\{}
[\![ 35, 11,  9 ]\!]_4 & 2 & 1 & 3 & 11 &             [\![ 52, 16, 13 ]\!]_4 & 1 & \omega^6 & 1 & 51 \\{}
[\![ 36,  4, 12 ]\!]_4 & 2 & 1 & 2 & 17 &             [\![ 52, 18, 12 ]\!]_4 & 1 & \omega^6 & 1 & 51 \\{}
[\![ 36,  12, 9 ]\!]_4 & 2 & 1 & 2 & 17 &             [\![ 52, 20, 11 ]\!]_4 & 1 & \omega^6 & 1 & 51 \\{}
[\![ 37,  7, 11 ]\!]_4 & 2 & 1 & 5 & 7 &              [\![ 52, 20, 11 ]\!]_4 & 1 & \omega^6 & 1 & 51 \\{}
[\![ 38,  2, 14 ]\!]_4 & 0 & 1 & 2 & 19 &             [\![ 52, 24, 10 ]\!]_4 & 1 & \omega^6 & 1 & 51 \\{}
[\![ 39,  1, 14 ]\!]_4 & 1 & 1 & 2 & 19 &             [\![ 52, 26, 9 ]\!]_4 & 1 & 1 & 3 & 17 \\{}
[\![ 39, 19, 8 ]\!]_4  & 1 & 1 & 2 & 19 &             [\![ 52, 30, 8 ]\!]_4 & 1 & \omega^6 & 1 & 51 \\{}
[\![ 42, 12, 11 ]\!]_4 & 0 & 1 & 2 & 21 &             [\![ 52, 34, 7 ]\!]_4 & 1 & \omega^6 & 1 & 51 \\{}
[\![ 43,  1, 15 ]\!]_4 & 1 & 1 & 2 & 21 &             [\![ 52, 36, 6 ]\!]_4 & 1 & \omega^6 & 1 & 51 \\{}
[\![ 43, 15, 10 ]\!]_4 & 1 & 1 & 2 & 21 &             [\![ 53, 21, 11 ]\!]_4 & 2 & \omega^6 & 1 & 51 \\{}
[\![ 44,  0, 16 ]\!]_4 & 2 & 1 & 2 & 21 &             [\![ 53, 25, 10 ]\!]_4 & 2 & \omega^6 & 1 & 51 \\{}
[\![ 44,  2, 15 ]\!]_4 & 0 & 1 & 4 & 11 &             [\![ 55, 27, 9 ]\!]_4 & 4 & 1 & 3 & 17 \\{}
[\![ 44, 22, 8 ]\!]_4 & 2 & \omega^{12} & 6 & 7  &     [\![ 55, 31, 8 ]\!]_4 & 4 & 1 & 3 & 17 \\{}
[\![ 45, 19, 9 ]\!]_4 & 0 & 1 & 5 & 9 &               [\![ 56, 22, 11 ]\!]_4 & 1 & 1 & 5 & 11 \\{}
[\![ 45, 23, 8 ]\!]_4  & 1 & 1 & 4 & 11 &             [\![ 59, 27, 10 ]\!]_4 & 3 & \omega^{12} & 8 & 7 \\{}
[\![ 46,  2, 16 ]\!]_4 & 0 & 1 & 2 & 23 &             [\![ 68, 48, 7 ]\!]_4 & 0 & 1 & 4 & 17 \\{}
[\![ 46,  6, 14 ]\!]_4 & 4 & 1 & 6 & 7 &              [\![ 70, 36, 11 ]\!]_4 & 0 & 1 & 2 & 35 \\{}
[\![ 46, 12, 12 ]\!]_4 & 2 & 1 & 4 & 11 &             [\![ 73, 43, 9 ]\!]_4 & 1 & 1 & 8 & 9 \\{}
[\![ 46, 24, 8 ]\!]_4  & 1 & 1 & 3 & 15 &             [\![ 77, 45, 10 ]\!]_4 & 0 & \omega^{12} & 11 & 7 \\{}
[\![ 46, 20,  9 ]\!]_4 & 1 & 1 & 3 & 15 &             [\![ 77, 47, 9 ]\!]_4 & 0 & 1 & 7 & 11 \\{}
[\![ 46, 26, 7 ]\!]_4 & 4 & \omega^{12} & 6 & 7 &      [\![ 78, 46, 10 ]\!]_4 & 1 & 1 & 7 & 11 \\{}
[\![ 47,  1, 16 ]\!]_4 & 1 & 1 & 2 & 23 &             [\![ 97, 73, 8 ]\!]_4 & 0 & 1 & 1 & 97 \\
\hline
          \end{array}$
        \end{center}
\end{table}

\begin{table}[ht!]
  \caption{Ququint codes from linear codes over $\F_{25}=\F_5(\omega)$
    with $\omega^2=\omega+3$\label{tab:ququint}}
   \def\arraystretch{1.145}
        \begin{center}
	$\begin{array}{|@{\ }l|c|ccc|@{\,}|@{\ }l|c|ccc|}
            \hline
		\multicolumn{1}{|c|}{\text{code}} & e & \lambda & \ell & m &
		\multicolumn{1}{c|}{\text{code}} & e & \lambda & \ell & m\\
                \hline&&&&&&&&&\\[-2.7ex]\hline
[\![ 18,  6, 6 ]\!]_5 & 2 & 1 & 2 & 8  &                [\![ 44, 24, 8 ]\!]_5 & 0 & 1 & 4 & 11 \\{}
[\![ 20,  8, 6 ]\!]_5 & 2 & 1 & 3 & 6 &                 [\![ 45, 13, 12 ]\!]_5 & 1 & -1 & 4 & 11 \\{}
[\![ 22,  2, 9 ]\!]_5 & 0 & 1 & 2 & 11  &               [\![ 45, 21, 9 ]\!]_5 & 1 & \omega^{16} & 2 & 22 \\{}
[\![ 22, 10, 6]\!]_5 & 0 & -1 & 2 & 11  &               [\![ 46,  0, 18 ]\!]_5 & 2 & 1 & 2 & 22\\{}
[\![ 24,  0, 11 ]\!]_5 & 0 & 1 & 2 & 12  &              [\![ 47, 11, 13 ]\!]_5 & 3 & 1 & 4 & 11 \\{}
[\![ 26,  4, 10 ]\!]_5 & 0 & \omega^{20} & 1 & 26  &     [\![ 48, 10, 14 ]\!]_5 & 0 & \omega^8 & 3 & 16 \\{}
[\![ 26,  6, 9 ]\!]_5 & 0 & 1 & 2 & 13  &               [\![ 52, 22, 11 ]\!]_5 & 0 & \omega^{16} & 1 & 52\\{}
[\![ 27,  9, 8]\!]_5 & 1 & \omega^8 & 2 & 13  &         [\![ 52, 28, 9 ]\!]_5 & 0 & \omega^{8} & 4 & 13\\{}
[\![ 28, 12, 7]\!]_5 & 2 & \omega^8 & 2 & 13  &         [\![ 53, 27, 9 ]\!]_5 & 1 & 1 & 4 & 13 \\{}
[\![ 33, 11, 9 ]\!]_5 & 0 & 1 & 3 & 11 &                [\![ 54, 22, 11 ]\!]_5 & 2 & 1 & 4 & 13 \\{}
[\![ 33, 13, 8 ]\!]_5 & 0 & -1 & 3 & 11 &               [\![ 54, 28, 9 ]\!]_5 & 2 & 1 & 4 & 13 \\{}
[\![ 34,  2, 13 ]\!]_5 & 0 & 1 & 2 & 17  &              [\![ 55, 29, 9 ]\!]_5 & 3 & 1 & 4 & 13 \\{}
[\![ 34, 18, 7 ]\!]_5 & 0 & 1 & 2 & 17 &                [\![ 56, 24, 11 ]\!]_5 & 4 & 1 & 4 & 13 \\{}
[\![ 36, 10, 10 ]\!]_5 & 3 & 1 & 3 & 11  &              [\![ 56, 34, 8]\!]_5 & 1 & -1 & 5 & 11 \\{}
[\![ 36, 16, 8 ]\!]_5 & 2 & \omega^{20} & 2 & 17  &      [\![ 58, 28, 11]\!]_5 & 0 & \omega^4 & 1 & 58\\{}
[\![ 37,  1, 15 ]\!]_5 & 0 & 1 & 1 & 37  &              [\![ 59, 31, 9 ]\!]_5 & 4 & 1 & 5 & 11 \\{}
[\![ 38,  0, 16 ]\!]_5 & 1 & -1 & 1 & 37  &             [\![ 66, 44, 8 ]\!]_5 & 0 & 1 & 3 & 22 \\{}
[\![ 38,  2, 14 ]\!]_5 & 0 & 1 & 2 & 19  &              [\![ 69, 35, 11 ]\!]_5 & 4 & 1 & 5 & 13 \\{}
[\![ 39,  7, 12 ]\!]_5 & 0 & 1 & 3 & 13  &              [\![ 69, 41, 9 ]\!]_5 & 4 & 1 & 5 & 13 \\{}
[\![ 39, 11, 11 ]\!]_5 & 0 & 1 & 3 & 13  &              [\![ 69, 45, 8 ]\!]_5 & 4 & \omega^8 & 5 & 13\\{}
[\![ 39, 13,10]\!]_5 & 0 & \omega^8 & 3 & 13  &         [\![ 72, 50, 7 ]\!]_5 & 0 & 1 & 9 & 8 \\{}
[\![ 39, 19, 8]\!]_5 & 0 & \omega^8 & 3 & 13  &         [\![ 74, 50, 8 ]\!]_5 & 2 & 1 & 9 & 8 \\{}
[\![ 40,  6, 13 ]\!]_5 & 1 & 1 & 3 & 13  &              [\![ 77, 57, 7 ]\!]_5 & 0 & 1 & 7 & 11 \\{}
[\![ 40, 14,10]\!]_5 & 1 & \omega^8 & 3 & 13  &         [\![ 78, 50, 9 ]\!]_5 & 0 & 1 & 6 & 13 \\{}
[\![ 40, 16, 9]\!]_5 & 1 & \omega^8 & 3 & 13  &         [\![ 78, 54, 8 ]\!]_5 & 0 & 1 & 6 & 13 \\{}
[\![ 41,  9, 12 ]\!]_5 & 2 & 1 & 3 & 13  &              [\![ 78, 58, 7 ]\!]_5 & 0 & 1 & 6 & 13 \\{}
[\![ 41, 17, 9]\!]_5 & 2 & \omega^8 & 3 & 13  &         [\![ 81, 59, 7]\!]_5 & 3 & \omega^8 & 6 & 13\\{}
[\![ 41, 21, 8]\!]_5 & 0 & 1 & 1 & 41  &                [\![ 82, 54, 9 ]\!]_5 & 4 & 1 & 6 & 13 \\{}
[\![ 42, 16, 10 ]\!]_5 & 3 & 1 & 3 & 13 &               [\![ 82, 60, 7 ]\!]_5 & 4 & \omega^8 & 6 & 13\\{}
[\![ 42, 20, 9]\!]_5 & 1 & 1 & 1 & 41  &                [\![ 88, 68, 7 ]\!]_5 & 0 & 1 & 4 & 22 \\{}
[\![ 43,  5, 14 ]\!]_5 & 4 & 1 & 3 & 13  &              [\![ 89, 65, 8 ]\!]_5 & 1 & 1 & 8 & 11 \\{}
[\![ 43,  7, 13 ]\!]_5 & 4 & 1 & 3 & 13  &              [\![ 91, 63, 9 ]\!]_5 & 0 & 1 & 7 & 13 \\{}
[\![ 44,  0, 16 ]\!]_5 & 0 & 1 & 2 & 22  &              [\![ 91, 67, 8 ]\!]_5 & 0 & 1 & 7 & 13 \\{}
[\![ 44,  2, 16 ]\!]_5 & 0 & 1 & 2 & 22  &              [\![ 91, 69, 7 ]\!]_5 & 0 & 1 & 7 & 13 \\{}
[\![ 44, 12, 12 ]\!]_5 & 0 & 1 & 4 & 11  &              [\![ 95, 65, 9 ]\!]_5 & 4 & 1 & 7 & 13 \\{}
[\![ 44, 14, 11 ]\!]_5 & 0 & -1 & 4 & 11 &              [\![ 100, 78, 7 ]\!]_5 & 1 & 1 & 9 & 11\\
\hline
	  \end{array}$
        \end{center}
\end{table}

\begin{table}[ht!]
  \caption{Qusept codes from linear codes over $\F_{49}=\F_7(\omega)$
    with $\omega^2=\omega+4$\label{tab:qusept}}
   \def\arraystretch{1.145}
        \begin{center}
	$\begin{array}{|@{\ }l|c|ccc|@{\,}|@{\ }l|c|ccc|}
            \hline
		\multicolumn{1}{|c|}{\text{code}} & e & \lambda & \ell & m &
		\multicolumn{1}{c|}{\text{code}} & e & \lambda & \ell & m\\
                \hline&&&&&&&&&\\[-2.7ex]\hline
[\![ 18,  4, 7 ]\!]_7 & 2 & -1 & 2 & 8 &            [\![ 36, 12, 10 ]\!]_7 & 0 & \omega^{36} & 3 & 12 \\{}
[\![ 25,  5, 9 ]\!]_7 & 0 & 1 & 1 & 25 &            [\![ 36, 14, 9 ]\!]_7 & 0 & 1 & 3 & 12 \\{}
[\![ 26,  2, 11 ]\!]_7 & 0 & 1 & 2 & 13 &           [\![ 36, 18, 8 ]\!]_7 & 0 & 1 & 3 & 12 \\{}
[\![ 26,  4, 10 ]\!]_7 & 1 & 1 & 1 & 25 &           [\![ 38,  0, 15 ]\!]_7 & 0 & 1 & 2 & 19 \\{}
[\![27, 9, 8 ]\!]_7 & 2 & \omega^{30} & 5 & 5 &      [\![ 40, 18, 9 ]\!]_7 & 0 & -1 & 2 & 20 \\{}
[\![ 28, 10, 8 ]\!]_7 & 3 & 1 & 5 & 5 &             [\![ 46, 26, 8 ]\!]_7 & 1 & \omega^{30} & 9 & 5 \\{}
[\![ 30, 10, 9 ]\!]_7 & 0 & 1 & 2 & 15 &            [\![ 47, 27, 8 ]\!]_7 & 2 & 1 & 5 & 9 \\{}
[\![ 31, 13, 8 ]\!]_7 & 1 & 1 & 2 & 15 &            [\![ 49, 29, 8 ]\!]_7 & 4 & 1 & 9 & 5 \\{}
[\![ 32, 14, 8 ]\!]_7 & 2 & \omega^{30} & 2 & 15  &  [\![ 50, 14, 14 ]\!]_7 & 0 & 1 & 2 & 25 \\{}
[\![ 33,  9, 10 ]\!]_7 & 3 & 1 & 6 & 5 &            [\![ 50, 18, 12 ]\!]_7 & 0 & 1 & 2 & 25 \\{}
[\![ 34,  2, 14 ]\!]_7 & 0 & 1 & 2 & 17 &           [\![ 50, 22, 11 ]\!]_7 & 0 & 1 & 2 & 25 \\{}
[\![ 34, 10, 10 ]\!]_7 & 4 & \omega^{30} & 2 & 15 &  [\![ 50, 26, 10 ]\!]_7 & 0 & \omega^{12} & 1 & 50\\{}
[\![ 36,  0, 15 ]\!]_7 & 0 & 1 & 2 & 18 &           [\![ 53, 27, 10 ]\!]_7 & 1 & 1 & 4 & 13 \\{}
[\![ 34, 16, 8 ]\!]_7 & 0 & 1 & 2 & 17 &            [\![ 57, 33, 9 ]\!]_7 & 0 & 1 & 3 & 19\\
\hline
	  \end{array}$
        \end{center}
\end{table}

\begin{table}[ht!]
  \caption{Quoct codes from linear codes over $\F_{64}=\F_2(\omega)$
    with $\omega^6=\omega^4+\omega^3+\omega+1$\label{tab:quoct}}
   \def\arraystretch{1.145}
        \begin{center}
	$\begin{array}{|@{\ }l|c|ccc|@{\,}|@{\ }l|c|ccc|}
            \hline
		\multicolumn{1}{|c|}{\text{code}} & e & \lambda & \ell & m &
		\multicolumn{1}{c|}{\text{code}} & e & \lambda & \ell & m\\
                \hline&&&&&&&&&\\[-2.7ex]\hline
[\![ 14, 4, 6 ]\!]_8 & 1 & 1 & 1 & 13 &                     [\![ 53, 29, 9 ]\!]_8 & 3 & 1 & 10 & 5 \\{}
[\![ 20, 6, 7 ]\!]_8 & 0 & 1 & 4 & 5 &                      [\![ 54, 30, 9 ]\!]_8 & 2 & 1 & 4 & 13 \\{}
[\![ 21, 7, 7 ]\!]_8 & 0 & 1 & 1 & 21 &                     [\![ 56, 32, 9 ]\!]_8 & 4 & 1 & 4 & 13 \\{}
[\![ 22, 6, 8 ]\!]_8 & 1 & 1 & 1 & 21 &                     [\![ 58, 30, 10 ]\!]_8 & 0 & \omega^{56} & 2 & 29 \\{}
[\![ 22, 8, 7 ]\!]_8 & 1 & 1 & 1 & 21 &                     [\![ 60, 34, 9 ]\!]_8 & 0 & 1 & 4 & 15 \\{}
[\![ 23, 9, 7 ]\!]_8 & 2 & \omega^{42} & 1 & 21 &            [\![ 62, 40, 9 ]\!]_8 & 1 & \omega^{49} & 1 & 61\\{}
[\![ 26, 2, 11 ]\!]_8 & 0 & 1 & 2 & 13 &                    [\![72, 54, 7 ]\!]_8 & 2 & \omega^{42} & 14 & 5 \\{}
[\![ 27, 5, 10 ]\!]_8 & 1 & 1 & 2 & 13 &                    [\![ 78, 52, 9 ]\!]_8 & 0 & 1 & 6 & 13 \\{}
[\![ 28, 4, 11 ]\!]_8 & 2 & \omega^{14}\!\! & 2 & 13 &       [\![ 78, 60, 7 ]\!]_8 & 0 & 1 & 6 & 13 \\{}
[\![ 28,  8, 9 ]\!]_8 & 2 & \omega^{14} & 2 & 13 &           [\![ 79, 53, 9 ]\!]_8 & 1 & 1 & 6 & 13 \\{}
[\![ 29, 11, 8 ]\!]_8 & 1 & \omega^{35} & 4 & 7 &            [\![ 79, 55, 8 ]\!]_8 & 4 & 1 & 15 & 5 \\{}
[\![ 34, 2, 14 ]\!]_8 & 0 & 1 & 2 & 17 &                    [\![ 80, 56, 8 ]\!]_8 & 5 & 1 & 15 & 5 \\{}
[\![ 36, 12, 10 ]\!]_8 & 1 & \omega^{42} & 7 & 5  &          [\![ 82, 60, 8 ]\!]_8 & 4 & 1 & 6 & 13 \\{}
[\![ 38,12, 11 ]\!]_8 & 1 & \omega^{35}\!\! & 1 & 37 &       [\![ 82, 62, 7 ]\!]_8 & 6 & \omega^{56} & 4 & 19 \\{}
[\![ 39, 15, 10 ]\!]_8 & 0 & \omega^{21} & 1 & 39 &          [\![ 84, 64, 7 ]\!]_8 & 4 & 1 & 16 & 5 \\{}
[\![ 40, 14, 11 ]\!]_8 & 1 & \omega^{14} & 3 & 13 &          [\![ 85, 61, 9 ]\!]_8 & 0 & \omega^{56} & 5 & 17 \\{}
[\![ 40, 16, 10 ]\!]_8 & 1 & \omega^{14} & 3 & 13 &          [\![ 85, 67, 7 ]\!]_8 & 0 & \omega^{56} & 5 & 17 \\{}
[\![ 40, 18, 9 ]\!]_8 & 1 & \omega^{21} & 1 & 39 &           [\![ 86, 68, 7 ]\!]_8 & 1 & \omega^{56} & 5 & 17 \\{}
[\![ 41, 17, 10 ]\!]_8 & 2 & \omega^{14} & 3 & 13 &          [\![ 88, 66, 8 ]\!]_8 & 3 & 1 & 5 & 17 \\{}
[\![ 42, 12, 12 ]\!]_8 & 3 & 1 & 3 & 13 &                   [\![ 89, 69, 7 ]\!]_8 & 4 & \omega^{42} & 17 & 5 \\{}
[\![ 44, 10, 13 ]\!]_8 & 2 & \omega^{42}\!\! & 2 & 21 &      [\![ 91, 67, 8 ]\!]_8 & 0 & 1 & 7 & 13 \\{}
[\![ 44, 14, 12 ]\!]_8 & 2 & 1 & 2 & 21 &                   [\![ 91, 71, 7 ]\!]_8 & 0 & \omega^{14} & 7 & 13 \\{}
[\![ 44, 16, 11 ]\!]_8 & 2 & 1 & 6 & 7 &                    [\![ 92, 68, 8 ]\!]_8 & 0 & 1 & 4 & 23 \\{}
[\![ 45, 19, 10 ]\!]_8 & 0 & \omega^{42} & 3 & 15 &          [\![ 92, 72, 7 ]\!]_8 & 2 & 1 & 18 & 5 \\{}
[\![ 45, 23, 9 ]\!]_8 & 0 & \omega^{42} & 3 & 15 &           [\![ 93, 69, 8 ]\!]_8 & 1 & 1 & 4 & 23 \\{}
[\![ 46, 20, 10 ]\!]_8 & 4 & 1 & 6 & 7 &                    [\![ 94, 70, 8 ]\!]_8 & 4 & 1 & 18 & 5 \\{}
[\![ 48, 20, 11 ]\!]_8 & 3 & 1 & 9 & 5 &                    [\![ 95, 71, 8 ]\!]_8 & 4 & \omega^{14} & 7 & 13 \\{}
[\![ 50, 14, 14 ]\!]_8 & 5 & \omega^{42}\!\! & 9 & 5 &       [\![ 95, 75, 7 ]\!]_8 & 4 & \omega^{14} & 7 & 13 \\{}
[\![ 50, 18, 12 ]\!]_8 & 5 & 1 & 9 & 5 &                    [\![ 96, 76, 7 ]\!]_8 & 1 & 1 & 19 & 5 \\{}
[\![ 50, 24, 10 ]\!]_8 & 0 & 1 & 10 & 5 &                   [\![ 97, 77, 7 ]\!]_8 & 2 & 1 & 19 & 5 \\{}
[\![ 51, 27, 9 ]\!]_8 & 0 & 1 & 3 & 17 &                    [\![ 99, 79, 7 ]\!]_8 & 4 & 1 & 19 & 5 \\{}
[\![ 52, 28, 9 ]\!]_8 & 0 & 1 & 4 & 1 &&&&&\\
\hline
\end{array}$
        \end{center}
\end{table}

\section{Final Remarks}\label{sec:final_remarks}
We have studied in detail the generalization of Quantum Construction X
by Li\v{s}onek and Singh to various inner products relevant for
quantum error-correcting codes. This also led to improved bounds on
the parameters of the resulting quantum codes.  At the same time, we
have characterized self-orthogonal and nearly self-orthogonal quasi-twisted
codes with respect to those inner products.  Using this
characterization, we found many quantum error-correcting codes with
good parameters.

Both the generalized construction for quantum codes and the
characterization of nearly self-orthogonal codes are of independent
interest on their own.

\appendices

\section{Relations Among Inner Products}\label{sec:IP_relations}
In this section, we consider relations among the various inner
products that are used in the construction of quantum stabilizer
codes from classical codes.

Instead of vectors over $\F_q\times\F_q$, one might consider vectors
over $\F_{q^2}$.  Fixing a basis $\mathcal{B}=\{\alpha,\beta\}$ of
$\F_{q^2}/\F_q$, we can identify $\F_q\times\F_q$ and $\F_{q^2}$ via
the $\F_q$-linear isomorphism
\begin{alignat}{5}
  \Phi\colon&& \F_q\times\F_q&{}\longrightarrow \F_{q^2}\\
    && (a|b) &{}\longmapsto a\alpha+b\beta.\label{eq:Phi_vector}
\end{alignat}
This isomorphism extends naturally to vectors, i.e.,
\begin{alignat}{5}
  \Phi\colon&& \F_q^{2n}&{}\longrightarrow \F_{q^2}^n\\
    && (\bm{a}|\bm{b}) &{}\longmapsto\alpha\bm{a}+\beta\bm{b}.\label{eq:map_q2}
\end{alignat}

On $\F_{q^2}^n$, we define the following trace-alternating form
\begin{alignat}{5}
  \langle \bm{v},\bm{w}\rangle_{\rm a}
     :={}& \tr\left(\frac{\langle\bm{v},\bm{w}\rangle_{\rm H}-\langle\bm{w},\bm{v}\rangle_{\rm H}}{\alpha\beta^q-\alpha^q\beta}\right)\\
      ={}& \tr\left(\frac{\bm{v}\cdot\bm{w}^q-\bm{v}^q\cdot\bm{w}}{\alpha\beta^q-\alpha^q\beta}\right),\label{eq:trace-alternating}
\end{alignat}
where $\bm{x}\cdot\bm{y}=\langle\bm{x},\bm{u}\rangle_{\rm E}$ denotes
the Euclidean inner product, $\bm{v}^q=(v_1^q,\ldots,v_n^q)$, and
$\tr$ denotes the absolute trace from $\F_q$ to $\F_p$. First note
that the argument of the trace is indeed an element of $\F_q$, as it
is invariant under the Galois automorphism $x\mapsto x^q$.  Moreover,
the denominator $\alpha\beta^q-\alpha^q\beta$ is non-zero. Otherwise,
assume that $\alpha\beta^q=\alpha^q\beta$. Both $\alpha$ and $\beta$
are non-zero. Then
$\alpha/\beta=\alpha^q/\beta^q=(\alpha/\beta)^q\in\F_q$, in
contradiction to the assumption that $\{\alpha,\beta\}$ form a basis.

The following lemma is similar to \cite[Lemma 14]{Ketkar}, in the
spirit of the remark at the end of \cite[Section IV.A]{Ketkar}.
\begin{lemma}
  For vectors $\bm{c},\bm{d}\in\F_q^{2n}$,
  \begin{alignat}{5}
    \langle\bm{c},\bm{d}\rangle_{\rm T}=\langle \Phi(\bm{c}),\Phi(\bm{d})\rangle_{\rm a},\label{eq:T_IP_alternating}
  \end{alignat}
  where $\langle\bm{c},\bm{d}\rangle_{\rm T}$ is the trace-symplectic
  inner product defined in \eqref{eq:tracesymplectic_IP2}.
\end{lemma}
\begin{IEEEproof}
  Let $\bm{c}=(\bm{a},\bm{b})$ and $\bm{d}=(\bm{a}',\bm{b}')$. We
  calculate
  \begin{alignat*}{5}
    \Phi(&\bm{c})\cdot\Phi(\bm{d})^q
      =(\alpha\bm{a}+\beta\bm{b})\cdot(\alpha\bm{a}'+\beta\bm{b}')^q\\
      &{} =\alpha^{q+1}\bm{a}\cdot\bm{a}'+\alpha\beta^q\bm{a}\cdot\bm{b}'
         +\alpha^q\beta\bm{b}\cdot\bm{a}'+\beta^{q+1}\bm{b}\cdot\bm{b}'\\
    \Phi(&\bm{c})^q\cdot\Phi(\bm{d})
       =(\alpha\bm{a}+\beta\bm{b})^q\cdot(\alpha\bm{a}'+\beta\bm{b}')\\
       &{} =\alpha^{q+1}\bm{a}\cdot\bm{a}'+\alpha^q\beta\bm{a}\cdot\bm{b}'
         +\alpha\beta^q\bm{b}\cdot\bm{a}'+\beta^{q+1}\bm{b}\cdot\bm{b}'
  \end{alignat*}
  and hence
  \begin{alignat}{5}
    \Phi(\bm{c})\cdot\Phi(&\bm{d})^q-\Phi(\bm{c})^q\cdot\Phi(\bm{d})\nonumber\\
       &{}=(\alpha\beta^q-\alpha^q\beta)\bm{a}\cdot\bm{b}'+(\alpha^q\beta-\alpha\beta^q)\bm{b}\cdot\bm{a}'\nonumber\\
       &{}=(\alpha\beta^q-\alpha^q\beta)(\bm{a}\cdot\bm{b}'-\bm{b}\cdot\bm{a}')\\
       &{}=(\alpha\beta^q-\alpha^q\beta)\bigl\langle(\bm{a}|\bm{b}),(\bm{a}'|\bm{b}')\bigr\rangle_{\rm S}.\label{eq:lemma_alt_form}
  \end{alignat}
  Dividing by $\alpha\beta^q-\alpha^q\beta$ and taking the trace,
  \eqref{eq:T_IP_alternating} follows.
\end{IEEEproof}

Clearly, two vectors $\bm{c},\bm{d}\in\F_{q^2}^n$ which are orthogonal
with respect to the Hermitian inner product \eqref{eq:Hermitian_IP},
are also orthogonal with respect to the trace-alternating form
\eqref{eq:trace-alternating}, independent of the chosen basis
$\mathcal{B}=\{\alpha,\beta\}$. The map $\Phi$ allows us to use
vectors over $\F_{q^2}$ and the Hamming weight instead of vectors over
$\F_q\times\F_q$ and the symplectic weight \eqref{eq:swt}. As also
shown in \cite[Lemma 18]{Ketkar}, for $\F_{q^2}$-linear codes the
various notions of duality coincide, \ie,
\begin{alignat}{5}\label{eq:dualities}
    C^\perpT = C^\perpS = \bigl(\Phi(C)\bigr)^\perpH.
\end{alignat}

In terms of duality, we can also reduce the trace-symplectic inner
product to the symplectic inner product over prime fields. Let
$q=p^m$, $p$ prime, and let $\mathcal{B}=\{\alpha_1,\ldots,\alpha_m\}$
be a basis for $\F_{p^m}/\F_p$. The dual basis
$\mathcal{B^\perp}=\{\beta_1,\ldots,\beta_m\}$ is defined via the
condition (see, \eg, \cite[Definition 2.30]{LidlNiederreiter})
\begin{alignat}{5}
  \tr(\alpha_i\beta_j)=\delta_{i,j},
\end{alignat}
where $\tr$ denotes the absolute trace of $\F_{p^m}/\F_p$ and
$\delta_{i,j}$ is the Kronecker delta. For $(a|b)\in\F_q\times\F_q$,
we define the following $\F_p$-linear map
\begin{alignat}{5}
  \PsiB\colon&& \F_{p^m}\times\F_{p^m}&{}\longrightarrow \F_p^m\times\F_p^m\label{eq:add_expansion}\\
    && (a|b) &{}\longmapsto (a_1,\ldots,a_m|b_1,\ldots,b_m),
\end{alignat}
where
\begin{alignat}{5}
  a={}\sum_{j=1}^m a_j\alpha_j,\quad b={}\sum_{i=1}^m
  b_j\beta_j,\qquad\text{with $a_j,b_j\in\F_p$}.
\end{alignat}
The map naturally extends to vectors, and we get the following.
\begin{lemma}\label{lemma:prime_field_IP}
  For vectors $\bm{c},\bm{d}\in\F_{p^m}^{2n}$,
  \begin{alignat}{5}
    \langle\bm{c},\bm{d}\rangle_{\rm T} =
    \langle\PsiB(\bm{c}),\PsiB(\bm{d})\rangle_{\rm S},
  \end{alignat}
  where $\PsiB$ is defined in \eqref{eq:add_expansion}.
\end{lemma}
\begin{IEEEproof}
  Let $\bm{c}=(\bm{a},\bm{b})$ and $\bm{d}=(\bm{a}',\bm{b}')$ with
  $\bm{a},\bm{b},\bm{a}',\bm{b}'\in\F_{p^m}^n$. We calculate
  \begin{alignat}{5}
    \langle\bm{c},\bm{d}\rangle_{\rm T}
      ={}& \sum_{i=0}^{n-1}\tr(a_ib_i'-a_i'b_i)\\
      ={}& \sum_{i=0}^{n-1}\tr\left(\sum_{j=1}^m a_{ij}\alpha_j\sum_{\ell=1}^m b'_{i\ell}\beta_j\right)\nonumber\\
         &  \quad-\sum_{i=0}^{n-1}\tr\left(\sum_{j=1}^m a'_{ij}\alpha_j\sum_{\ell=1}^m b_{i\ell}\beta_j\right)\\
      ={}& \sum_{i=0}^{n-1}\sum_{j=1}^m a_{ij} b'_{ij} - a_{ij}' b_{ij}\\
      ={}& \langle\PsiB(\bm{c}),\PsiB(\bm{d})\rangle_{\rm S}.
  \end{alignat}
\end{IEEEproof}

\section{Gram-Schmidt Type Orthonormalization}\label{sec:Gram-Schmidt}

In this section, we present Gram-Schmidt type algorithms that allow
us to compute suitable bases of vector spaces with respect to certain
inner products.

First, we present an efficient algorithm for the Hermitian inner product.
\begin{lemma}\label{lemma:HermitianGramSchmidt}
Let $1\leq e \leq n$ be an integer and let $V$ be an $\F_{q^2}$-linear
subspace of $\F_{q^2}^n$ with $V\cap V^\perpH=\{\bm{0}\}$. Let
$\{\bm{x}_1,\ldots,\bm{x}_e\}$ be a basis of $V$ over
$\F_{q^2}$. Then, there exists an efficient algorithm to find a
Hermitian orthonormal basis $\{\bm{z}_1,\ldots,\bm{z}_e\}$ of
$V$ over $\F_{q^2}$ such that, for  $1\leq i, j \leq e$,
\begin{equation}\label{GSH}
  \langle \bm{z}_i, \bm{z}_j \rangle_{\rm H}
  = \begin{cases}
    1, &\text{if $i = j$;}\\
    0, &\text{if $i\neq j$.}
  \end{cases}
\end{equation}
\end{lemma}
\begin{IEEEproof}
Assume first that $e=1$. Then $\langle\bm{x}_1,\bm{x}_1\rangle_{\rm
  H}\neq 0$ as $\bm{x}_1\notin V\cap V^\perpH$. Note that $\langle
\bm{x}_1,\bm{x}_1 \rangle_{\rm H}\in \F_q^*$. Choose
$\alpha\in\F_{q^2}^*$ such that $\langle\bm{x}_1,\bm{x}_1\rangle_{\rm
  H}=\alpha^{q+1}$. Putting $\bm{z}_1=\frac{1}{\alpha}\bm{x}_1$ yields
$\langle\bm{z}_1,\bm{z}_1\rangle_{\rm H}=1$.

Next, assume that $e>1$. If there exists an $i$ in the range $1\le
i\le e$ such that $\langle\bm{x}_i,\bm{x}_i \rangle_{\rm H}\neq 0$,
then the arguments in the previous paragraph allow us to define
$\bm{z}_1\in V$ such that $\langle\bm{z}_1,\bm{z}_1\rangle_{\rm H}=
1$. Relabeling the indices of the basis, we can assume without loss of
generality that $i=1$.

Otherwise, assume that $\langle\bm{x}_i,\bm{x}_i\rangle_{\rm H}=0$ for
$1\le i\le e$. Then there exists an integer $2\leq i \leq e$ such that
$\langle\bm{x}_1,\bm{x}_i \rangle_{\rm H}\neq 0$. Indeed, otherwise we
would have $\bm{x}_1\in V\cap V^\perpH$, which is a contradiction. Let
$2\leq i' \leq e$ be the smallest integer such that
$\langle\bm{x}_1,\bm{x}_{i'} \rangle_{\rm H}=c\neq 0$.  For
$a\in\F_{q^2}$, we observe that
\begin{equation}
\langle\bm{x}_1+a\bm{x}_{i'},\bm{x}_1+a\bm{x}_{i'}\rangle_{\rm H}=a^q c + a c^q.\label{eq:poly_a}
\end{equation}
This is a polynomial in $a$ of degree $q$ since $c\neq 0$.  Hence, it
has at most $q$ roots in $\F_{q^2}$. In particular, there exists
$a\in\F_{q^2}$ such that the right hand side of \eqref{eq:poly_a} is
non-zero. Using such an $a\in\F_{q^2}$, let
$\bm{y}_1=\bm{x}_1+a\bm{x}_{i'}$ and $\beta\in\F_{q^2}^*$ such that
$\beta^{q+1}=\langle\bm{y}_1,\bm{y}_1 \rangle_{\rm H}$. Then,
$\bm{z}_1=\frac{1}{\beta}\bm{y}_1$ satisfies the condition
$\langle\bm{z}_1,\bm{z}_1\rangle_{\rm H}= 1$.

Given $\bm{z}_1$ with $\langle\bm{z}_1,\bm{z}_1\rangle_{\rm H}= 1$,
for $2\leq i\leq e$, let
\[
\bm{y}_i = \bm{x}_i - \langle\bm{x}_i,\bm{z}_1\rangle_{\rm H}\,\bm{z}_1.
\]
Note that
\begin{alignat*}{5}
\langle\bm{y}_i,\bm{z}_1 \rangle_{\rm H}
 &{}=\bigl\langle\bm{x}_i-\langle\bm{x}_i,\bm{z}_1 \rangle_{\rm H}\,\bm{z}_1, \bm{z}_1 \bigr\rangle_{\rm H}\nonumber\\
 &{}=\langle\bm{x}_i,\bm{z}_1\rangle_{\rm H} - \langle\bm{x}_i,\bm{z}_1\rangle_{\rm H}\,\langle\bm{z}_i,\bm{z}_1 \rangle_{\rm H} =0,
\end{alignat*}
for all $2\leq i\leq e$. Let
$\widetilde{V}=\Span_{\F_{q^2}}\{\bm{y}_2,\ldots,\bm{y}_e\}$. We conclude that
$V=\Span_{\F_{q^2}}\{\bm{z}_1\}\oplus \widetilde{V}$ and $\widetilde{V}\cap \widetilde{V}^\perpH=\{\bm{0}\}$
with $\dim_{\F_{q^2}}(\widetilde{V})=e-1$. Applying the same method on $\widetilde{V}$
recursively, we obtain an efficient algorithm giving a Hermitian
orthonormal $\F_{q^2}$-basis $\{\bm{z}_1,\ldots,\bm{z}_e\}$ of $V$ with the
property \eqref{GSH}.
\end{IEEEproof}

Next, we present an efficient algorithm for the symplectic inner product.
\begin{lemma}\label{lemma:SymplecticGramSchmidt}
Let $e$ be an integer such that $2\leq 2e \leq 2n$ and let $V$ be an
$\F_{q}$-linear subspace of $\F_{q}^{2n}$ with $V\cap
V^\perpS=\{\bm{0}\}$. Then the dimension of $V$ is even.  Let
$\{\bm{x}_1\ldots,\bm{x}_{2e}\}$ be a basis of $V$ over $\F_{q}$. Then
there exists an efficient algorithm to find a basis
$\{\bm{z}_0,\ldots,\bm{z}_{2e-1}\}$ of $V$ over $\F_{q}$ such that
\begin{equation}\label{eq:GSS}
\langle\bm{z}_{2i_1+i_0},\bm{z}_{2j_1+j_0}\rangle_{\rm S}
= \begin{cases}
   \hfil 0, &\text{if $i_1\neq j_1$;}\\
   \hfil 0, &\text{if $i_1=j_1$, $i_0=j_0$;}\\
   \hfil 1, &\text{if $i_1=j_1$, $i_0=0$, $j_0=1$;}\\
   -1, &\text{if $i_1=j_1$, $i_0=1$, $j_0=0$,}
  \end{cases}
\end{equation}
where $0\leq i_1, j_1 \leq e-1$, $0\le i_0,j_0\le 1$. We refer to the
vectors $\{\bm{z}_{2i_1},\bm{z}_{2i_1+1}\}$ as \emph{symplectic pairs}.
\end{lemma}
\begin{IEEEproof}
  Let $\{\bm{x}_1,\ldots,\bm{x}_\ell\}$ be a basis of $V$.  Then there
  exists an integer $2\leq i \le \ell$ such that
  $\langle\bm{x}_1,\bm{x}_i\rangle_{\rm S}\neq 0$. Indeed, otherwise
  $\bm{x}_1\in V\cap V^\perpS$, which is a contradiction.  This also
  implies that the dimension of $V$ is at least two.  The recursive
  algorithm below shows that the dimension $\ell=2e$ is even.

  Assume first that $e=1$. Then $\langle\bm{x}_1,\bm{x}_2\rangle_{\rm
    S}\neq 0$ as $\bm{x}_1\notin V\cap V^\perpS$.  Let $\alpha =
  \langle\bm{x}_1,\bm{x}_2\rangle_{\rm S} \in \F_q^*$,
  $\bm{z_0}=\bm{x}_1$ and $\bm{z}_1=\frac{1}{\alpha}\bm{x}_2$. Note
  that the symplectic pair $\{\bm{z_0},\bm{z}_1 \}$ is a basis of $V$
  satisfying property \eqref{eq:GSS} when $e=1$.

  Next, assume that $e>2$.  By relabeling the vectors $\bm{x}_i$,
  without loss of generality we can assume that
  $\langle\bm{x}_1,\bm{x}_2\rangle_{\rm S}=\beta\neq 0$. Let
  $\bm{z}_0=\bm{x}_1$ and $\bm{z}_1=\frac{1}{\beta}\bm{x}_2$. Note
  that $\langle\bm{z}_0,\bm{z}_1\rangle_{\rm S}=1$. Then,
  $\{\bm{z}_0,\bm{z}_1\} \cup \{\bm{x}_3,\ldots,\bm{x}_{2e}\}$ is a
  basis of $V$ over $\F_q$.

  For $3\leq i\leq 2e$, define
  \[
  \bm{y}_i = \bm{x}_i - \langle\bm{x}_i,\bm{z}_1\rangle_{\rm S}\,\bm{z}_0
               + \langle\bm{x}_i,\bm{z}_0\rangle_{\rm S}\,\bm{z}_1.
  \]
  Note that
  \[
    \langle\bm{y}_i,\bm{z}_0\rangle_{\rm S} =0
      \quad\text{and}\quad
    \langle\bm{y}_i,\bm{z}_1\rangle_{\rm S} =0
  \]
  for all $3\leq i\leq 2e$. Let
  $\widetilde{V}=\Span_{\F_{q}}\{\bm{y}_3,\ldots,\bm{y}_{2e}\}$. We conclude
  that $V=\Span_{\F_{q}}\{\bm{z}_0,\bm{z}_1\}\oplus \widetilde{V}$ and $\widetilde{V}\cap
  \widetilde{V}^\perpS=\{\bm{0}\}$ with $\dim_{\F_{q}}(\widetilde{V})=2e-2$. Applying the
  same method on $\widetilde{V}$ recursively, we obtain an efficient algorithm
  giving an $\F_{q}$-basis $\{\bm{z}_0\ldots,\bm{z}_{2e-1}\}$ of $V$
  with property \eqref{eq:GSS}.
\end{IEEEproof}

\section*{Acknowledgments}
M. Grassl acknowledges support by the Foundation for Polish Science
(IRAP project, ICTQT, contracts no. 2018/MAB/5 and 2018/MAB/5/AS-1,
co-financed by EU within the Smart Growth Operational Programme; IRA
Programme, project no. FENG.02.01-IP.05-0006/23, financed by the FENG
program 2021-2027, Priority FENG.02, Measure FENG.02.01). He also
acknowledges the hospitality during his visit in Ankara, supported by
the T\"UB\.{I}TAK 2221 Program.
M. F. Ezerman and S. Ling are supported by Nanyang Technological
University Research Grant Number 04INS000047C230GRT01.


\begin{IEEEbiographynophoto}{Martianus Frederic Ezerman}
  received the double B.A. in philosophy and B.Sc. in mathematics
  degrees in 2005, and the M.Sc. degree in mathematics in 2007, all
  from Ateneo de Manila University, Philippines. He obtained the
  Ph.D. degree in mathematical sciences from Nanyang Technological
  University (NTU), Singapore, in 2011.

  He is currently an Adjunct Assistant Professor at NTU and serves as
  the director of a company that develops core cryptographic products
  in trusted execution environments. His research interests include
  coding theory, cryptography, holographic data representations, and
  quantum information theory.
\end{IEEEbiographynophoto}

\begin{IEEEbiographynophoto}{Markus Grassl}
  received his diploma degree in Computer Science in 1994 and his
  doctoral degree in 2001, both from the Fakult\"at f\"ur Informatik,
  Universit\"at Karlsruhe (TH), Germany. From 1994 to 2007 he was a
  member of the Institut f\"ur Algorithmen und Kognitive Systeme,
  Fakult\"at f\"ur Informatik, Universit\"at Karlsruhe (TH), Germany.

  From 2007 to 2008 he was with the Institute for Quantum Optics and
  Quantum Information of the Austrian Academy of Sciences in
  Innsbruck. From 2009 to 2014, he was a Senior Research Fellow at the
  Centre for Quantum Technologies at the National University of
  Singapore. In 2014, he joined the Friedrich-Alexander-Universit\"at
  Erlangen-N\"urnberg and the Max Planck Institute for the Science of
  Light (MPL), Erlangen. Since 2019, he has been a Senior Scientist at
  the International Centre for Theory of Quantum Technologies,
  University of Gdansk.

  His research interests include quantum computation, focusing on
  quantum error-correcting codes, and methods of computer algebra in
  algebraic coding theory.  He maintains tables of good block quantum
  error-correcting codes as well as good linear block codes.

  Dr. Grassl served as Associate Editor for Quantum Information Theory
  of the IEEE Transactions on Information Theory from 2015 till 2017.
\end{IEEEbiographynophoto}

\begin{IEEEbiographynophoto}{San Ling}
  received the B.A. degree in mathematics from the University of
  Cambridge in 1985, and the Ph.D. degree in mathematics from the University of
  California, Berkeley in 1990.

  He is currently President's Chair in Mathematical Sciences, at the
  School of Physical and Mathematical Sciences, Nanyang Technological
  University, Singapore, which he joined in April 2005. Prior to that,
  he was with the Department of Mathematics, National University of
  Singapore. Since August 2022, he also holds the concurrent
  appointment as the Chief Scientific Advisor of the National Research
  Foundation, Singapore.

  His research fields include: arithmetic of modular curves and
  application of number theory to combinatorial designs, coding
  theory, cryptography and sequences.
\end{IEEEbiographynophoto}

\begin{IEEEbiographynophoto}{Ferruh \"Ozbudak}
  received the B.S. degree in electrical and electronics engineering
  and the Ph.D. degree in mathematics from Bilkent University, Ankara,
  Turkey, in 1993 and 1997, respectively.  Since 2006, he has been a
  professor of mathematics.  Currently, he is affiliated with
  Sabanc{\i} University, Istanbul, and Middle East Technical
  University, Ankara. His research interests include algebraic
  curves, codes, sequences, cryptography, finite fields, and finite
  rings.
\end{IEEEbiographynophoto}

\begin{IEEEbiographynophoto}{Buket \"Ozkaya}
  received the B.S. degree in mathematics from Bo{\u{g}}azi{\c{c}}i
  University, Istanbul, in 2006, the M.S. degree in mathematics from
  Georg-August-Universit\"at G\"ottingen, Germany, in 2009, and the
  Ph.D. degree from Sabanc{\i} University, Istanbul, in 2014, under
  the supervision of Cem G\"uneri.

  She held several post-doctoral positions at Middle East Technical
  University, Ankara, at Tel\'ecom ParisTech, France, at Sabanc{\i}
  University, Istanbul, and at Nanyang Technological University,
  Singapore. Since 2023, she has been an Assistant Professor at the
  Institute of Applied Mathematics, Middle East Technical University,
  Ankara. She is interested in algebraic coding theory and its
  applications.
\end{IEEEbiographynophoto}

\vfill
\phantom{.}

\end{document}